\documentclass[journal]{IEEEtran}
\usepackage{amsmath}
\usepackage[table,xcdraw]{xcolor}
\usepackage{subfig}
\usepackage[T1]{fontenc}
\usepackage[utf8]{inputenc}
\usepackage[english]{babel}
\usepackage{xcolor}
\usepackage[normalem]{ulem}
\usepackage{multirow}
\usepackage{tabu}
\ifCLASSINFOpdf
 \usepackage[pdftex]{graphicx}
   \graphicspath{{../}}
  \DeclareGraphicsExtensions{.pdf,.jpeg,.png}
\else
   \usepackage[dvips]{graphicx}
  \DeclareGraphicsExtensions{.eps}
\fi
\usepackage{epstopdf}

\usepackage{url}
\begin{document}
\title{Suitability Analysis of Holographic vs Light Field and 2D Displays for Subjective Quality Assessment of Fourier Holograms}	
\author{Ayyoub~Ahar,~\IEEEmembership{Member,~IEEE,}
        Maksymilian~Chlipala,
        Tobias~Birnbaum,~\IEEEmembership{Member,~IEEE,}
        Weronika~Zaperty,
        Athanasia~Symeonidou,~\IEEEmembership{Member,~IEEE,}
        Tomasz~Kozacki,
        Malgorzata~Kujawinska,
        and~Peter~Schelkens,~\IEEEmembership{Member,~IEEE}
\thanks{A. Ahar, T.~Birnbaum, A.~Symeonidou and P.~Schelkens are with the Dept. of Electronics and Informatics (ETRO), 
	Vrije Universiteit Brussel (VUB),Pleinlaan 2, B-1050 Brussels, Belgium and imec, Kapeldreef 75, B-3001 Leuven, Belgium
	 e-mail: ayyoub.ahar@vub.be}
\thanks{M.~Chlipala, W.~Zaperty, T.~Kozacki and M.~Kujawinska are with Warsaw University of Technology, Institute of Micromechanics and Photonics,
	 8 Sw. A.~Boboli St., 02-525 Warsaw, Poland.}
\thanks{}}

%
\markboth{Submitted to the Journal of IEEE Transactions On Multimedia,20~September~2019}%
{Shell \MakeLowercase{\textit{et al.}}: Bare Demo of IEEEtran.cls for IEEE Journals}
%
\maketitle
\begin{abstract}
	Visual quality assessment of digital holograms is facing many challenges. Main difficulties are related to the limited spatial resolution and angular field of view of holographic displays in combination with the complexity of steering and operating them for such tasks. Alternatively, non-holographic displays -- and in particular light-field displays -- can be utilized to visualize the numerically reconstructed content of a digital hologram. However, their suitability as alternative for holographic displays has not been validated. In this research, we have investigated this issue via a set of comprehensive experiments. We used Fourier holographic principle to acquire a diverse set of holograms, which were either computer-generated from point clouds or optically recorded from real macroscopic objects. A final public data set comprising 96 holograms was created using three compression methods which encoded the holograms at four bit-depths. Three separate subjective-tests were conducted using a holographic display, a light field display and a 2D display. For these subjective experiments, a double stimulus, multi-perspective, multi-depth subjective testing methodology was designed and implemented. 
	The tests show that the non-holographic displays indicate a higher sensitivity to artifacts than the holographic display, though at the same time it is demonstrated they are highly correlated. This indicates that the numerically reconstructed holograms rendered on a light field or 2D display have a high predictive value for the perceived quality on holographic display.
\end{abstract}
\IEEEpeerreviewmaketitle
\section{Introduction} 
\label{sec:Intro}
\IEEEPARstart{D}{igital} holography, in theory, is considered to be the holy grail of 3D imaging solutions~\cite{lee2013three}. While the concept and its initial realizations has been around for almost half a century, only recently it is gaining again interest for 3D visualization. This is due to steady growth of available computational power and significant improvements in nano-electronics, optical hardware and photonics technologies. However, quite a few hardware and signal processing challenges are yet to be addressed in order to facilitate an immersive 3D experience via a complete pipeline for high-quality dynamic holography with full-parallax and wide field of view (FoV)~\cite{blinder_signal_2018}.

In this regard, one of the core challenges is modeling the perceived visual quality of the rendered holograms, which has a vital impact on steering the other components of the holographic imaging pipeline. While the design of highly efficient numerical methods in Computer-Generated Holography (CGH)~\cite{Symeonidou:15,park_recent_2017,pan_review_2016, sugie_high-performance_2018, shimobaba_tomoyoshi_computer_2019} and efficient encoders for holographic content \cite{blinder_signal_2018,el_rhammad_color_2018, peixeiro_holographic_2018, bernardo_holographic_2018, schelkens_jpeg_2019} is gaining momentum, Visual Quality Assessment (VQA) of holograms has a rather long way to reach its primary milestones due to various open problems along the way \cite{blinder_signal_2018, schelkens_jpeg_2019}. Indeed, conducting a systematic subjective test and creating a scored database from a diverse set of holograms is the very first step. But this by itself reveals to be a challenging task. Not only there is no widely-accepted standard methodology for plenoptic content and especially for holographic data, but also holographic displays with acceptable visual characteristics are still scarce. Moreover, configuring and operating such displays requires advanced technical skills. Often, researchers have been rendering numerically reconstructed holograms on non-holographic displays, including regular 2D displays or more recently multi-view light-field displays, to alleviate this problem \cite{ahar_subjective_2015, symeonidou_three-dimensional_2016, Symeonidou:18}. However, potential perceptual differences between visualization on holographic displays and numerical reconstructions rendered on non-holographic displays to our knowledge have not been investigated thoroughly before. Some of the most evident issues include loss of visual cues related to the depth perception on 2D displays and light-field displays, FoV and appearance of different types of speckle noise.

These issues are inter-connected with the chosen display for visualizing the holographic content. For example for a 2D display, only a specific focus plane and perspective of the hologram can be rendered. For a light-field display, for each view the hologram needs to be reconstructed for a particular focus plane utilizing a suitable aperture. As such, only a section of the 3D scene volume described by the hologram is rendered properly. Nevertheless, both displays can support high spatial resolution and large display sizes. On the other hand, holographic displays can render the complete plenoptic scene, but their resolution and overall size are currently limited, which in practice results in only a tiny viewing window(VW) to explore the visualized hologram. These fundamentally diverse properties require different strategies and procedures per display type to conduct a subjective test.

The main objectives and novelties of this manuscript include:
\begin{enumerate}
\item Comparing holographic versus non-holographic displays based on the visual appearance of same set of holograms;
\item Designing and implementing a test methodology for subjective testing of holograms;
\item Creating the first publicly available database of optically recorded and computer-generated holograms annotated with subjective test results;
\item Evaluation of computer-generated against optically recorded Fourier holograms.

\end{enumerate}

In section~\ref{sec:Disp}, the holographic, light field and regular 2D display are described that are being compared in this test to assess their suitability to evaluate the visual quality of holograms. 
The details about the numerical and optical methods to produce the holograms used in this experiment are provided in section~\ref{sec:data} as well as the content preparation. 
Section~\ref{sec:Method} introduces the subjective test methodology including its details for each setup and training of the test subjects.   
In section~\ref{sec:experimental-results}, we explain the statistical post-processing of the experimental results and provide the analysis and discussion of the outcomes. 
Finally, section~\ref{sec:conclusion} presents the concluding remarks.

\section{Display systems }
\label{sec:Disp}


\subsection{Holographic display}
\label{sec:HoloDisp}

\begin{figure} 
	\centering
	\includegraphics[width=8.2cm]{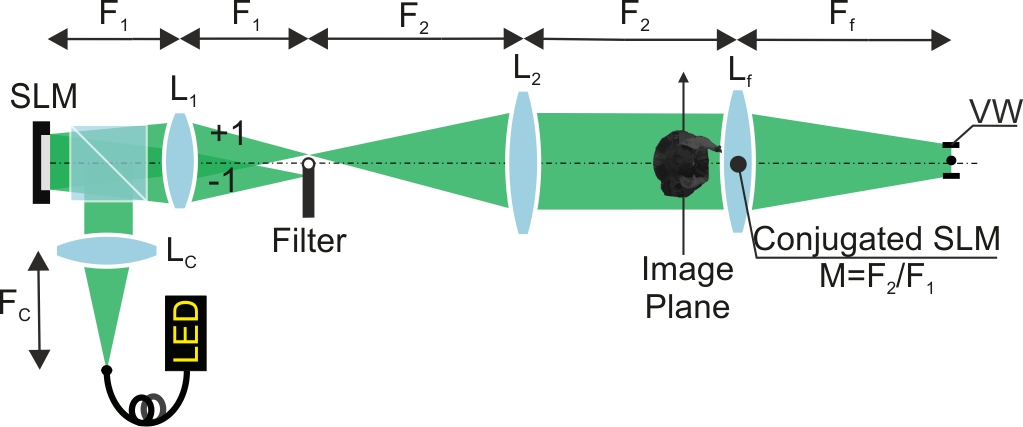} 
	\caption{Fourier holographic display setup.}
	\label{fig:Fourier_disp}
\end{figure}

In this work, a Fourier holographic display with an incoherent LED source is employed. The system provides high-quality orthoscopic reconstructions of large objects~\cite{Kozacki2018color}, which can be viewed with a naked eye. Also, it facilitates a stable performance through very deep scenes~\cite{KozackiSpeckle}. The display setup is presented in the Fig.~\ref{fig:Fourier_disp}. In this system, a phase-only spatial light modulator (SLM) (Holoeye 1080P, 1920$\times$1080 pixels, pixel pitch 8~$\mu$m) is illuminated by a normal plane wave, which is formed by an LED source (Doric Lenses, center wavelength $\lambda_G$ = 515 nm and fiber core of 960~$\mu$m) and a collimating lens L$_C$ (F$_C$ = 400 mm). The SLM is conFigd to display the object wave with removed spherical phase factor. 
Next, the reflected beam passes through the imaging module, which introduces a magnification and facilitates the complex wave coding. The first imaging element is realized by a 4F afocal imaging system composed of the lenses L$_1$ (F$_1$ = 100 mm) and L$_2$ (F$_2$ = 600 mm) with magnification ratio $M$ = -6. The 4F system and the field lens $L_f$ conjugate the SLM plane with a 3D hologram reconstruction volume focused on the VW. The complex coding scheme is experimentally supported with the absorbing cut-off filter in the Fourier plane of the 4F system \cite{li20133d}. 

In this experiment, all the optical components on the optical table were covered using black colored barriers such that no environmental light would enter the black box, i.e. the display setup. A small slit was carved into the box and a metal chinrest and forehead holder were put in front of the slit such that all subjects could easily observe the displayed holograms as soon as they would position their head accordingly. 

The Fourier holography enables reconstruction of a 1:1 orthoscopic copy of the 3D object with no visible distortions for the holographic display~\cite{FourierOE} described above. The full object is viewed by the naked eye and objects with a maximum size of 107 mm can be observed from a distance of 700 mm. With the available Space Bandwidth Product (SBP)~\cite{lohmann_spacebandwidth_1996,goodman_introduction_2004} of the SLM this results in an angular FoV = 8.8\textordmasculine~and an angular resolution of display which is comparable to the resolution of the human eye for dark observation conditions. The 2D and light field displays discussed below are based on 2D reconstructions of a single or multiple views, respectively. Imaging on both displays benefits from the convention of Fourier holography as well since it provides full use of the SBP and thereby achieves the highest quality during the recording/generation process. 

\subsection{2D display} 
\label{sec:2DDisp}

The issued 2D display is a professional Eizo CG318-4K monitor with 4K UHD resolution (3840$\times$2160 pixels) and 10-bit color depth, which is recommended for use in visual test laboratories~\cite{eizo}. The color representation mode was set to ITU-R BT.709-6. The monitor was calibrated using the build-in sensor on the monitor, operated by the ColorNavigator-7 Color Management Software. The calibration was done according to the following profile: sRGB Gamut, D65 white point, 120 cd/$m^2$ brightness, and minimum black level of 0.2 cd/$m^2$. On this display, numerical reconstructions in the object plane were rendered for a particular reconstruction distance and perspective.

\subsection{Light field display}
\label{sec:LFDisp}

The light field display system is a HoloVizio-722RC by  Holografika~\cite{holografika}. This is a 72 inch display having an horizontal angular FoV of 70\textordmasculine~with a total 3D resolution of 73 Mpixel. It provides a 2D equivalent resolution of 1280$\times$768 pixels for each of the 72 views. It provides a 24-bit RGB color system with a brightness of $\approx$1000 cd/$m^{2}$. Holograms are rendered on this display by calculating numerically the reconstructions for a particular reconstruction distance for each view supported by the display. 

\begin{figure*} 
	\centering
	\includegraphics[width=0.95\textwidth]{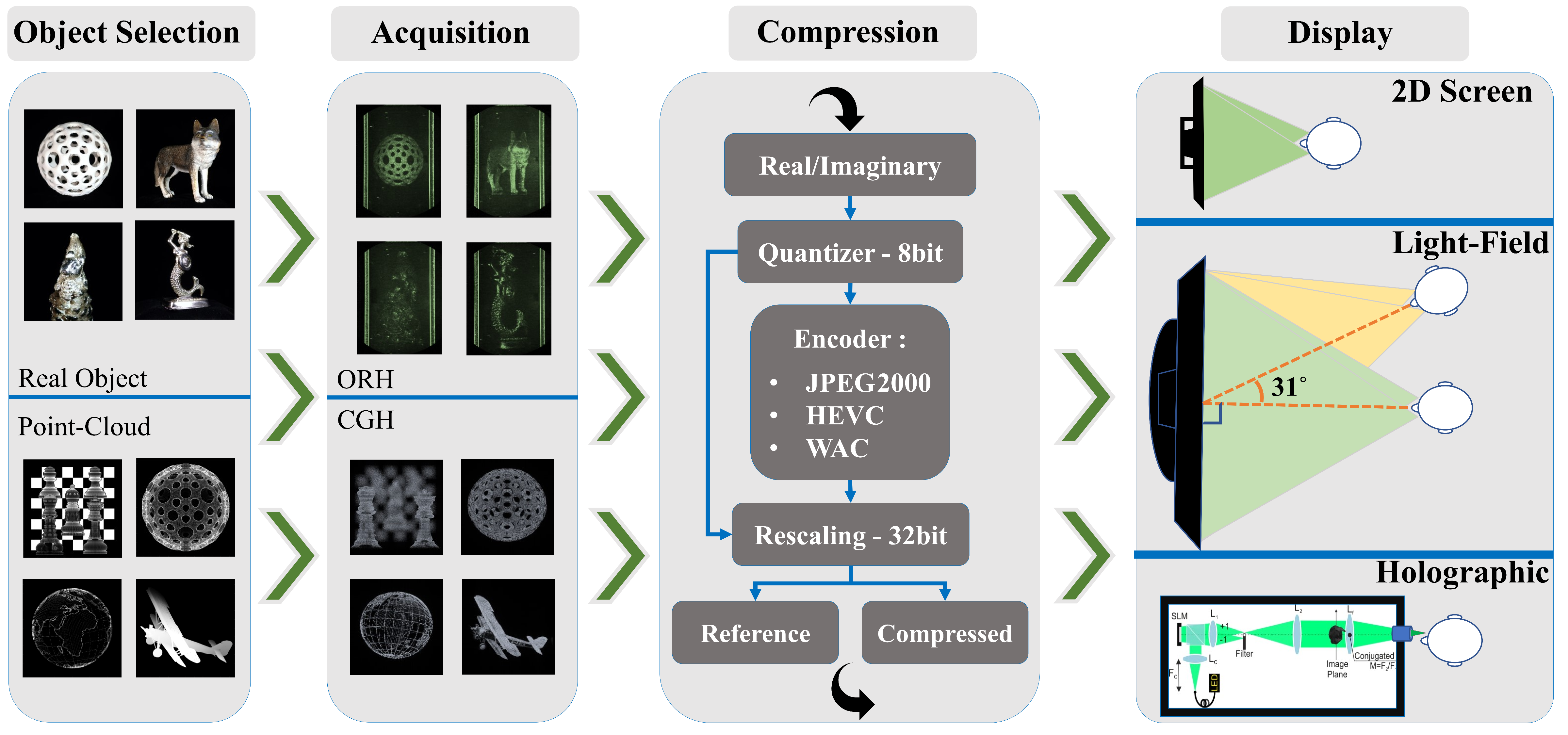} 
	\caption{An abstract schematic of the experimental pipeline.}
	\label{fig:pipeline}
	\vspace*{-1em}
\end{figure*}


\section{Test Data}
\label{sec:data}
For a successful subjective experiment the test data need to provide sufficient diversity in terms of the features of the represented 3D scenes, production of the holograms and distortions introduced.
Having a set of holograms obtained from a diverse set of objects is vital to avoid any bias in the results. This is particularly important in holography since each hologram is an interferogram. Therefore, characteristics of the recorded scene (e.g. object positions, their distances to the recording plane, surface properties and occlusions) will affect the entire interference footprint on the recorded hologram. 

On top of the scene characteristics, the method which produces the hologram must be taken into account. Holograms can either be optically recorded or numerically computed. The latter category, CGHs, can nowadays be calculated at very high spatial resolutions allowing support for high SBPs, efficient occlusion handling and Bidirectional Reflectance Distribution Functions (BRDFs). However, photo-realistic quality is difficult to achieve. Therefore, it is also important to include Optically Recorded Holograms (ORHs) in the test data. Moreover, ORHs have  different characteristics such as the presence of incoherent measurement noise. 


Four ORHs from objects of various dimensions, surface characteristics and capture distances, were selected, which are shown in Fig.~\ref{fig:pipeline}. The first object is a “Mermaid” figurine with small depth, while the second object is a “Squirrel” figurine with larger depth. Both objects are characterized by a glossy, metallic surface. The third and the fourth objects, “Wolf”, a rubber toy, and “Sphere”, the 3D printed model based on the input content of the CGH "Ball", respectively have diffuse surfaces and rather large depths. The holograms of these real objects were recorded using a lensless Fourier holographic capture system~\cite{LenslessSA}, described in section~\ref{sec:OptRec}.
The holograms of the synthetic objects, represented as point clouds, are generated with a multiple Wavefront Recording Plane (WRP) method~\cite{Symeonidou:15} shortly discussed in section~\ref{sec:CGH}.

All considered holograms are created with respect to a spherical reference wave with focal point in the scene center. In the Fourier holographic capture system a spherical reference point source is placed at the center of the object plane. In the CGH calculation framework a demodulation with a Fresnel approximated spherical phase factor is performed numerically after initial propagation to the hologram plane. In this way the advantage of the lensless Fourier holographic capture system in terms of SBP \cite{claus2011quantitative} is utilized in both scenarios. This means that the hologram pixel count does not limit the maximal object dimensions but instead the maximal FoV. ORHs and CGHs are obtained for a high resolution of 16384$\times$2048 pixels. Table \ref{tab:Hols} summarizes the important characteristics of the holograms produced for this experiment.

\subsection{Optical acquisition} 
\label{sec:OptRec}

\begin{figure}
	\centering
	\includegraphics[width=8.8cm]{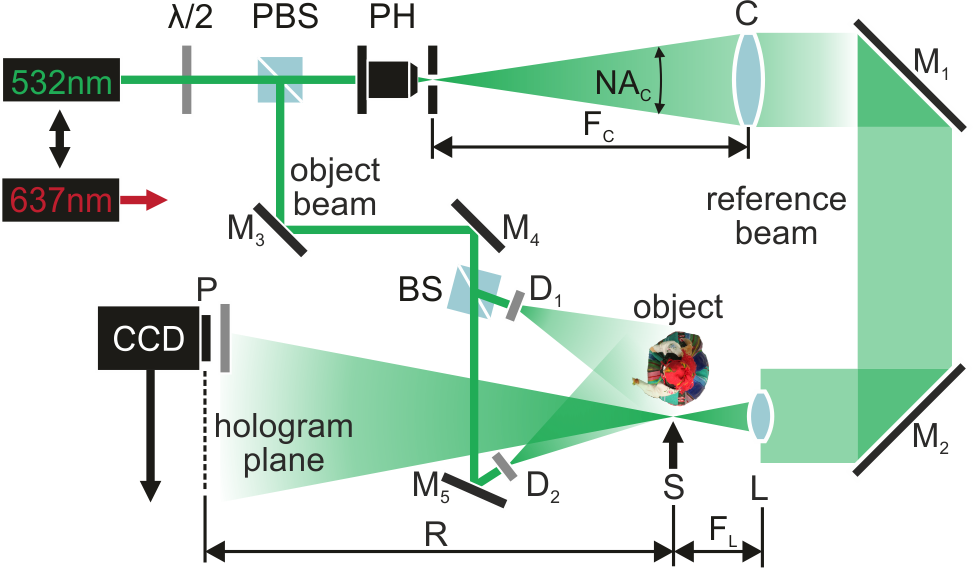} 
	\caption{Lensless Fourier synthetic aperture holographic capture system.}
	\label{fig:OPTRecodSetup}
	\vspace*{-1em}
\end{figure}

For optical acquisition a lensless, Fourier synthetic aperture holographic capture system \cite{LenslessSA,Stroke} is employed (Fig.~\ref{fig:OPTRecodSetup}). The laser beam is divided into a reference and an object beam by a polarizing beam splitting cube PBS. The intensity ratio of both beams is adjusted with an achromatic half-wave plate $\lambda$/2, to obtain a high contrast for the interference fringes of a given scene. The reference beam is formed and directed by the following set of elements: a pinhole PH, an achromatic collimating lens C (F$_C$ = 300 mm, NA$_C$ = 0.13), and mirrors M$_1$ and M$_2$. The reference point source S is generated at the object plane by an achromatic objective L(F$_L$ = 60 mm, NA$_L$ = 0.21). The lenses C and L are selected such that they cover the entire area of the synthetic aperture hologram capture. The diffusers D$_1$ and D$_2$ create a double-sided illumination with the help of the mirrors M$_3$, M$_4$, M$_5$ and another beam splitting cube BS. The analyzer A, placed in front of the camera, improves the hologram contrast by filtering out non-interfering light. The hologram is recorded by a charge-coupled device (CCD) camera (Basler piA2400-12gm) with a pixel pitch of 3.45~$\mu$m and a resolution of 2448$\times$2050 pixels. To realize the synthetic aperture, the CCD is translated in the horizontal direction with the use of motorized linear stage with steps of 2.85 mm over a range of 60 mm. This results in an overlap of $60\%$ between adjacent captured sub-holograms and enables data stitching with sub-pixel precision using a correlation-based routine \cite{Stitching}. The obtained off-axis, synthetic aperture lensless Fourier holograms are composed of 20 sub-holograms that have a physical size of approximately 56.5 mm$\times$7.1 mm with a corresponding spatial resolution of 16384$\times$2048 pixels. The maximum scene size is reduced by half in the horizontal direction, due to the presence of a twin image in the hologram. Holograms were recorded at distances R$_m$ adapted to the scene size and with either 532 nm or 632.8 nm laser beam wavelengths $\lambda_n$: “Mermaid”, R$_1$ = 450 mm, $\lambda_1$ = 532 nm; “Squirrel”, R$_2$ = 500 mm, $\lambda_2$ = 632.8 nm; “Wolf”, R$_3$ = 780 mm, $\lambda_1$ = 532 nm; and “Sphere”, R$_4$ = 960 mm with $\lambda_1$ = 532 nm. 

\subsection{Computer-generated holograms}
\label{sec:CGH}

\begin{figure}[htb] 
  \centering
    \includegraphics[width=0.4\textwidth]{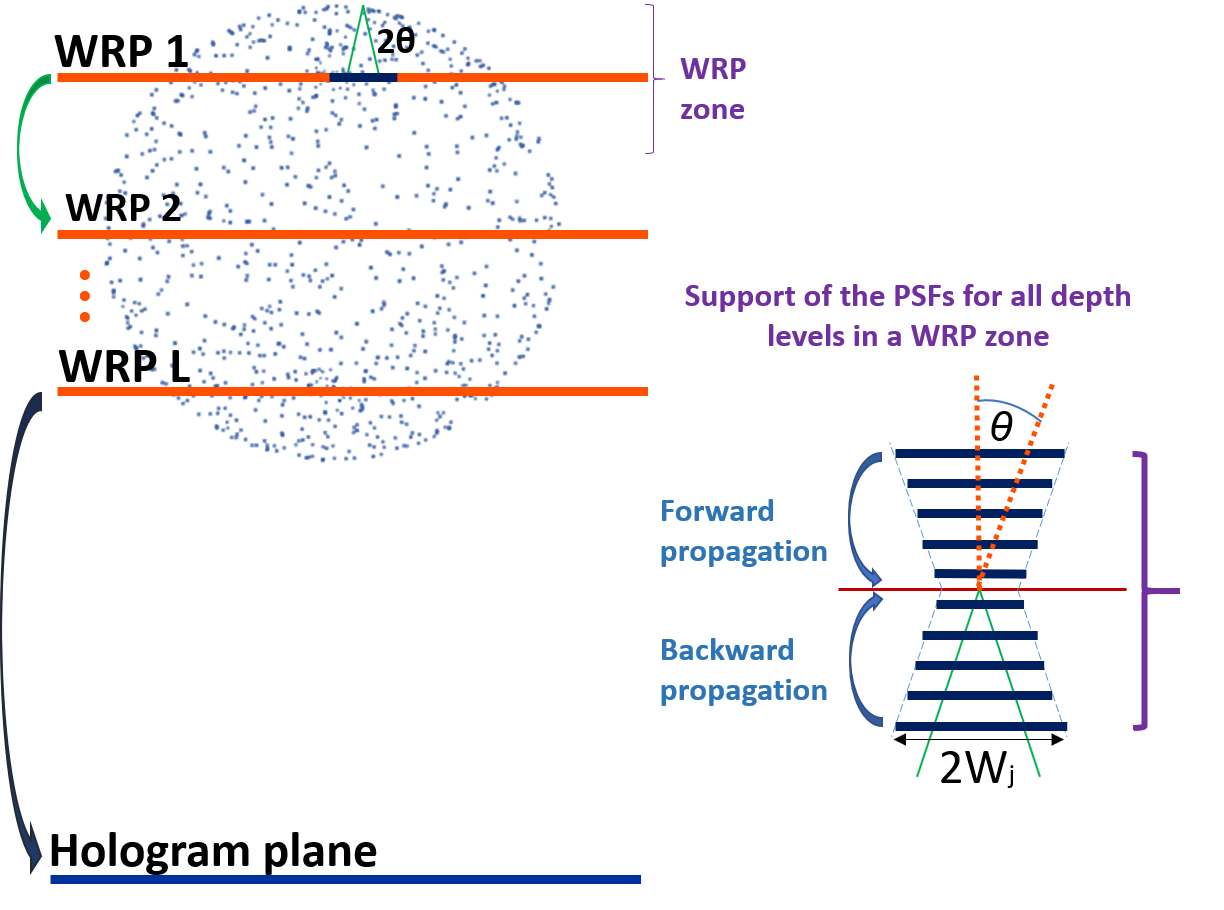} 
\caption{Illustration of the multiple-WRP CGH method used for the generation of the CGHs~\cite{Symeonidou:15}. Additionally, the variation of the support of the PSF per depth level of the LUT is shown, which is determined by the distance to the WRP and the maximum diffraction angle.}
\label{fig:cgh_method}
\vspace*{-1em}
\end{figure}

The CGHs, used for the subjective tests, were generated from point clouds with an extension of the WRP method~\cite{Symeonidou:15}. This method employs multiple parallel wavefront recording planes and pre-computed look-up tables. Moreover, it includes an occlusion handling technique. As shown in Fig.~\ref{fig:cgh_method}, the contribution of each point -- starting from the point furthest away to the hologram plane to the closest point -- is added to the respectively closest WRP. When all the points that belong to the current WRP are accounted for the wavefield is propagated to the next WRP and so forth. To simulate diffuse reflection we assign a random phase to the point spread function during the calculation of the LUTs, as presented in~\cite{Symeonidou:16a}. 
However, there is a very important difference, compared to the previously published methods. To exploit the SBP advantage of the Fourier holographic approach, the wavefield at the last WRP plane is converted to comply with the Fourier hologram configuration, contrary to the in-plane configuration that it supported before. This is done in two steps. First,  hologram is propagated to its proper viewing distance using the angular spectrum method\cite{ goodman2005introduction, Matsushima:10} and subsequently demodulated with a quadratic Fresnel phase kernel corresponding to the axial distance between the hologram and the last WRP. The second step approximates a spherical wavefront with focus in the center-plane of the object by using the Fresnel approximation.

The four CGHs have the same setup parameters: the pixel pitch is 3.45~$\mu$m, the wavelength of the reference beam is 532 nm and the scene center plane were located 700 mm from the hologram plane. 

\begin{table*}
	\caption{ Characteristics of the objects utilized to generate the holograms.}
	\begin{tabular}{|l|c|c|c|c|c|c|}
		\hline
		\multicolumn{1}{|c|}{\textbf{Hologram}} & \textbf{Aquisition Method} & \textbf{PC density/Material} & \textbf{No. WRP} & \textbf{Recording Dist(mm)} & \textbf{Obj. Size: W$\times$H$\times$D(mm)} & \textbf{Rec.Dist/Depth} \\ \hline\hline
		\rowcolor[HTML]{F0F0F0} 
		\textbf{OR-Mermaid}                        & ORH                        & Polished Metal                        & -              & 450                              & 27 $\times$ 53 $\times$ 5                   & 90.00       \\ \hline
		\rowcolor[HTML]{F0F0F0} 
		\textbf{OR-Ball}                         & ORH                        & 3D-Print                              & -              & 960                              & 65 $\times$ 65 $\times$ 65                   & 14.76      \\ \hline
		\rowcolor[HTML]{F0F0F0} 
		\textbf{OR-Squirrel}                       & ORH                        & Brushed Metal                         & -              & 500                              & 43 $\times$ 85 $\times$ 70                   & 7.14       \\ \hline
		\rowcolor[HTML]{F0F0F0} 
		\textbf{OR-Wolf}                           & ORH                        & Plastic                               & -              & 780                              & 50 $\times$ 60 $\times$ 80                    & 9.12       \\ \hline		
		\textbf{CG-Ball}                           & CGH                        & 1.313.280                               & 101              & 700                              & 50 $\times$ 50 $\times$ 50                   & 14.00       \\ \hline
		\textbf{CG-Chess}                          & CGH                        & 219.100                                & 200              & 491                              & 38 $\times$ 43 $\times$ 310                   & 1.58    \\ \hline
		\textbf{CG-Earth}                          & CGH                        & 306.372                                & 101              & 706                              & 46 $\times$ 46 $\times$ 46                   & 15.34    \\ \hline
		\textbf{CG-Plane}                          & CGH                        & 9.999.079                               & 200              & 716                              & 53 $\times$ 46 $\times$ 71                   & 10.08    \\ \hline

	\end{tabular}
	\label{tab:Hols}
	\vspace*{-1em}
\end{table*}

\begin{figure*}
	\centering
	\includegraphics[width=0.8\textwidth]{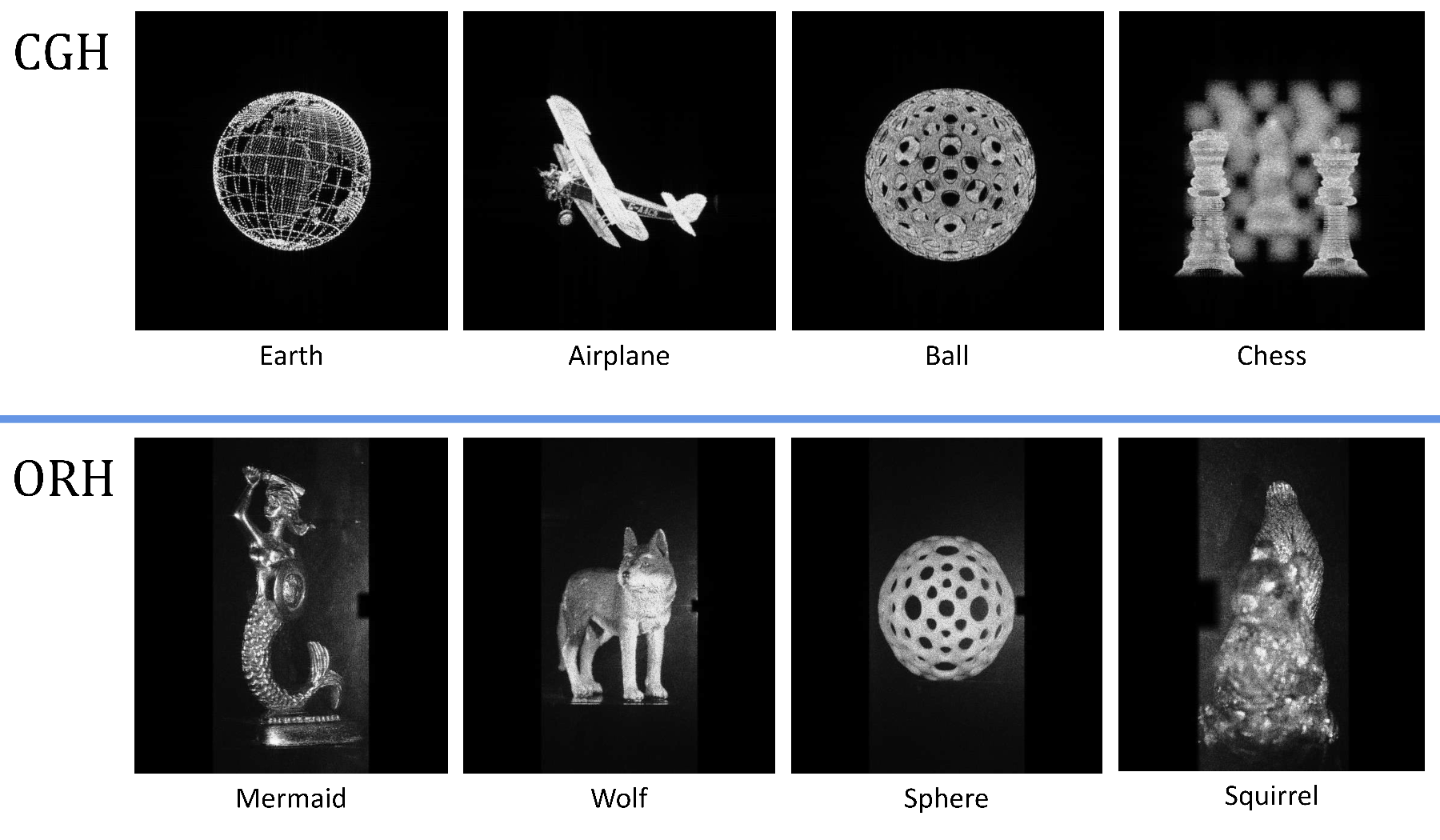} 
	\caption{ Center-views of numerical reconstructions for the reference holograms generated and utilized for this experiment. The top row contains the 4 CGHs from point-clouds and the bottom row shows the ORHs from real objects. The "Sphere" hologram was recorded from the 3D print of the "Ball" Point-Cloud.}
	\label{fig:Holos}
	\vspace*{-1em}
\end{figure*}

\subsection{Content preparation} 

Finally, to facilitate subjective testing and to examine the suitability of each used display, the produced holograms have to be processed such that they are available at different visual quality levels. This enables testing the sensitivity of each display for quality degradation of the holographic content. An added complexity here is the selection of suitable distortion types. 
Nonetheless, compression artifacts are a good starting point for a holographic dataset, both from a practical point of view, comparing the performance of the available compression methods, and also based on the fact that their artifacts normally stem from a combination of multiple distortion types as a result of different processes undergone inside the encoders.  Though all classically used distortions in visual quality testing could be considered, it is important to realize that the end-user will observe the reconstructed hologram in the object plane and not in the hologram plane. During the reconstruction or back-propagation process the propagated data from each point on the fringe pattern updates each and every point of the reconstructed scene. Hence, the reconstructed scene is particularly resilient to local artifacts or even complete loss of information in some small regions of the hologram. As an example, salt and pepper noise, which in regular imaging significantly degrades the visual quality, almost completely vanishes after reconstruction of the hologram. 
Therefore, in this experiment we constrained the distortions to those that have a more global impact on the hologram, more particularly compression distortions. We employed three coding engines: JPEG~2000~\cite{taubman_jpeg2000:_2002, schelkens_jpeg_2009}, intra H.265/HEVC~\cite{sullivan_overview_2012} and wave atom coding (WAC) ~\cite{birnbaum_wave_2019}. 

The IRIS-JP3D software package was deployed to implement the JPEG~2000 compression~\cite{JP2Kpackage, bruylants_wavelet_2015}. The default configuration for JPEG~2000 was utilized using a 4-level Mallat decomposition and CDF 9/7 wavelets with 64$\times$64-pixel sized code blocks.

For the experiments, revision HM-16.18 \cite{HEVCPackage} was used as implementation of the H.265/HEVC compression standard. Since all tested images were in grayscale format, we used 4:0:0 subsampling and fed the images as an 8-bit luminance channel with empty chrominance channels. Hence, cross-component prediction and motion search settings were disabled. "Frame rate" and "frames to be encoded" were set to 1. The desired compression level was achieved by tuning the quantization parameter (QP). All other parameters were set to their default values.

The WAC leverages the orthonormal wave atom transform. This non-adaptive multi-resolution transform has good space-frequency localization and its orthonormal basis is suitable for sparsifying holographic signals. Basically this codec is based on a JPEG~2000 coding architecture where the CDF 9/7 wavelet transform is replaced by the 2D wave atom transform where the spatial footprint of each atom scales parabolically across resolutions, while the quantization and Embedded Block Coding by Optimizated Truncation (EBCOT)~\cite{taubman_high_2000} are further deployed. EBCOT code blocks of size 128$\times$128 pixels are issued.


All three encoders compressed the 8-bit quantized real and imaginary parts of each hologram separately. The holograms were compressed at bitrates 0.25 bpp, 0.5 bpp, 0.75 bpp and 1.5 bpp. These bitrates were determined via a series of mock-up tests where the holograms were compressed at 9 different bit-depths between 0.15 bpp to 4 bpp and their visual appearance were tested on 2D and light field setups. Also on the holographic display, 3 sample holograms were tested for all the 9 bit-depths and the others were verified for the chosen bit depths. The goal was to ensure that the distortion levels resulted in broad range of visual quality levels ranging from very poor to imperceptible.

\begin{table*}
	\caption{Details of the test conditions and the gathered scores per display setup.(* The hologram "Mermaid" was reconstructed at 1 focal distance and "Chess" at 3. Although the average number of reconstructions per hologram is equal to 2.)}
	\begin{tabular}{|l|c|c|c|c|c|c|c|}
		\hline
		\multicolumn{1}{|c|}{\textbf{Setup}} & \textbf{No. Tested Objects} & \textbf{Distortions} & \textbf{Dist. Levels} & \textbf{Perspective} & \textbf{Recon. Distance} & \textbf{Total Conditions} & \multicolumn{1}{l|}{\textbf{Scores per Condition}} \\ \hline
		\textbf{Optical}                     & 8                           & 3                    & 4                     & 2                    & 1                        & 192                       & 20                                                 \\ \hline
		\textbf{Light field}                 & 8                           & 3                    & 4                     & 2                    & 2*       & 384                       & 20                                                 \\ \hline
		\textbf{Regular 2D}                  & 8                           & 3                    & 4                     & 2                    & 2*       & 384                       & 20                                                 \\ \hline
	\end{tabular}
	\label{tab:TestConds}
	\vspace*{-1em}
\end{table*}

The full set of the test holograms along with their acquired quality scores and other related data to these experiment are publicly available at: \url{http://data.etrovub.be/holodb}. 

\section{Test methodology}
\label{sec:Method}

\subsection{Generic procedure for subjective quality assessment} 
\begin{figure*}
	\centering
	\includegraphics[width=0.8\textwidth]{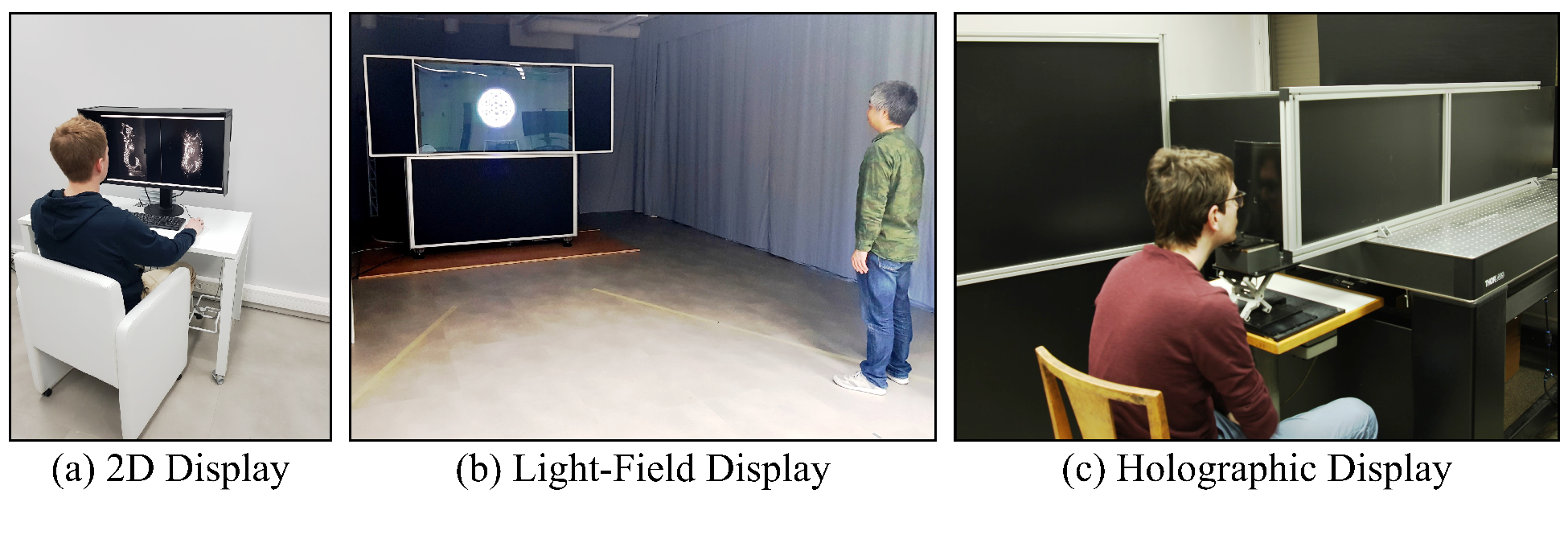}
	\caption{Subjective testing experimental setups for 2D display (a), light-field display(b) and the holographic display setup (c)}
	\label{fig:ExperiSetup}
	\vspace*{-1em}
\end{figure*}
Holographic modalities and in extension, plenoptic modalities, pose specific challenges as it concerns evaluating their visual quality. This is due to the fact that the plenoptic function "allows for the reconstruction of every possible view, at every moment, from every position, at every wavelength within the space-time wavelength region under consideration"~\cite{adelson_plenoptic_1991}.
As a consequence, to meet its time and resource constraints, a subjective experiment has to be limited to evaluate only a necessary subset of this 7D space. For instance, several subjective quality assessment approaches have been reported for 4D light fields. At IEEE~ICME~2016 a Grand Challenge on Light Field Compression was organized~\cite{viola_objective_2016} deploying a double stimulus continuous quality scale (DSCQS) methodology~\cite{ITU-R_recommendation_2012}. In this solution uncompressed and decoded views are shown side-by-side on a high-end monitor while selecting a limited set of views and focus points per light field. In the context of the JPEG Pleno Light Field Coding standardization effort and a associated Grand Challenge on Light Field Coding organized at IEEE~ICIP~2017, a double stimulus comparison scale (DSCS) methodology was employed with side-by-side rendering of the light field as a pseudo video sequence and using a discrete quality scale ranging from -3 to 3~\cite{viola_quality_2018,iso/iec_jtc1/sc29/wg1_jpeg_2017}. 
Viola et al.~\cite{viola_impact_2017} assessed the impact of an interactive approach to determine perceived quality as such enabling to evaluate a larger fraction of the light field compared to the passive approach of~\cite{iso/iec_jtc1/sc29/wg1_jpeg_2017}. The same authors applied the advocated solution also to evaluate a larger set of light field compression techniques~\cite{viola_comparison_2017}.
For point clouds data few subjective quality assessment experiments have been also conducted. To evaluate point cloud compression techniques, Javaheri et al.~\cite{javaheri_subjective_2017} rendered a video sequence by having a virtual camera spiralling around the point cloud object. The Double Stimulus Impairment Scale (DSIS)~\cite{ITU-R_recommendation_2012} methodology was adopted and the video sequences of the impaired and original point cloud were shown sequentially. A similar procedure was applied for the evaluation of point cloud denoising algorithms~\cite{javaheri_subjective_2017b}. For holographic data few earlier efforts took place, several open access test data bases have been proposed; such as: the B-Com Repository~\cite{gilles2016hybrid,gilles2016computer}, ERC Interfere I~\cite{blinder_open_2015}, II~\cite{Symeonidou:16a} and III~\cite{Symeonidou:18}, and EmergImg-HoloGrail v1 and v2~\cite{bernardo_holographic_2018}. Recently, Amirpourazarian et al. presented a methodology to evaluate perceptual quality of compressed holograms on a 2D display ~\cite{pinheiro_quality_2019}.

As mentioned in section~\ref{sec:Intro}, the testing method should be adapted to the specific limitations and different technical requirements of each display type. 
The holograms were shown mainly following the procedure for DSIS. In our method the reference and distorted stimuli are sequentially shown to the subject and then the subject scores the second stimulus (impaired version) based on the first (reference). The hologram sequences were shown in a fully randomized order. The presentation order was also randomized for each subject.

The scoring procedure was followed by the standard one providing 5 quality scales. Depending on the perceived mismatch, subject chooses a quality number from 1 to 5 representing one of the impairment scales: Very Annoying, Annoying, Slightly Annoying, Perceptible but not Annoying, and Imperceptible.  The testlab conditions corresponded to ITU-R BT.500-13 recommendations~\cite{ITU-R_recommendation_2012} and recommendations described in Annex B of ISO/IEC 29170-2 (AIC Part-2).

\subsection{Subjective quality assessment on holographic display}
\label{sec:subholo}

The subjective test on the holographic display was conducted in the photonics laboratory of the Institute of Micromechanics and Photonics of Warsaw University of Technology. The holograms were shown mainly following the DSIS procedure. From each synthetic aperture hologram a sub-holograms of 2048$\times$2048 pixels were used to visualize and score the center and right-corner views. In this particular case, the room lighting condition did not meet the standard ITU-R BT.500-13 recommendations. However, the environmental luminance does not impact the visibility of the stimuli because, as explained in section~\ref{sec:HoloDisp}, the subject has to position his eye on the watching-slit. Since subjects watch the stimuli with one eye only (due to the limited resolution of the screen), we preferred to use a dark room to guarantee the repeatability of the experiment. This way subjects keep the other eye open without getting any effect from the environmental light. During the mock-up test session we realized that for holographic displays, subjects felt fatigued significantly sooner than on non-holographic displays due to higher concentration of incident light to their eye and the bigger effort required to focus the eye on the content. Hence, the test was divided into 4 sessions of at most 10 minutes each. The experiment was performed in two days such that each subject participated in only 2 sessions per day. A compulsory minimum of 5 minute rest was facilitated by the test operator before starting the next session. The maximum rest time was not limited and subjects had the freedom to take larger recuperation periods in case they felt it to be necessary. 

\subsection{Subjective quality assessment on light field display}
\label{sec:sublf}

For the Light-field display, again the DSIS method was implemented following the ITU-BT.500-13 recommendations~\cite{ITU-R_recommendation_2012}. Although, in this case, the display provides a wide angle, simultaneous rendering of multiple views and each subject was required to watch and score the center and right-corner view of each displayed hologram-pair. To facilitate a repeatable procedure, the places where subjects have to stand to see the required views, were marked on the ground. The distance from the screen was chosen $3.2$ times the height of the screen. For each tested hologram, the subject starts standing in the center-position and the operator displayed the reference and impaired holograms sequentially. After recording the score, the subject moved to the right-corner position and again both reference and impaired holograms were displayed by the operator followed by the scoring. According to Table \ref{tab:TestConds} the number of test-conditions per subject was twice the number of test-conditions per subject in the holographic setup. This is due to the fact that for each hologram, test subjects scored the visual quality at two different reconstructions distances for the light field display. The test in this setup was conducted in 2 sessions with a target duration of 20 minutes. Since the subjects were required to stand and move multiple times to designated positions during the test, at least 1 hour rest was considered before starting the second session.               

\subsection{Subjective quality assessment on 2D display}
\label{sec:sub2D}

For the 2D setup, see Fig.~\ref{fig:ExperiSetup}.a, each reconstructed hologram was shown for the 2 perspective positions corresponding to the one for the light field display and holographic display and two reconstruction distances corresponding to the light field display test. The reconstructed reference and impaired holograms were displayed side by side reducing the test time per subject by half. 

\subsection{Training of test subjects} 
For each setup, 40 subjects participated. From the total of 120 participants, the number of female and male participants were 54 and 66 respectively. Their age was between 18 to 30 years old. Prior to the test, subjects were required to pass the Snellen visual acuity test. Though, all the content shown to the subjects was monochromatic, the Ishihara test to detect the colorblindness was performed as well. Prior to the first test session in each setup, a 5 minute training session was conducted where the test and scoring procedure was explained and rehearsed.  

\section{Results and Analysis}
\label{sec:experimental-results}
In this section, we provide the results of our subjective experiments and further investigate various aspects of the outcomes, potential similarities and correlations among the gathered scores from the three testing setups.   
\begin{figure}
	\centering
	\subfloat[Light-Field display]{\includegraphics[width=0.48\textwidth]{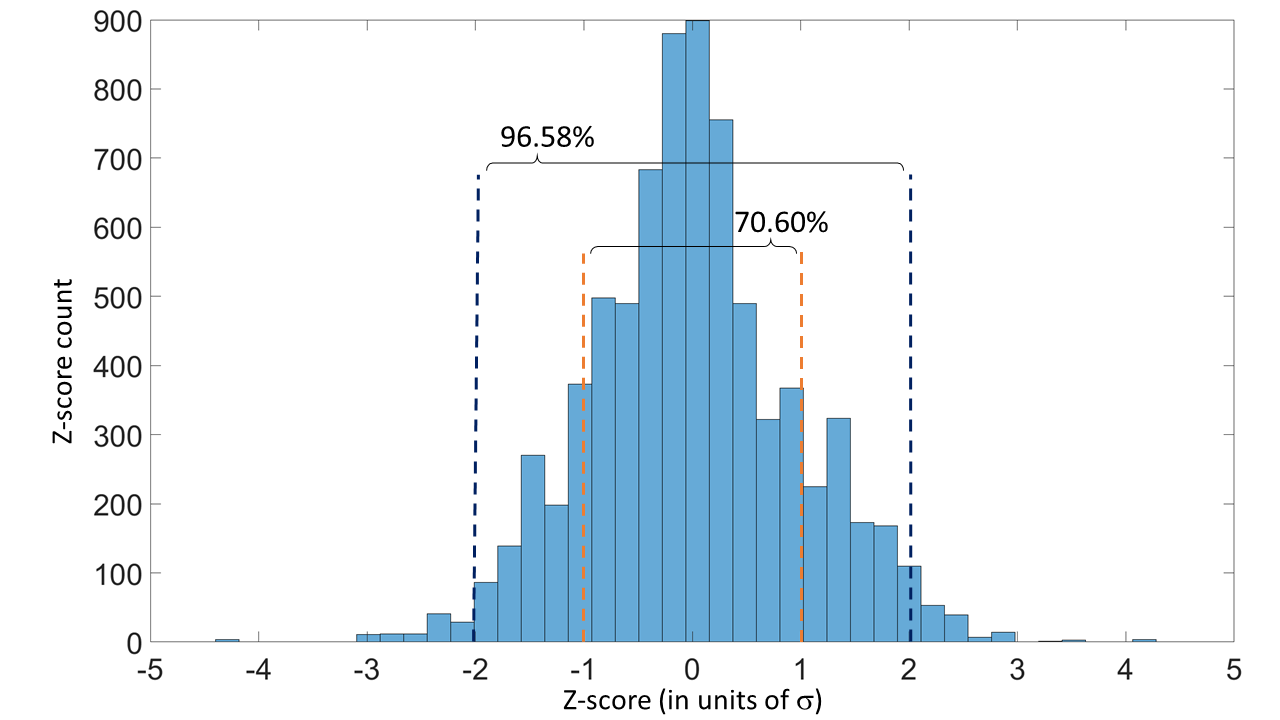}}
	
	\subfloat[2D display]{\includegraphics[width=0.48\textwidth]{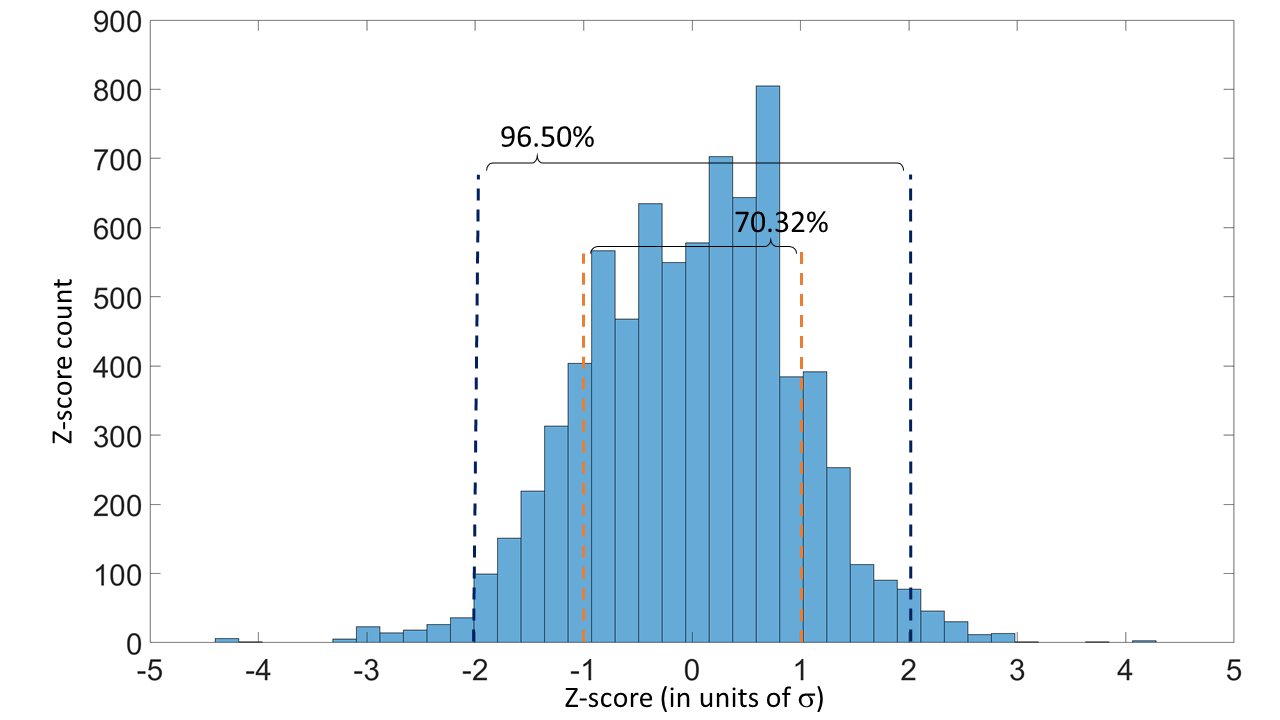}}
	
	\subfloat[Holographic display]{\includegraphics[width=0.48\textwidth]{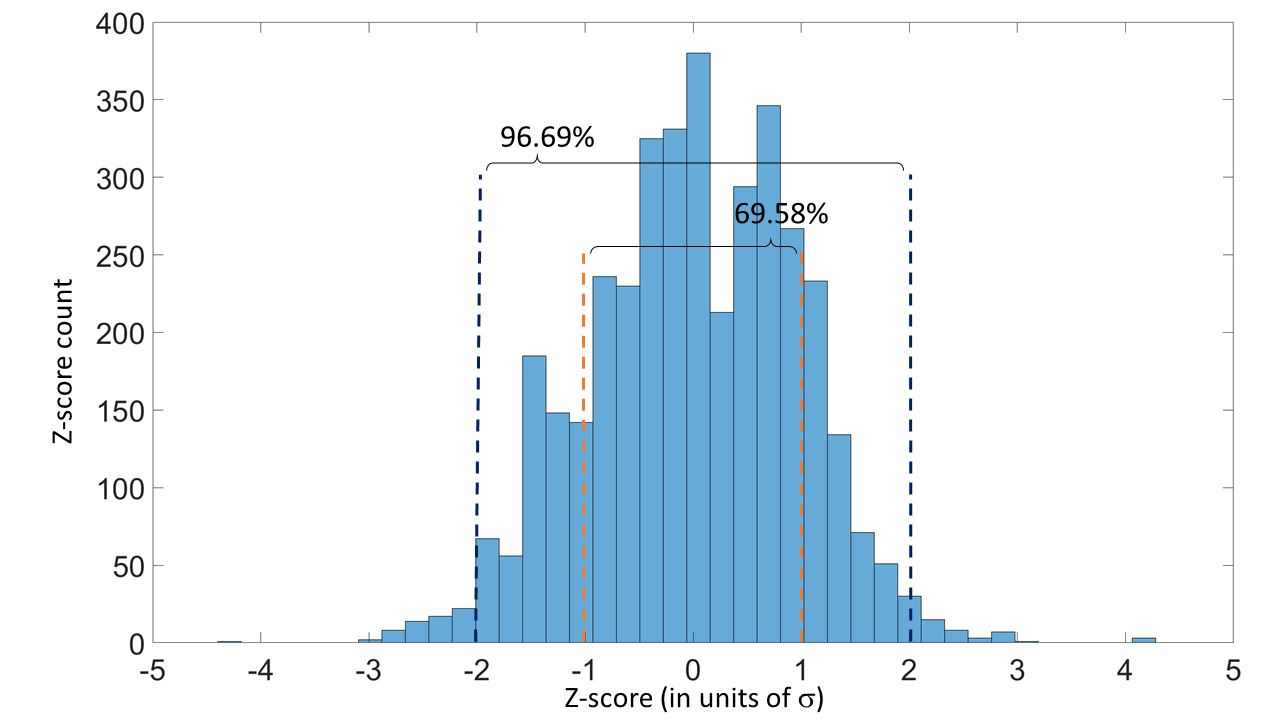}}
	\caption{Histogram of Z-scores per display setup - calculated per condition from the raw scores before outlier removal and averaging. The histograms represent the underlying distributions of raw scores and are thus directly comparable. The percentage of Z-scores, which falls within 1 and 2 standard deviation(s) from their mean (their MOS after outlier removal) is shown on the graphs.}
	\label{fig:ZscoreHists}\vspace*{-2em}
\end{figure}
\subsection{Reliability analysis of the obtained MOS} 
First, a reliability analysis is performed on the acquired opinion scores for the three setups. Before performing any post-processing on the scores and calculating the Mean Opinion Score (MOS) for each test condition (see procedure described in Sec. \ref{sec:statistical postproc}), it is important to check whether the average is a reliable representative of the underlying distribution per condition. To determine the MOS reliability, one should ideally identify the distribution model of the data. Though, considering our limited sample size (20 scores per condition, from each setup) conventional statistical modeling may not necessarily reach to a conclusive result. Instead, a kurtosis analysis has been recommended in standards like ITU.BT500.13~\cite{ITU-R_recommendation_2012}, where a score distribution with a kurtosis value of 2 to 4 is interpreted as a representative of the normal probability model. However, this is a vague and flawed assumption. It is indeed correct that the kurtosis of a normal distribution model is equal to 3, though mathematically this is a necessary but not sufficient condition. Moreover, by definition its only unambiguous interpretation is in terms of distribution tail extremity~\cite{westfall2014kurtosis}. Nonetheless, no score set (per condition) in our dataset showed any irregular kurtosis value.

Next, we seek to answer two questions: (1) Whether the subject scores for a specific test condition reach a consensus about the visual quality score for this test condition or not? (2) If the answer to the first question is positive, to what extend can that consensus be represented by the mean of these scores? To compactly address both, first we standardize the scores per condition. Z-scores are calculated where each score is normalized by the mean and standard deviation of the scores for the same test condition. The advantage of Z-scores is that their normalization enables direct comparison of individual scores across all conditions and even different setups. Nonetheless, the Z-score value does not provide any information about the actual visual quality level. It gives the distance of each individual score from the average opinion score (in units of standard deviation). This way a histogram of all scores for a particular setup (Fig.~\ref{fig:ZscoreHists}) gives an abstract view on the overall agreement of test subjects. Notably, for all setups, a significantly good agreement is available around the mean opinion values, such that more than $69.5\%$ and $96.5\%$ of the individual scores fall within only 1 and 2 standard deviations(s) away from their corresponding mean, respectively. (The corresponding values for a perfect normal distribution are $68.27$ and $95.45\%$). Based on this and the fact that no specific skewness can be seen around the tails of shown distributions, we believe our MOS values can appropriately represent opinions of the majority of tested subjects. 

\subsection{Statistical processing of results} 
\label{sec:statistical postproc}
The distributions of Fig.~\ref{fig:ZscoreHists} shows that a very small portion of scores per setup are more than 4 standard deviations away from the average scores per condition. Therefore, an outlier detection and removal was performed on the test results. Following the procedure used in \cite{hanhart2012subjective} and \cite{ahar_subjective_2015},  the $25th$ ($Q1$) and $75th$ ($Q3$) percentiles were calculated. A score $u$ was considered as an outlier if $u > Q3 +w(Q3-Q1)$ or $u < Q1-w(Q3-Q1)$, where $w$ was the maximum whisker length. $w = 1.5$ for normally distributed data corresponds to $99.3\%$ coverage, which was utilized in the experiment. Our results also showed that no test subject had more than $15\%$ outlier scores. Consequently,  no test subjects were removed from the dataset. After removing the outlier scores, the average of the remaining scores for a particular test condition was combined into the final MOS.

\subsection{MOS analysis based on reconstruction focal-point } 
\begin{figure}
\centering
	\subfloat[ $MOS_{LFfront}$ vs $MOS_{LFback}$ ]
	{\includegraphics[width=0.25\textwidth]{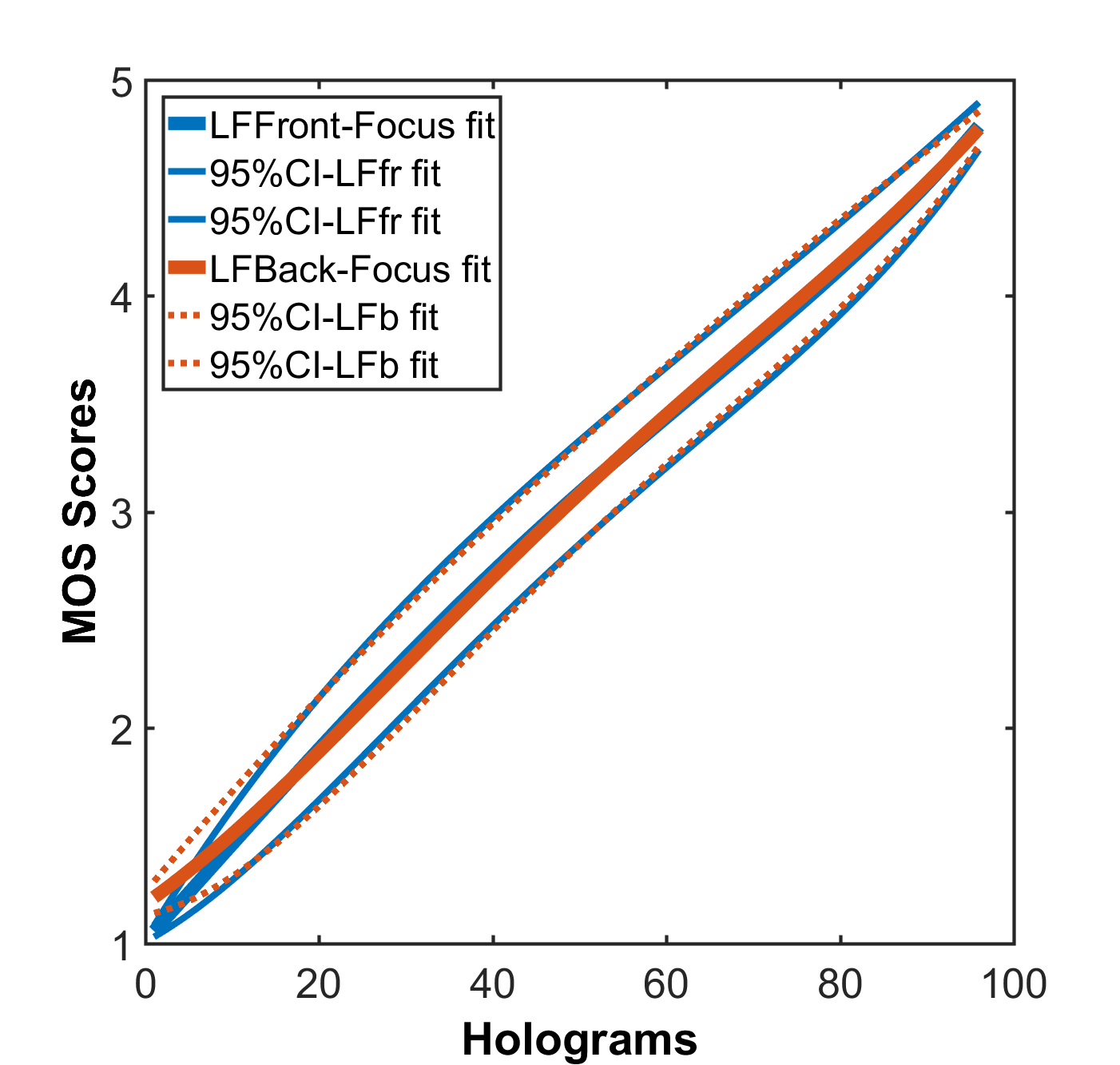}}
	\subfloat[ $MOS_{2Dfront}$ vs $MOS_{2Dback}$ ]
	{\includegraphics[width=0.25\textwidth]{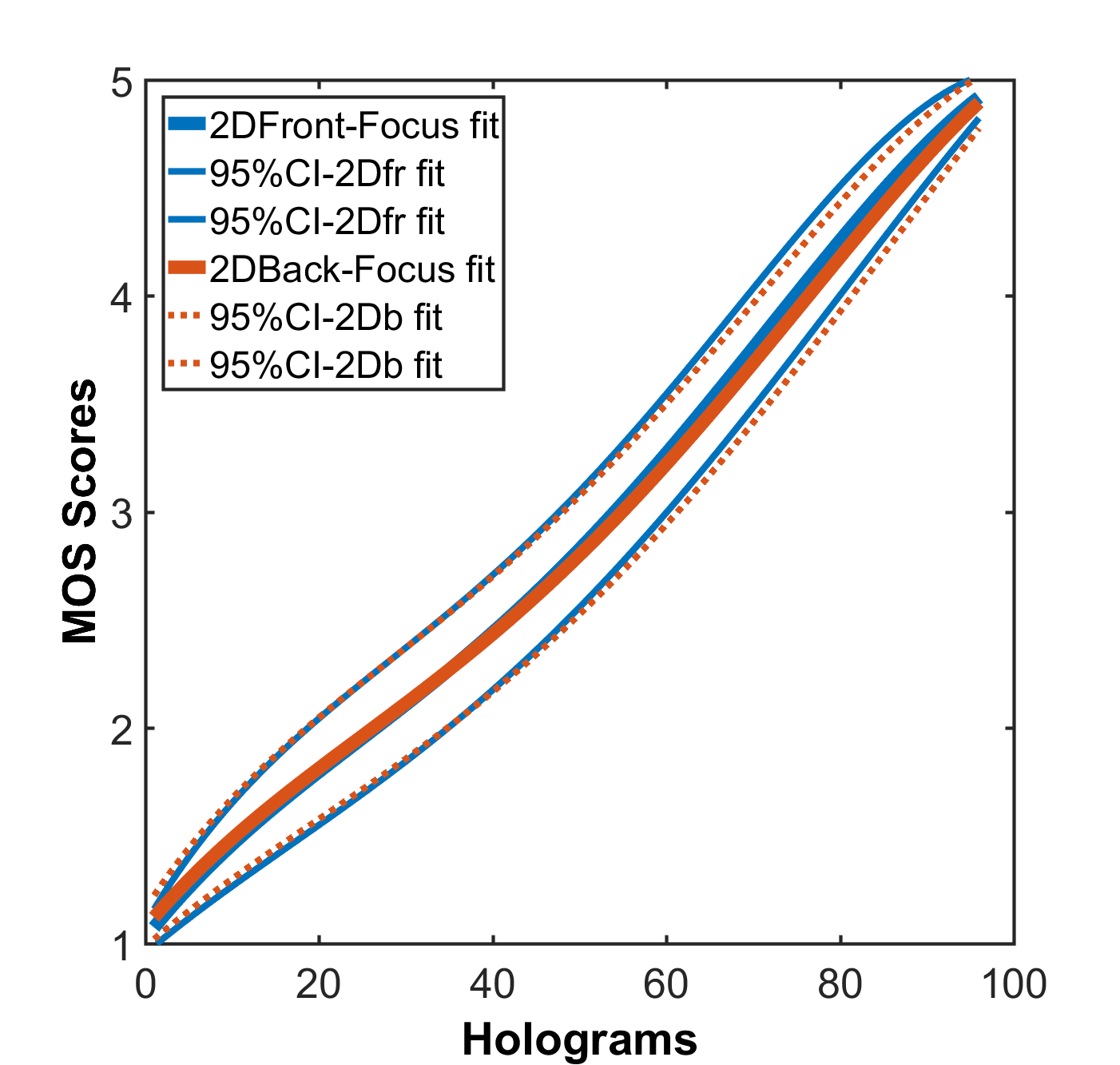}}
	\vspace*{-1em}
	
	\subfloat[ $MOS_{LFfront}$ vs $MOS_{LFback}$ ]
	{\includegraphics[width=0.25\textwidth]{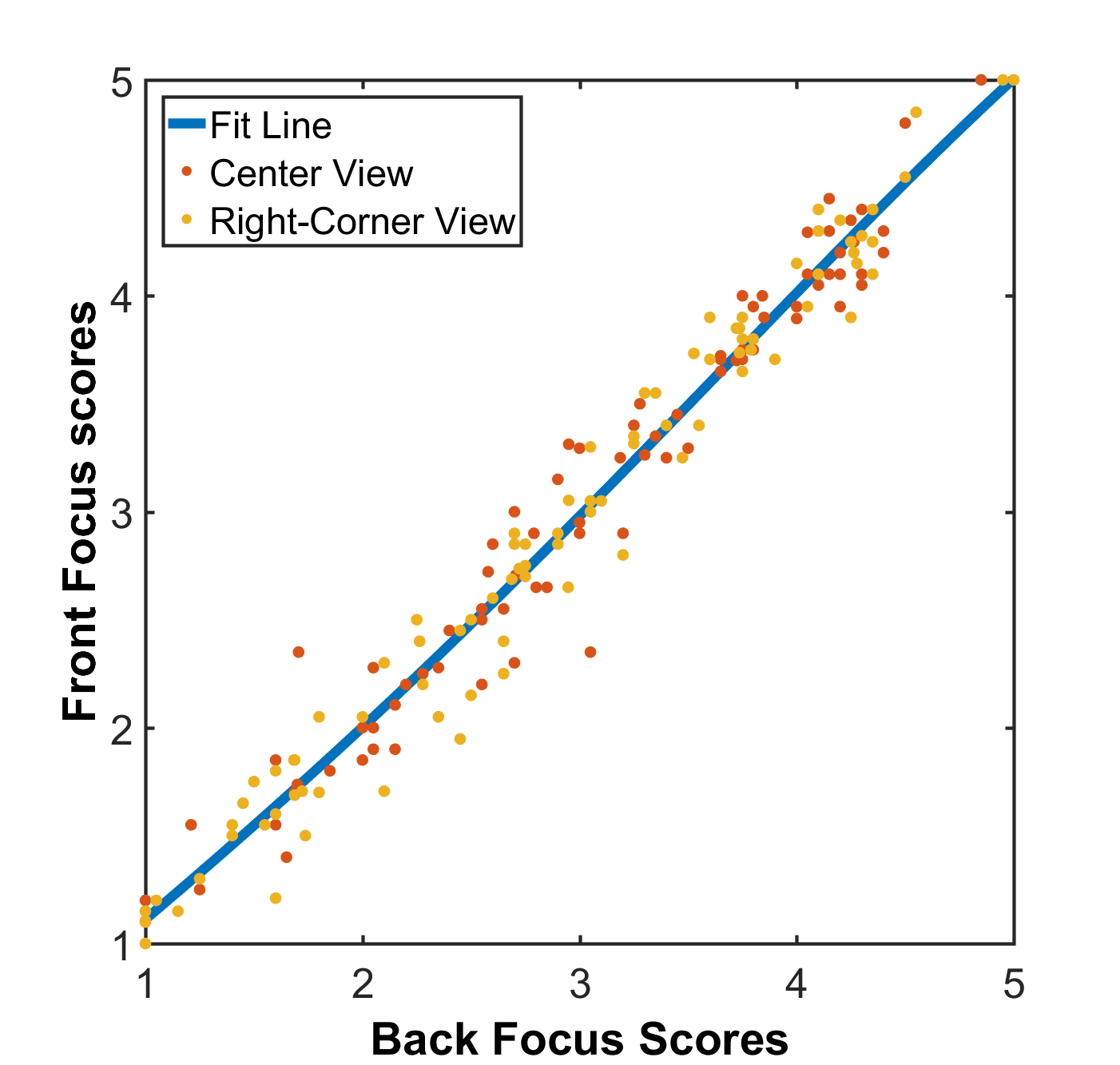}}
	\subfloat[ $MOS_{2Dfront}$ vs $MOS_{2Dback}$ ]
	{\includegraphics[width=0.25\textwidth]{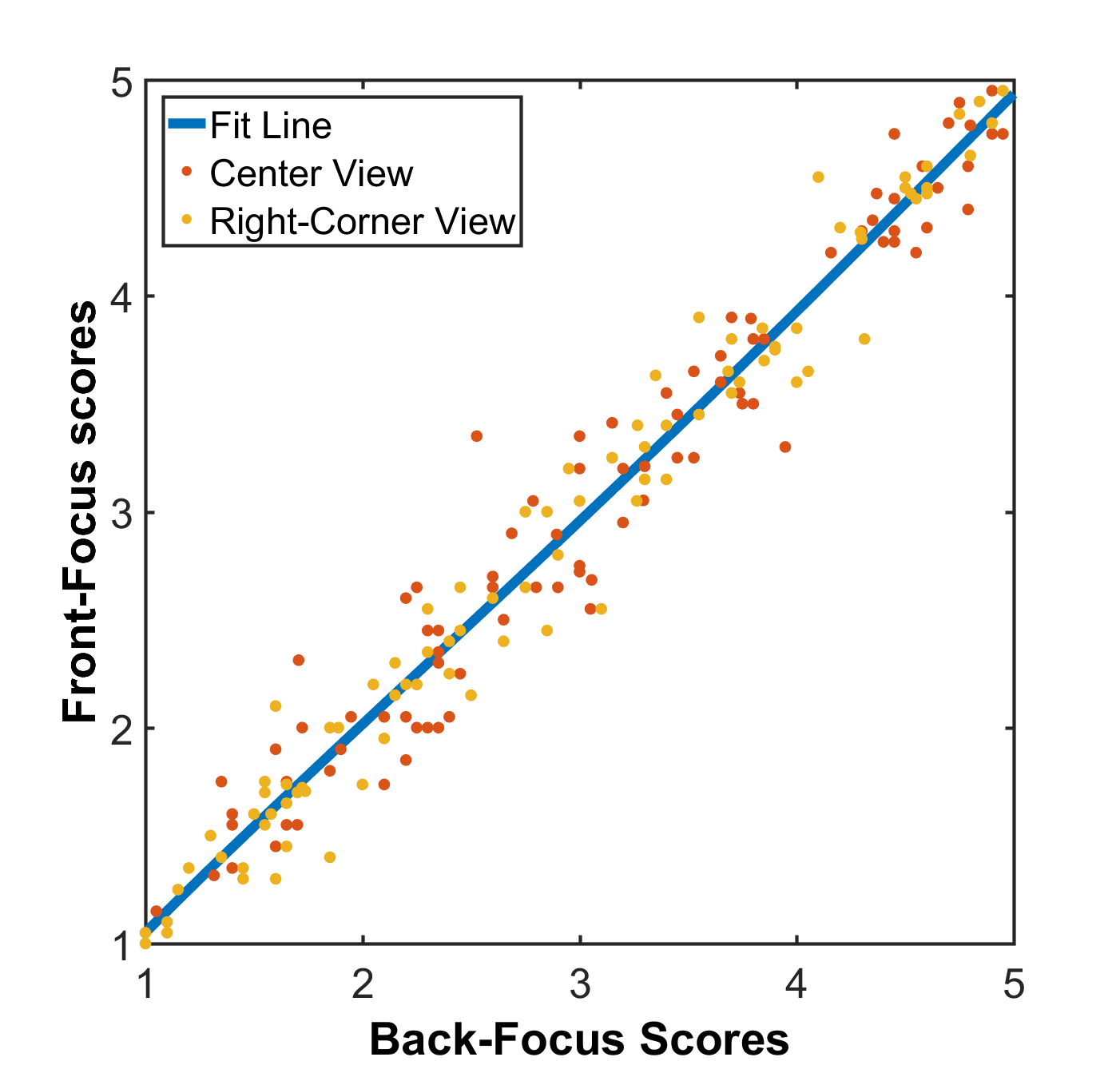}}
	
\caption{Overall comparison of the Front-Focus MOS versus Back-Focus MOS for the light field display, (a) and (c), and the  2D display, (b) and (d). The  results shown in (a) and (b) for each depth are averaged over center and corner perspectives. The raw data is shown color coded for both cases in (c) and (d). The indicated lines in (a) and (b) are 4th-degree polynomial fit lines for the data with indices of the sorted front focus MOS and sorted back focus MOS, respectively.}
\label{fig:allMOS_D1D2-comp}
\vspace*{-1em}
\end{figure}

First, the MOS values at different reconstruction distances were evaluated for light field and 2D displays. Fig.~\ref{fig:allMOS_D1D2-comp} shows the overall comparison between front and back focus MOS, while each MOS is averaged between the two perspectives (Center and Right-Corner views). It is obvious that the MOS from both depths are very closely following the same trend. Nonetheless, the non-averaged MOS are also visualized in the scatter plots of Fig.~\ref{fig:allMOS_D1D2-comp}(c, d). Therein points are colored differently by perspective. At this point, results do not show any meaningful differences. Therefore, to limit the degrees of freedom for our analysis, we use in the next subsections the MOSes, which have been averaged over the focal distances. This means the number of MOSs for light field and 2D setups will be equal to the ones from the holographic setup (96 scores per perspective and a total of 192 scores per setup).

\begin{figure*}
\centering
	\subfloat[ $MOS_{OPTc}$ vs $MOS_{OPTr}$\label{fig:allMOS_CR-comp_OPT}]
	{\includegraphics[width=0.3\textwidth]{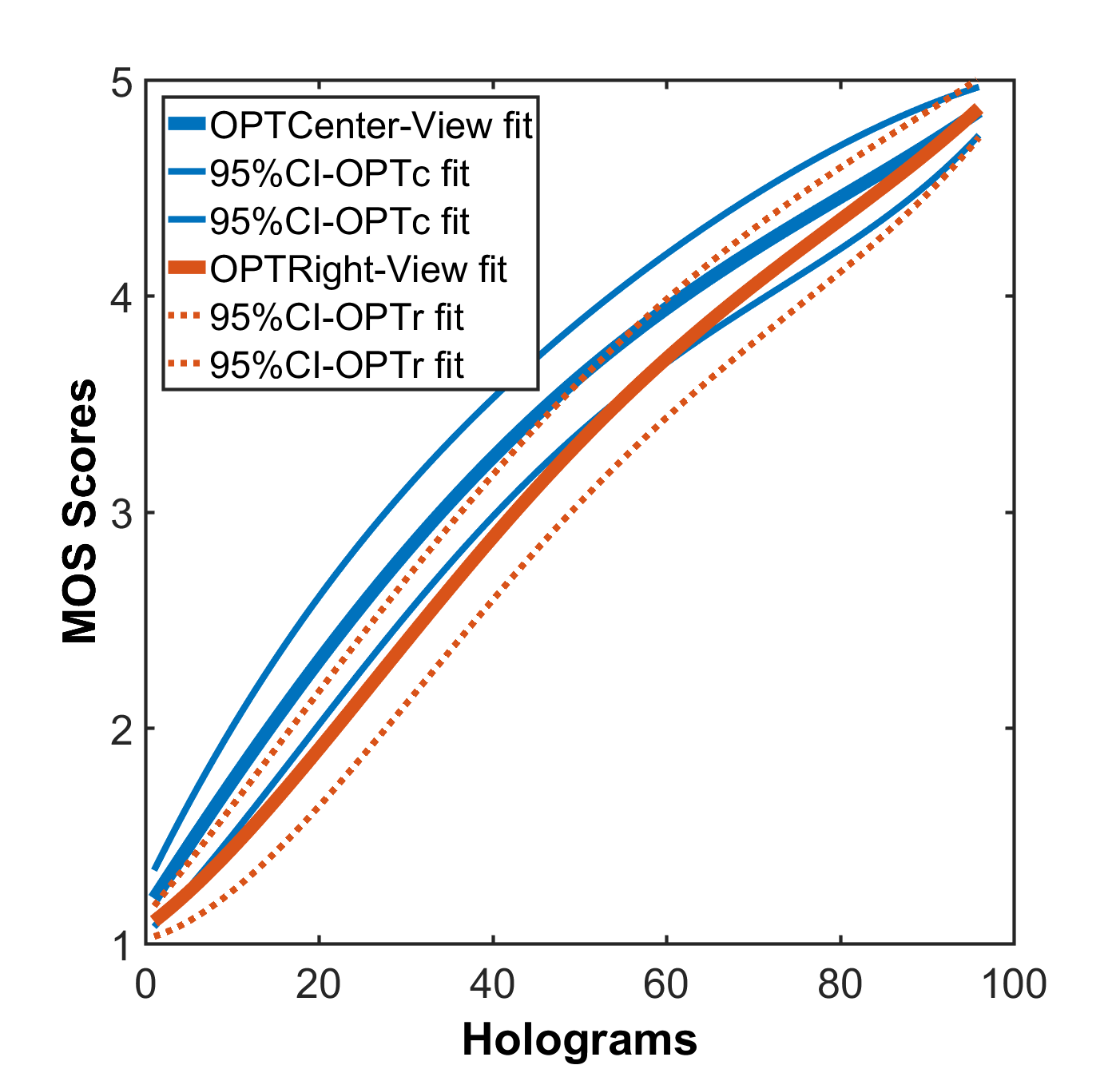}}
	\subfloat[ $MOS_{LFc}$ vs $MOS_{LFr}$ ]
	{\includegraphics[width=0.3\textwidth]{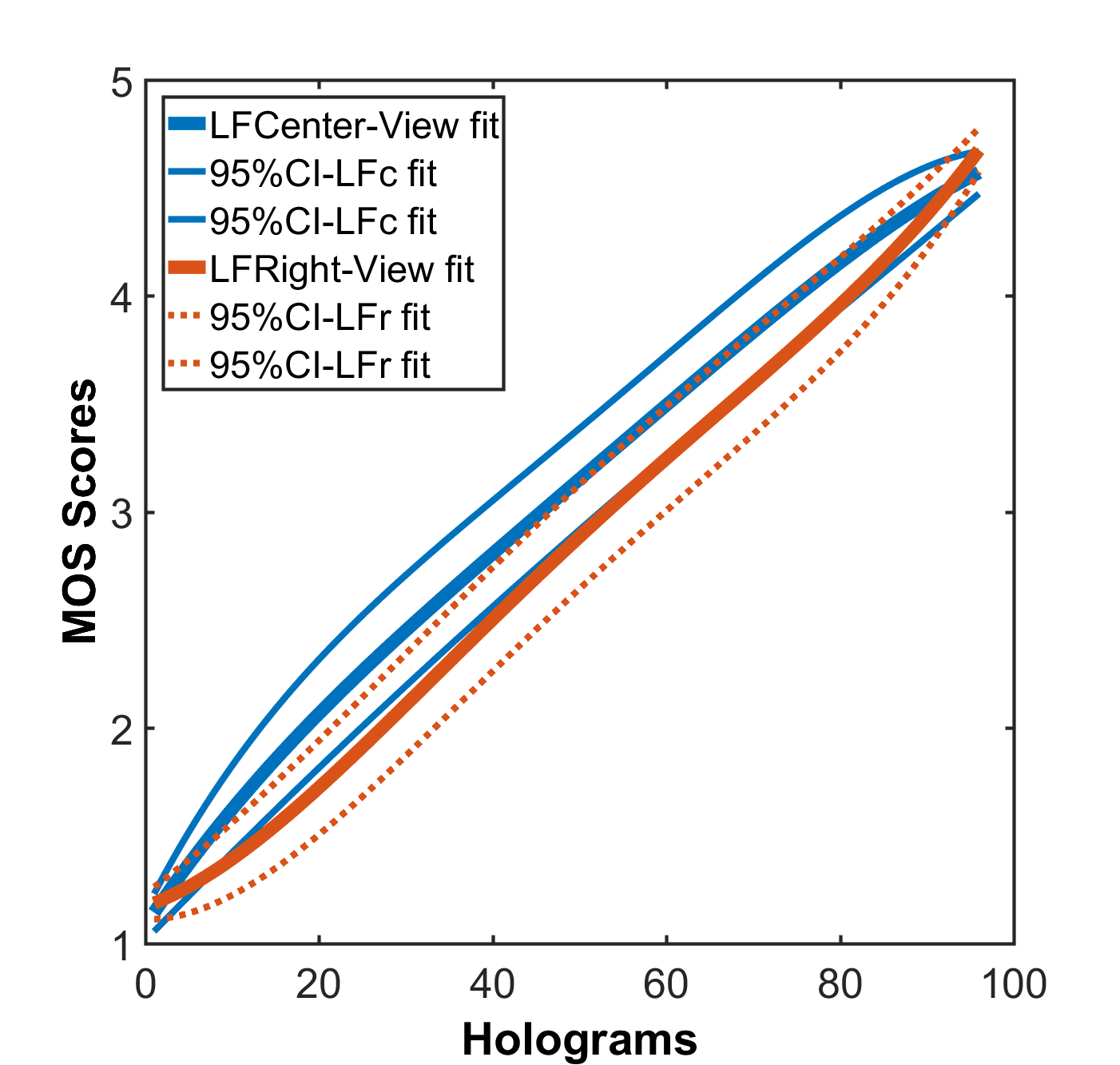}}
	\subfloat[ $MOS_{2Dc}$ vs $MOS_{2Dr}$ ]
	{\includegraphics[width=0.3\textwidth]{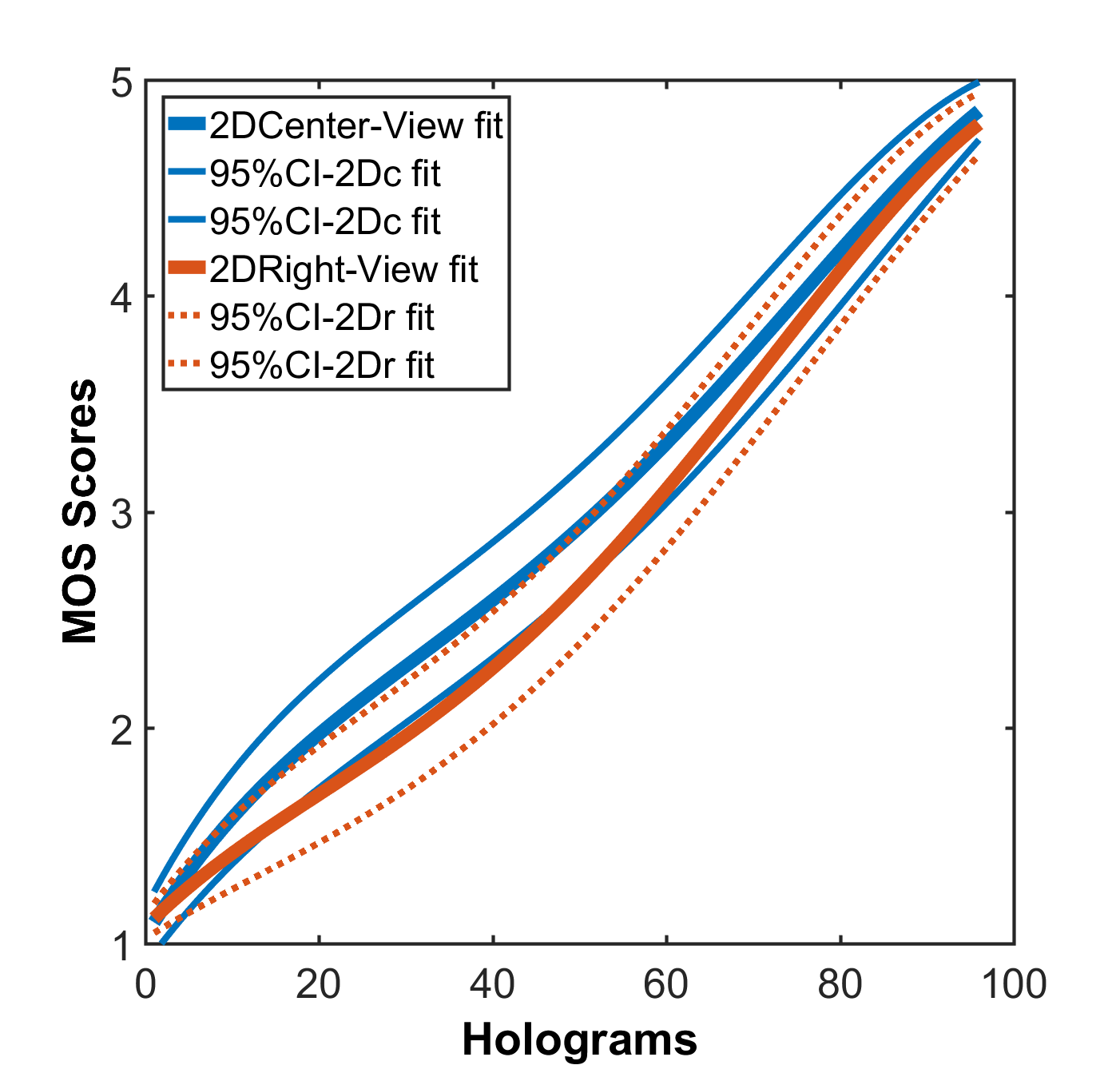}}
	\vspace*{-1em}
	\subfloat[ $MOS_{OPTc}$ vs $MOS_{OPTr}$]
	{\includegraphics[width=0.3\textwidth]{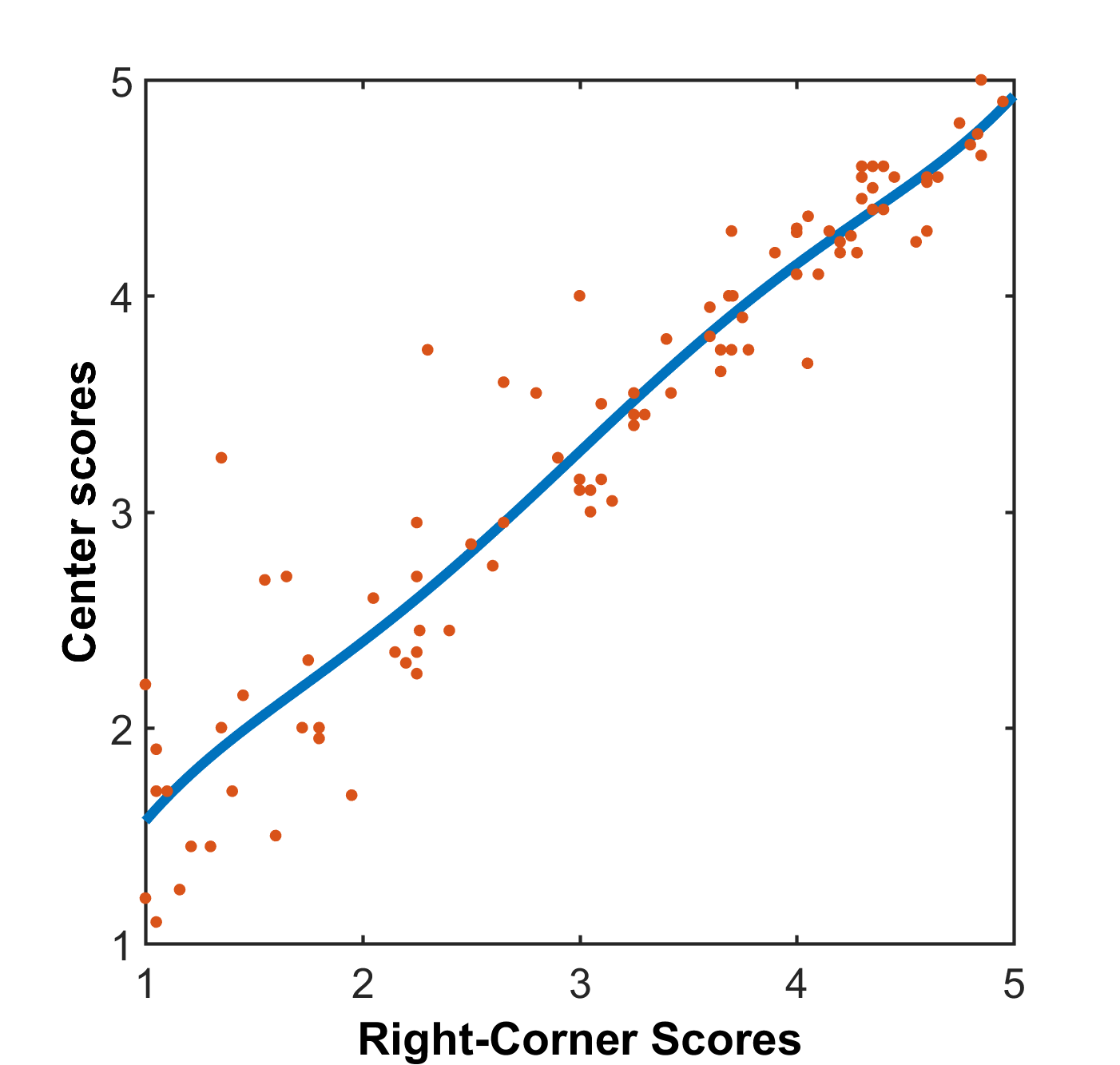}}
	\subfloat[ $MOS_{LFc}$ vs $MOS_{LFr}$ ]
	{\includegraphics[width=0.3\textwidth]{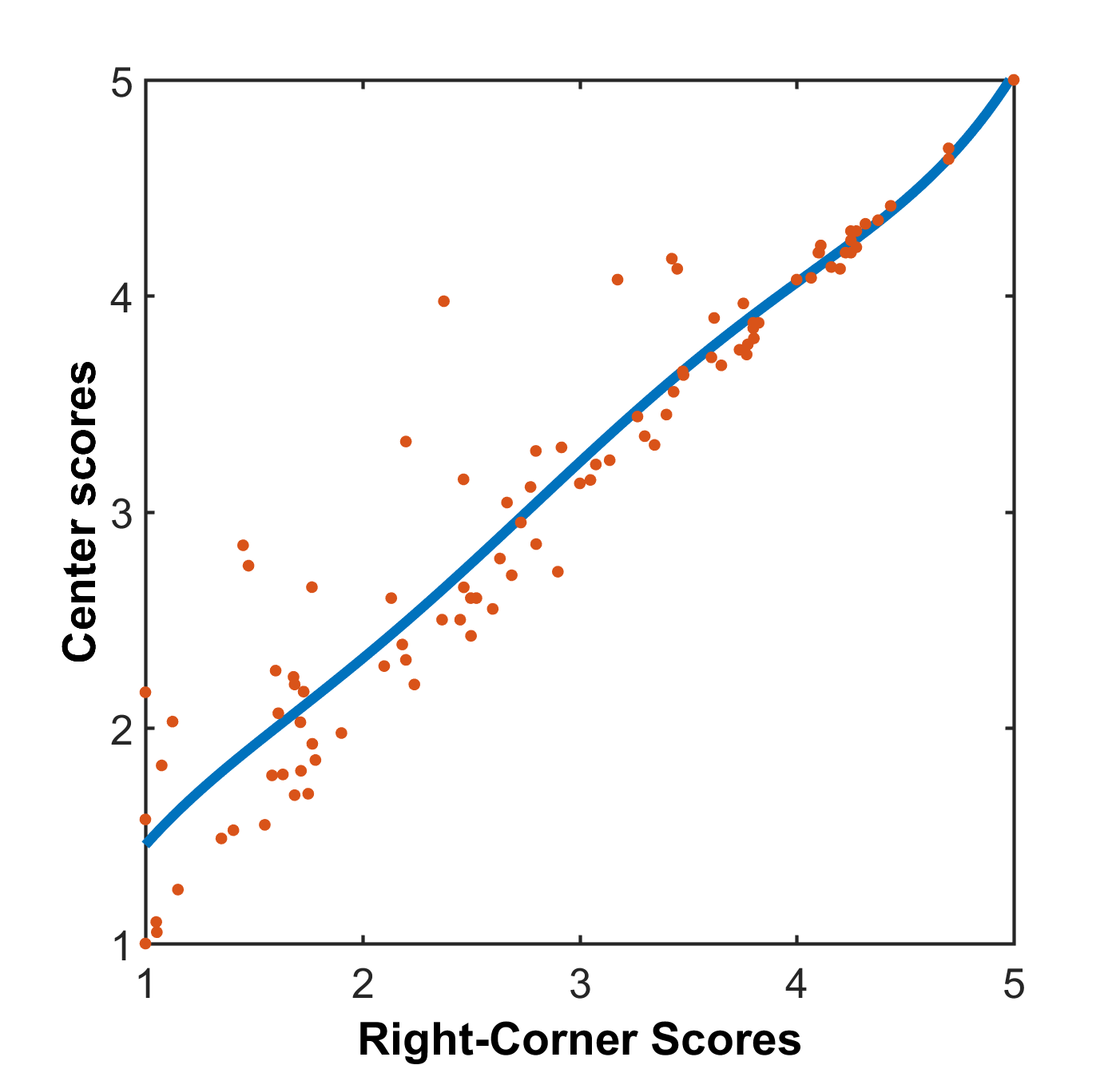}}
	\subfloat[ $MOS_{2Dc}$ vs $MOS_{2Dr}$ ]
	{\includegraphics[width=0.3\textwidth]{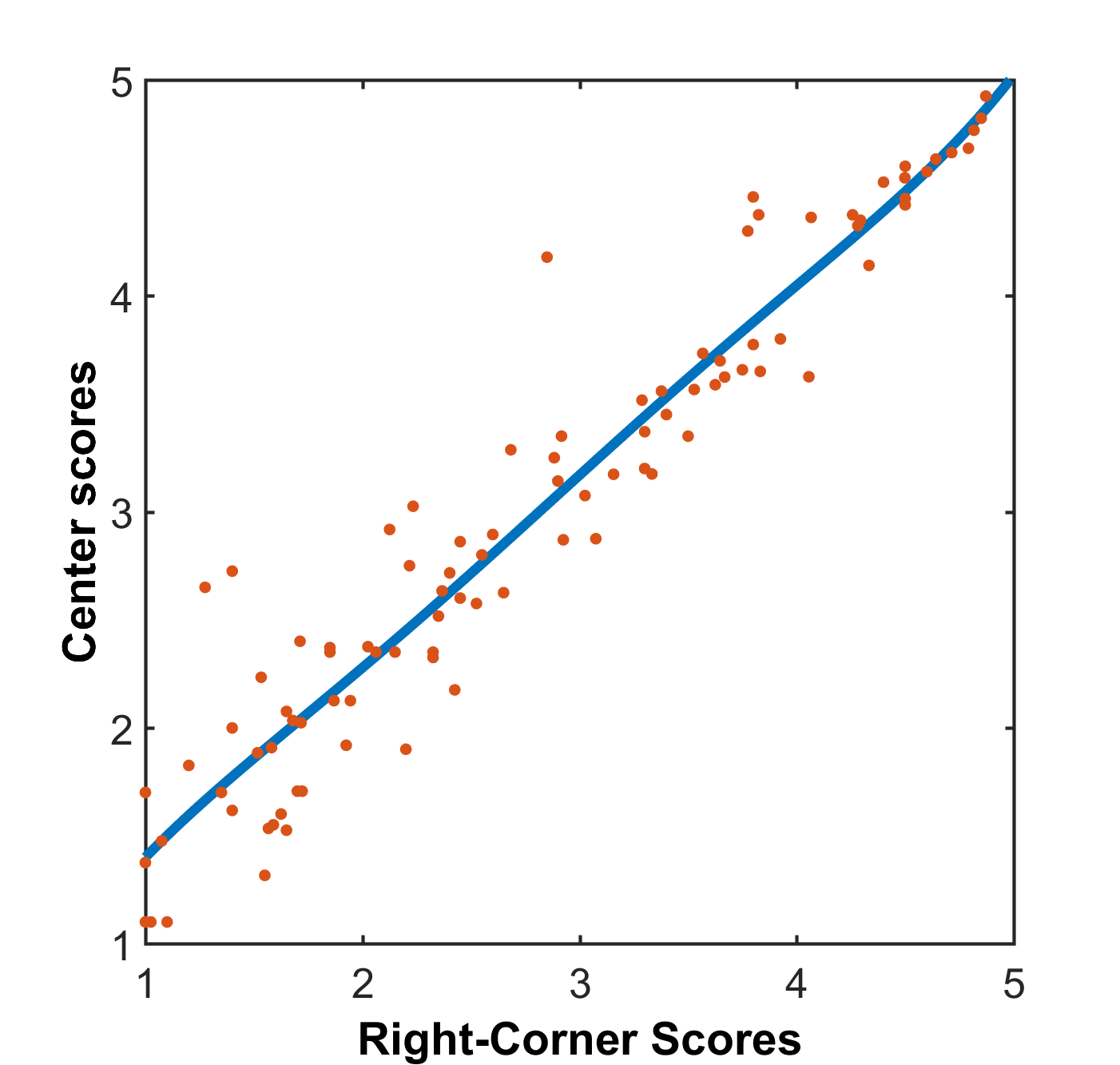}}
	
\caption{Overall comparison of the center view MOS versus right corner view MOS for (\ref{fig:allMOS_CR-comp_OPT}, d) Holographic display, (b,e) Light-field display and (c,f) 2D regular display. In (a,b,c) additionally 4th-degree polynomial fit lines are shown for the data with indices of the sorted center view MOS.}
\label{fig:allMOS_CR-comp}%
\vspace*{-1em}
\end{figure*}

\subsection{MOS analysis based on perspective } 
Next, the correlation between the scores for the two tested perspectives are evaluated. Fig.~\ref{fig:allMOS_CR-comp} shows per setup the comparison of the MOS from the center view with the right-corner view. For each setup, first the center scores were sorted and the sorting indices were used to plot the corner view MOS. The $95\%$ confidence intervals for each perspective are shown as well. To avoid clutter and to further clarify the trend, only 4th degree polynomial fit lines for the mentioned data are shown in Fig.~\ref{fig:allMOS_CR-comp} (a,b,c). To provide more detail, scatter plots of the center vs right corner MOSes are shown in Fig.~\ref{fig:allMOS_CR-comp} (d,e,f). The score plots clearly show a distinct trend across the setups where central views regularly obtain a higher MOS compared to the corner views of the same hologram. However, the score difference evolves across the quality range. More specifically, for all setups, the corner view MOS for high quality holograms (holograms with center view MOS higher than 3.5) remains within the confidence interval fits of the center view MOS. On the other hand for holograms in the lower end of quality range, the differences increase. This is perfectly in line with the expected behaviour of how some encoders compress the holograms. When performing lossy compression the general objective is to weight the transform components in the space-frequency domain higher, which carry more visually important information. However, if very high compression ratios are demanded, this will translate into complete elimination of the weakest coefficients. This leads, in the case of the chosen WAC variant, to an introduction of overlapping first diffraction orders by imperfect coefficient cancellation, which is more pronounced away from the center. In the case of the other selected methods, it leads to an elimination of high frequency components, which correspond to high diffraction angles (corner view information). The MOS variations experimentally reveals this shortcoming of the current holographic encoders. The scatter plots show furthermore that there are some cases that do not follow this difference trend. In some extreme cases the center MOS is 1.5 points higher than the corner MOS. 
\begin{figure*}
\centering
	\subfloat[ $MOS_{OPTc}$ vs $MOS_{LFc}$]
	{\includegraphics[width=0.3\textwidth]{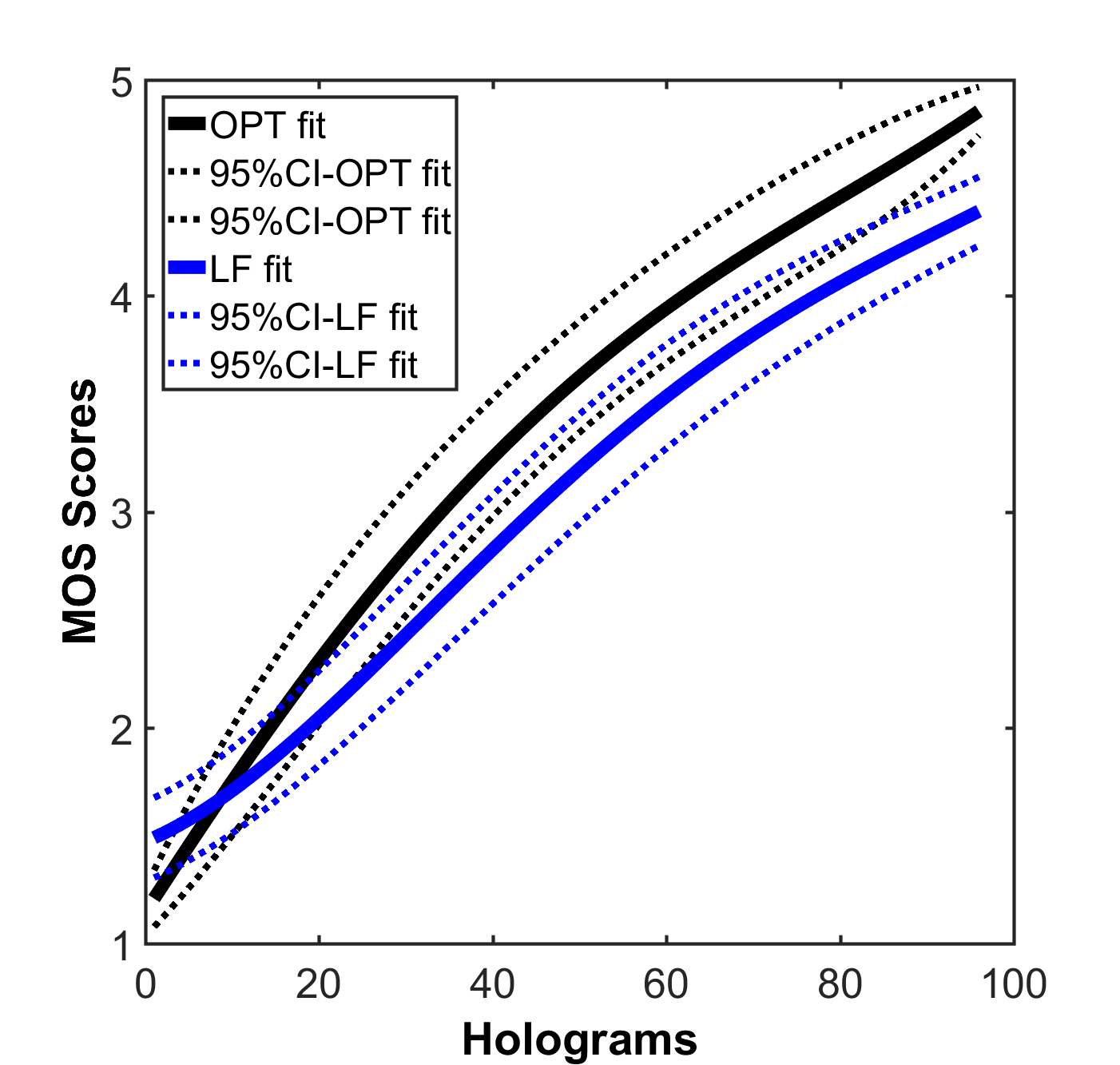}}
	\subfloat[ $MOS_{OPTc}$ vs $MOS_{2Dc}$ ]
	{\includegraphics[width=0.3\textwidth]{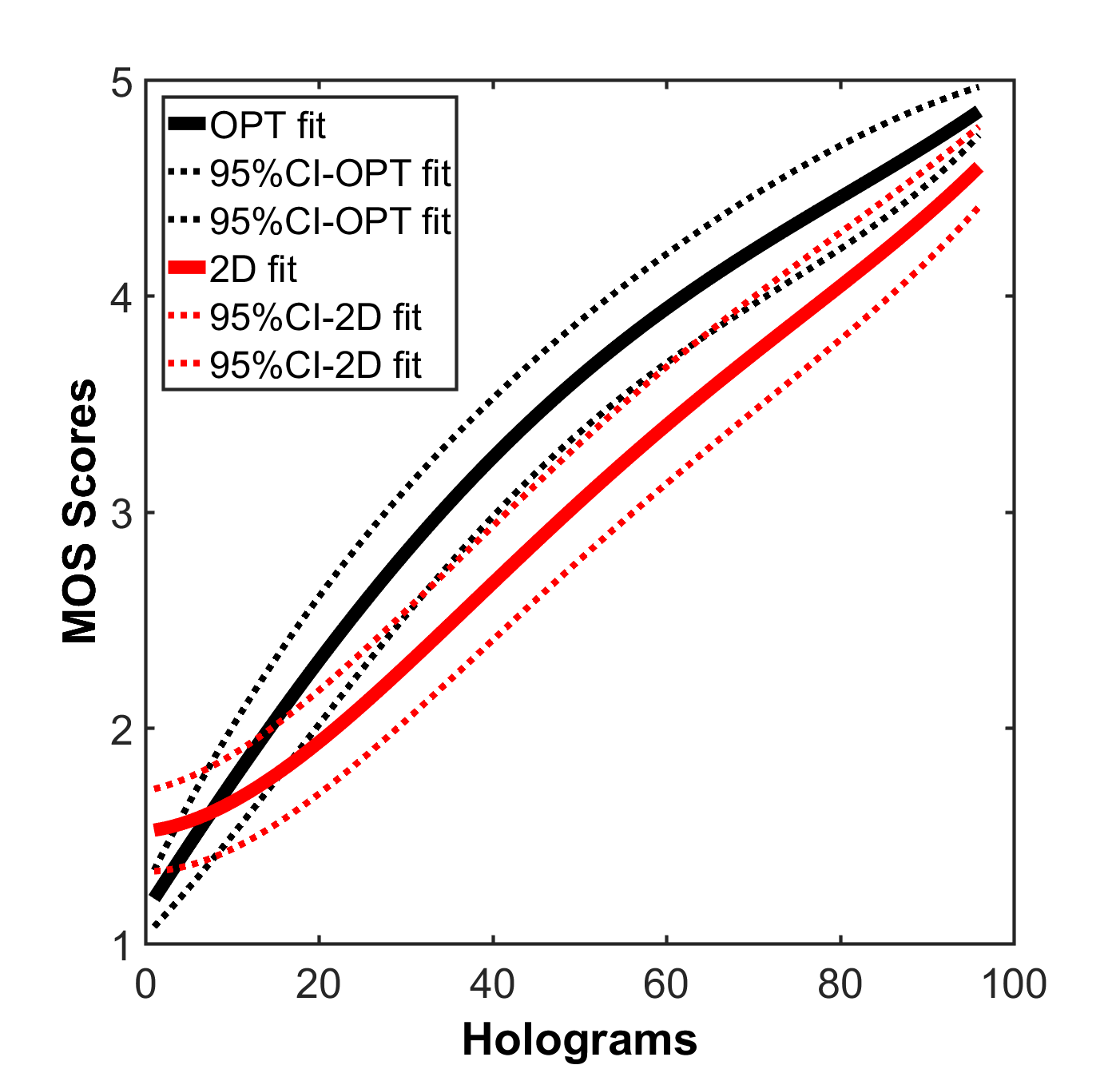}}
	\subfloat[ $MOS_{LFc}$ vs $MOS_{2Dc}$ ]
	{\includegraphics[width=0.3\textwidth]{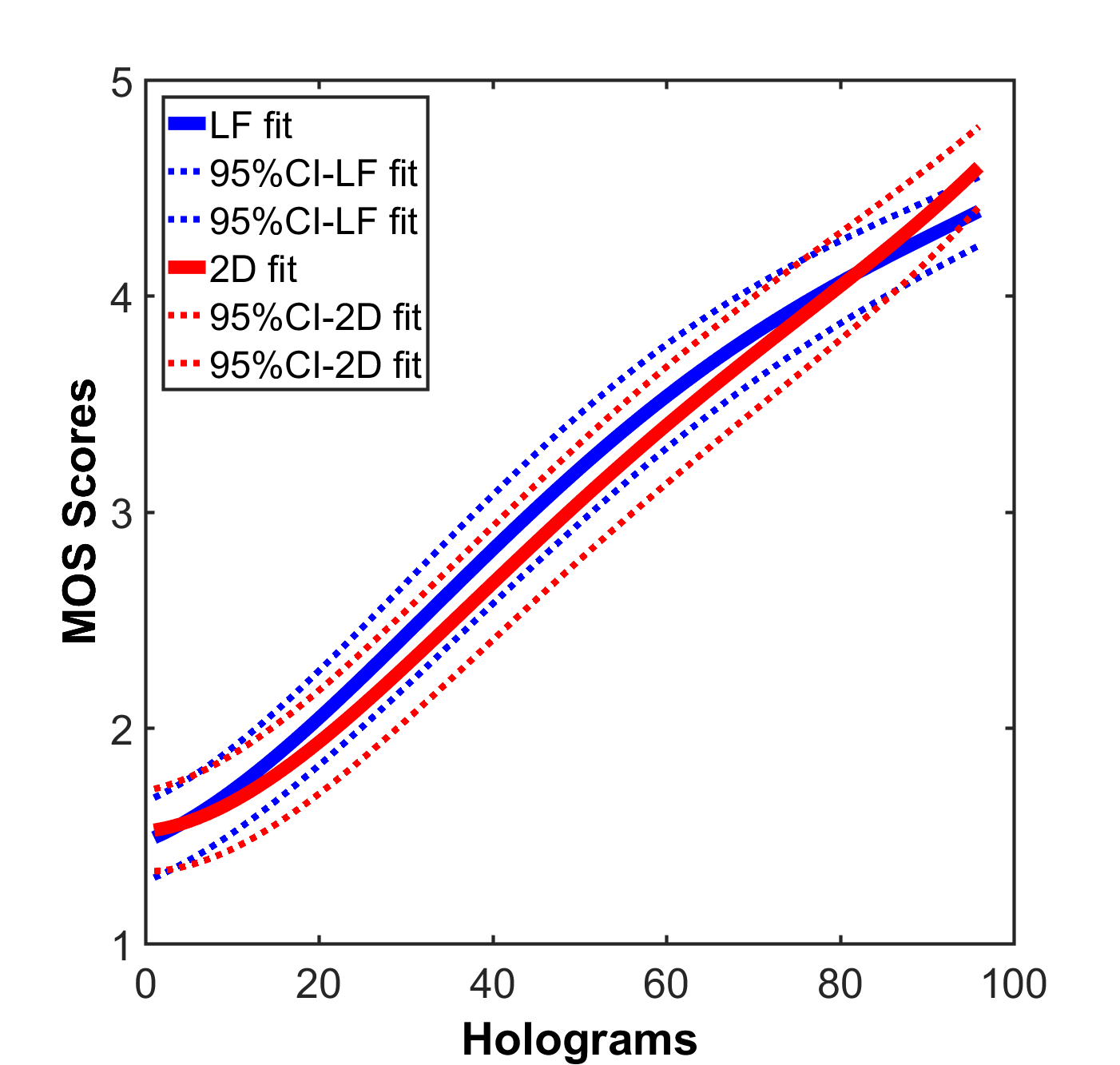}}
	\vspace*{-1em}
	\subfloat[ $MOS_{OPTc}$ vs $MOS_{LFc}$]
	{\includegraphics[width=0.3\textwidth]{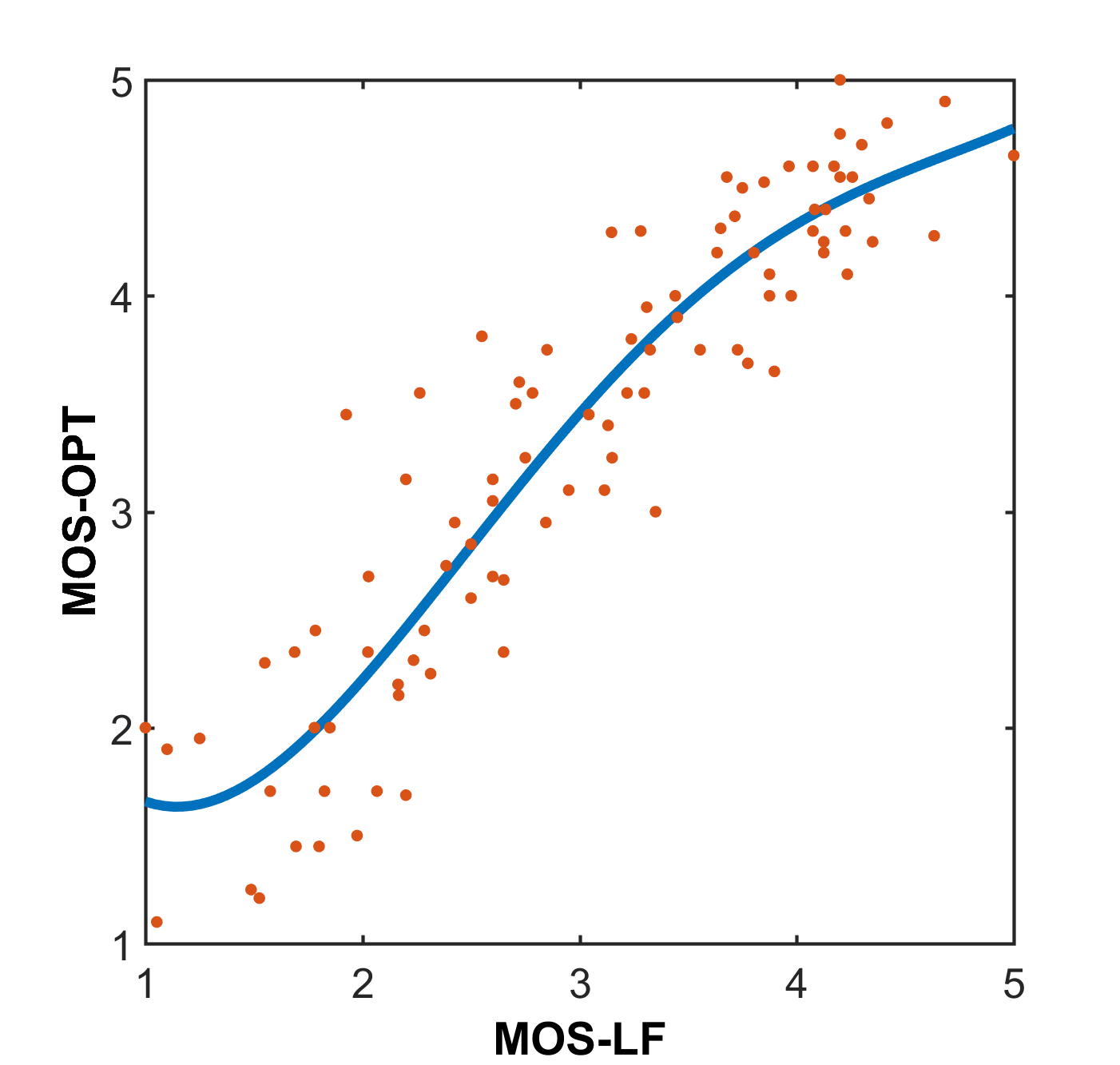}}
	\subfloat[ $MOS_{OPTc}$ vs $MOS_{2Dc}$]
	{\includegraphics[width=0.3\textwidth]{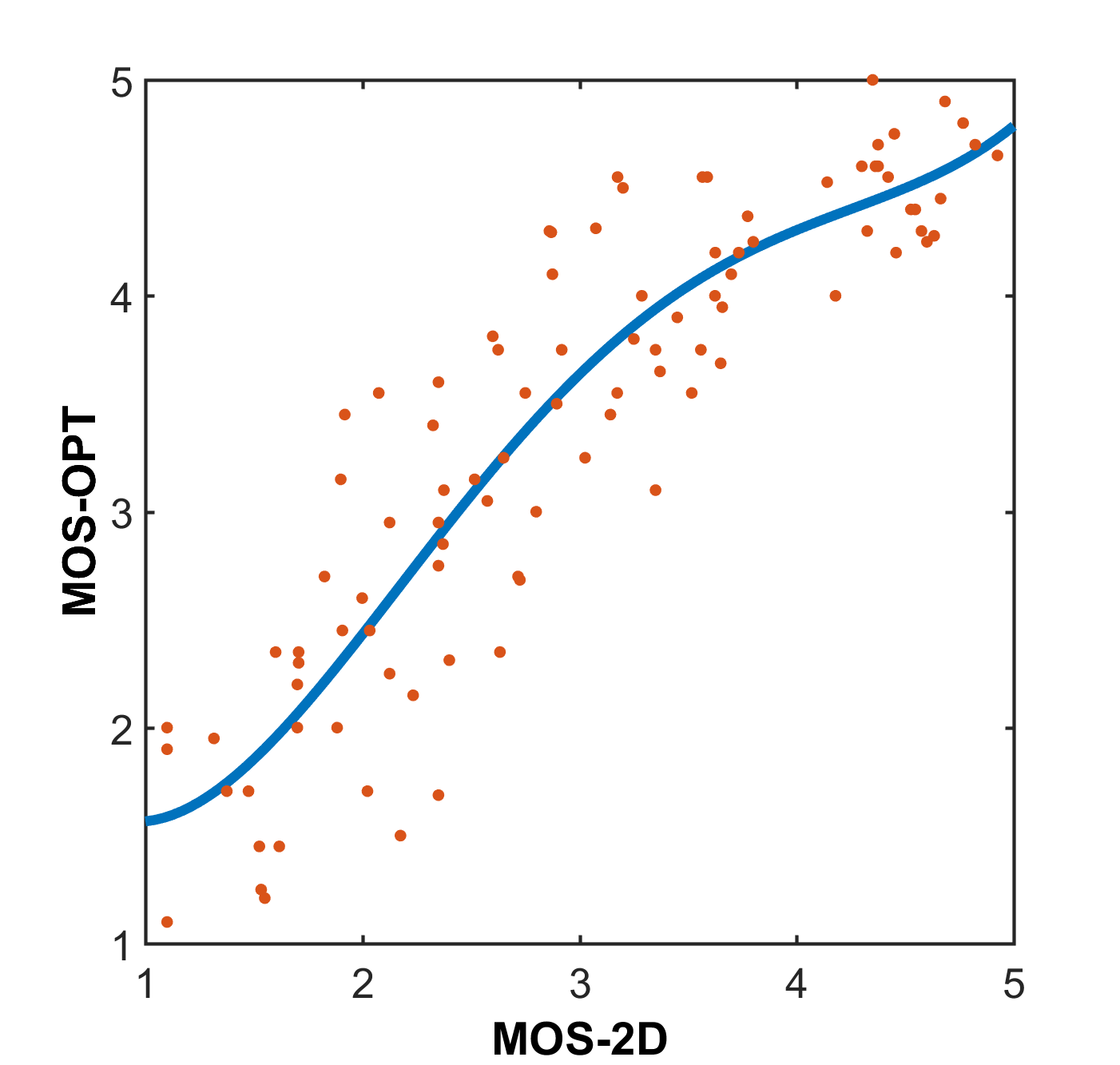}}
	\subfloat[ $MOS_{LFc}$ vs $MOS_{2Dc}$]
	{\includegraphics[width=0.3\textwidth]{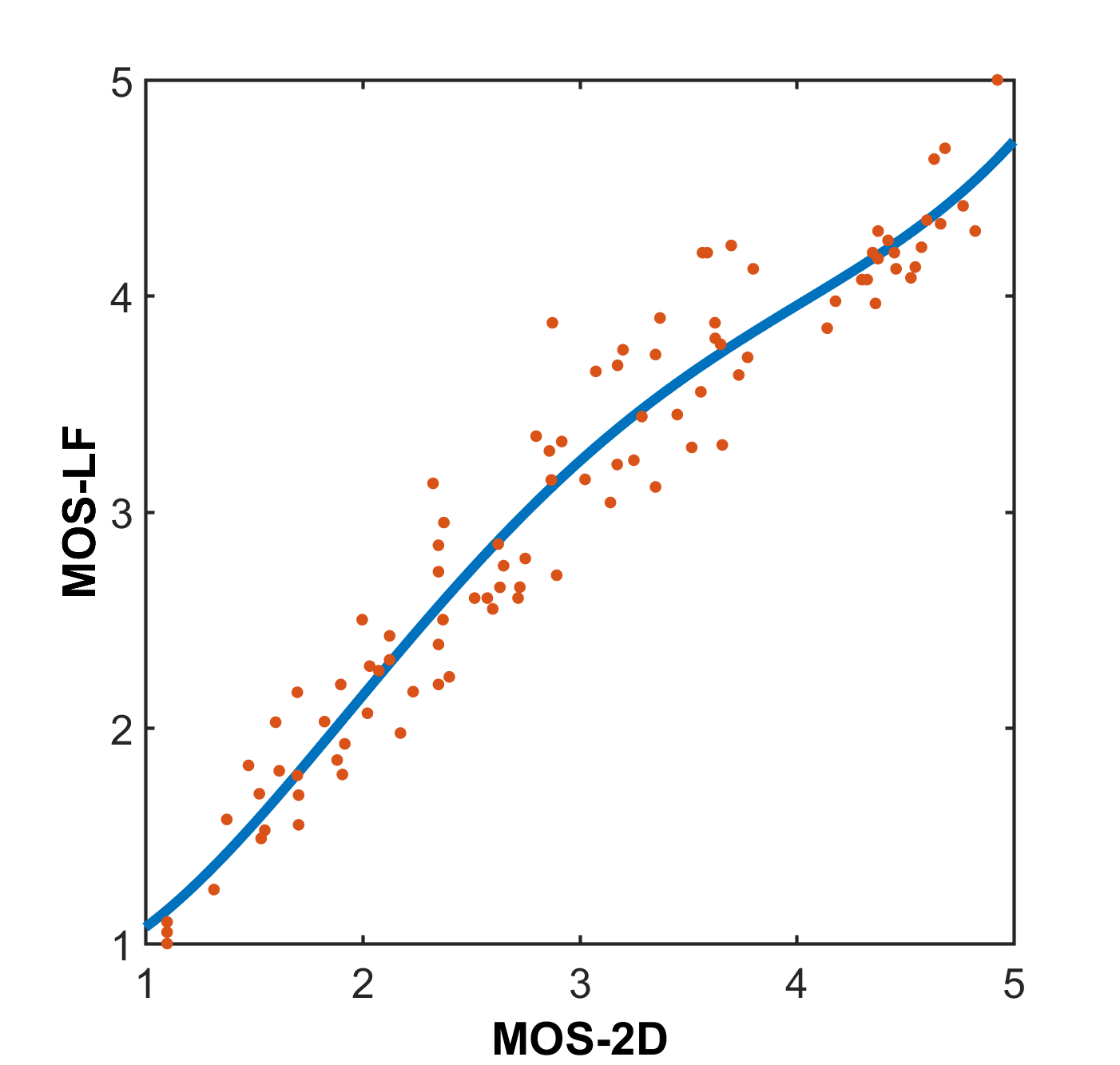}}
	
\caption{Center-view MOS comparison for holographic versus Light Field display (a,d), holographic versus 2D display (b,e) and Light Field versus 2D display (c,f).}
\label{fig:allMOSc_OPT-LF-2D}
\end{figure*}

\subsection{Inter-setup comparison of results} 
In this section, the MOS results obtained with the different display systems are compared. First, the overall trend of the scores is evaluated. Thereafter, a more detailed analysis is performed related to the influence of the characteristics of the encoded objects and the bit-depths used to encode them in this experiment. 
\begin{table*}[t]
	\centering
	\caption{Calculated coefficients of the fit functions, which enable to map the gathered MOS from different setups into another for the center and the corner views. Pearson and Spearman correlation coefficients are provided to evaluate the accuracy of the functions. As for the robustness, the last column shows the maximum absolute error in unit score for the predicted fit, if one of the test subjects changes a score for a condition with $\pm1$ unit score.}
\begin{tabu}{|l|l|c|c|c|c|c|[2pt]c|c|c|c|[2pt]c|}
\hline
\multicolumn{2}{|l|}{\multirow{2}{*}{}}                 & \multicolumn{5}{l|[2pt]}{$ p(x)= p_{1}x^{4}+p_{2}x^{3}+p_{3}x^{2}+p_{4}x+p_{5}.$} & \multicolumn{2}{c|}{\textbf{Before Fit}}                                       & \multicolumn{2}{c|[2pt]}{\textbf{After Fit}}                                        & \multirow{2}{*}{\textbf{Error}} \\ \cline{3-11}
\multicolumn{2}{|l|}{}                                  & $p_{1}$       & $p_{2}$        & $p_{3}$      & $p_{4}$       & $p_{5}$      & \multicolumn{1}{l|}{\textbf{Pearson}} & \multicolumn{1}{l|}{\textbf{Spearman}} & \multicolumn{1}{l|}{\textbf{Pearson}} & \multicolumn{1}{l|[2pt]}{\textbf{Spearman}} &                                 \\ \hline
\multirow{3}{*}{\textbf{Center View}} & \textbf{LF $\rightarrow$ OPT} & 0.03923       & -0.56276       & 2.72965      & -4.27371      & 3.72549      & 0.9179                                & 0.9210                                 & 0.9873                                & 0.9992                                 & 0.0273                          \\ \cline{2-12} 
& \textbf{2D $\rightarrow$ OPT}  & 0.05112       & -0.65713       & 2.8338       & -3.80237      & 3.14057      & 0.8824                                & 0.8975                                 & 0.9946                                & 0.9998                                 & 0.0279                          \\ \cline{2-12} 
& \textbf{2D $\rightarrow$ LF}   & 0.03264       & -0.38953       & 1.52786      & -1.27431      & 1.17860      & 0.9587                                & 0.9650                                 & 0.9938                                & 0.9999                                 & 0.0299                          \\ \hline
\multirow{3}{*}{\textbf{Corner View}} & \textbf{LF $\rightarrow$ OPT}  & 0.00585       & -0.13114       & 0.74093      & -0.41638      & 1.06515      & 0.9342                                & 0.9357                                 & 0.9968                                & 0.9998                                 & 0.0259                          \\ \cline{2-12} 
& \textbf{2D $\rightarrow$ OPT}  & 0.02748       & -0.31877       & 1.12589      & -0.21345      & 0.47994      & 0.9257                                & 0.9405                                 & 0.9958                                & 0.9998                                 & 0.0221                          \\ \cline{2-12} 
& \textbf{2D $\rightarrow$ LF}   & 0.02725       & -0.30645       & 1.12026      & -0.54832      & 0.78461      & 0.9531                                & 0.9538                                 & 0.9936                                & 0.9999                                 & 0.0349                          \\ \hline
\end{tabu}
\label{tab:FitFuns_Corrs}
\vspace*{-1em}
\end{table*}

\begin{figure*}
\centering
	\subfloat[ $MOS_{OPTr}$ vs $MOS_{LFr}$]
	{\includegraphics[width=0.3\textwidth]{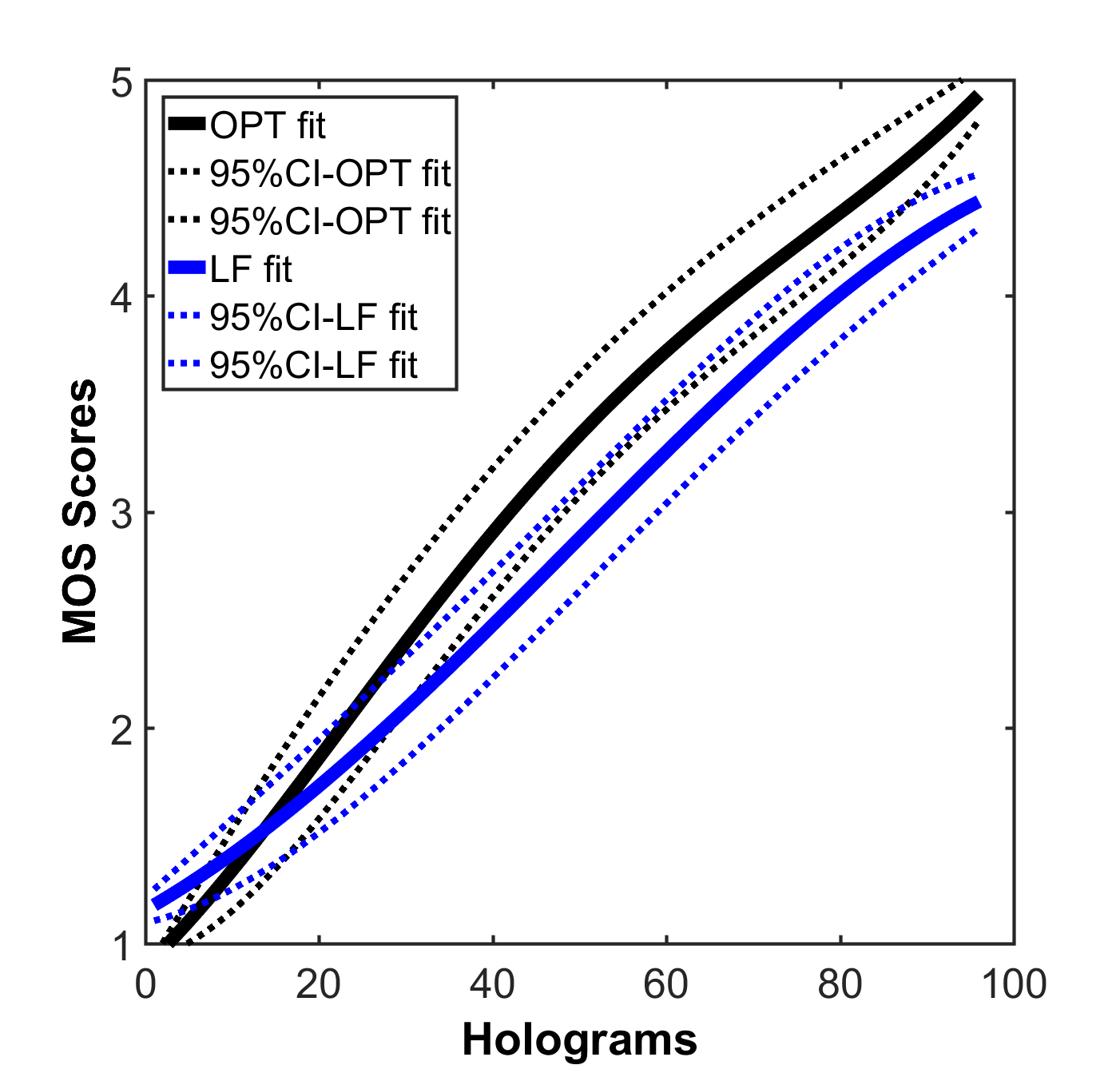}}
	\subfloat[ $MOS_{OPTr}$ vs $MOS_{2Dr}$ ]
	{\includegraphics[width=0.3\textwidth]{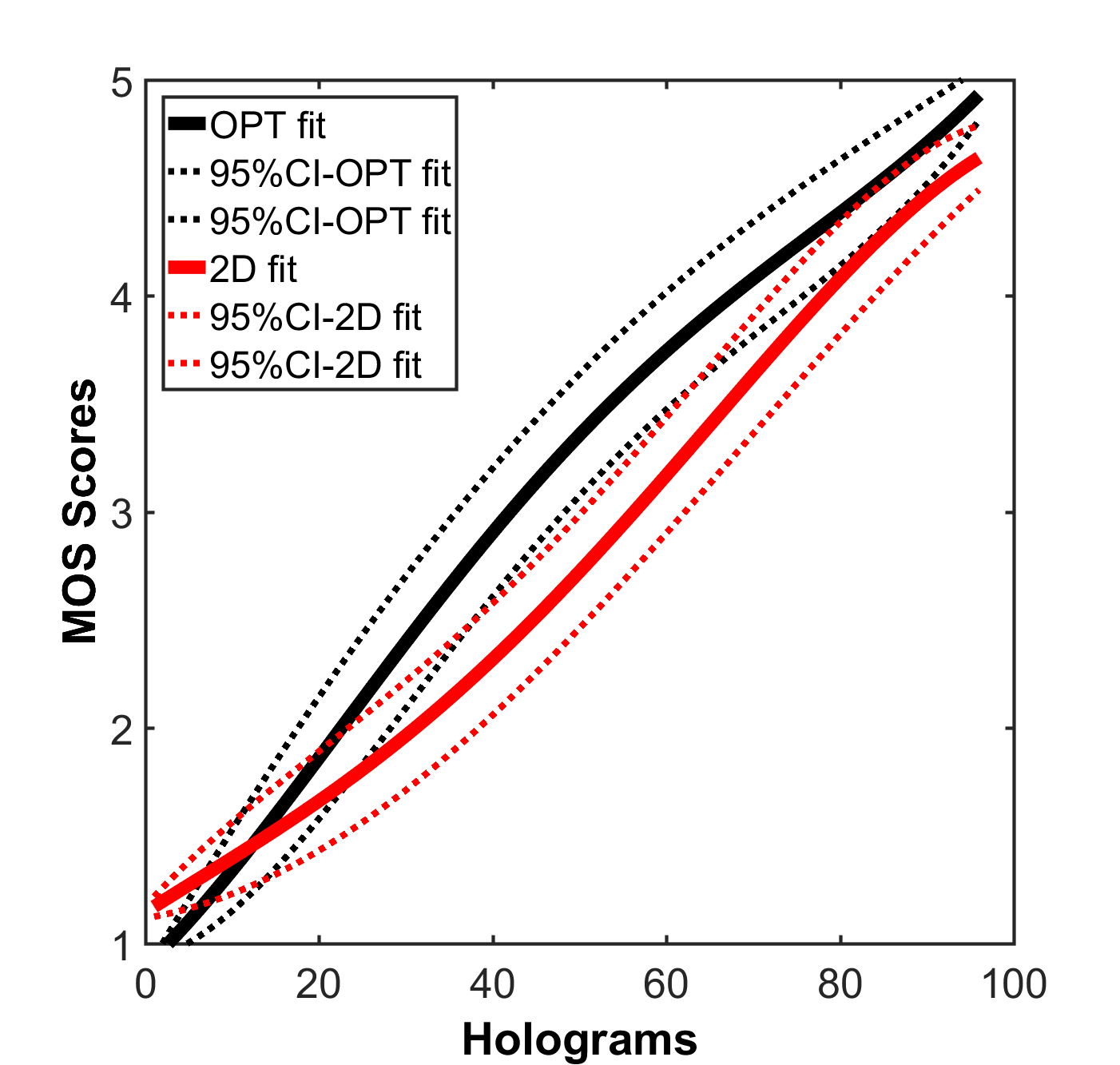}}
	\subfloat[ $MOS_{LFr}$ vs $MOS_{2Dr}$ ]
	{\includegraphics[width=0.3\textwidth]{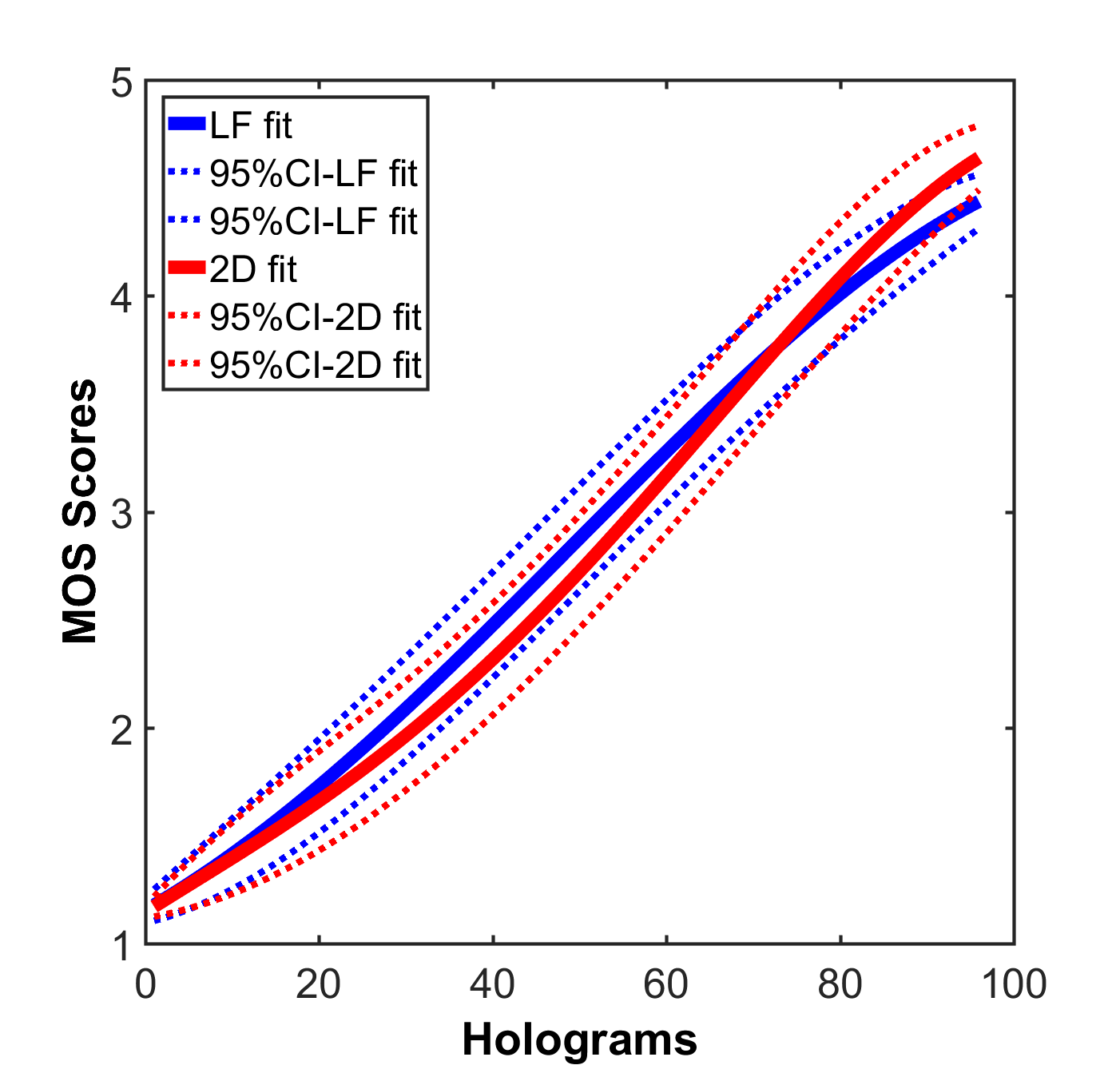}}
	\vspace*{-1em}
	\subfloat[ $MOS_{OPTr}$ vs $MOS_{LFr}$]
	{\includegraphics[width=0.3\textwidth]{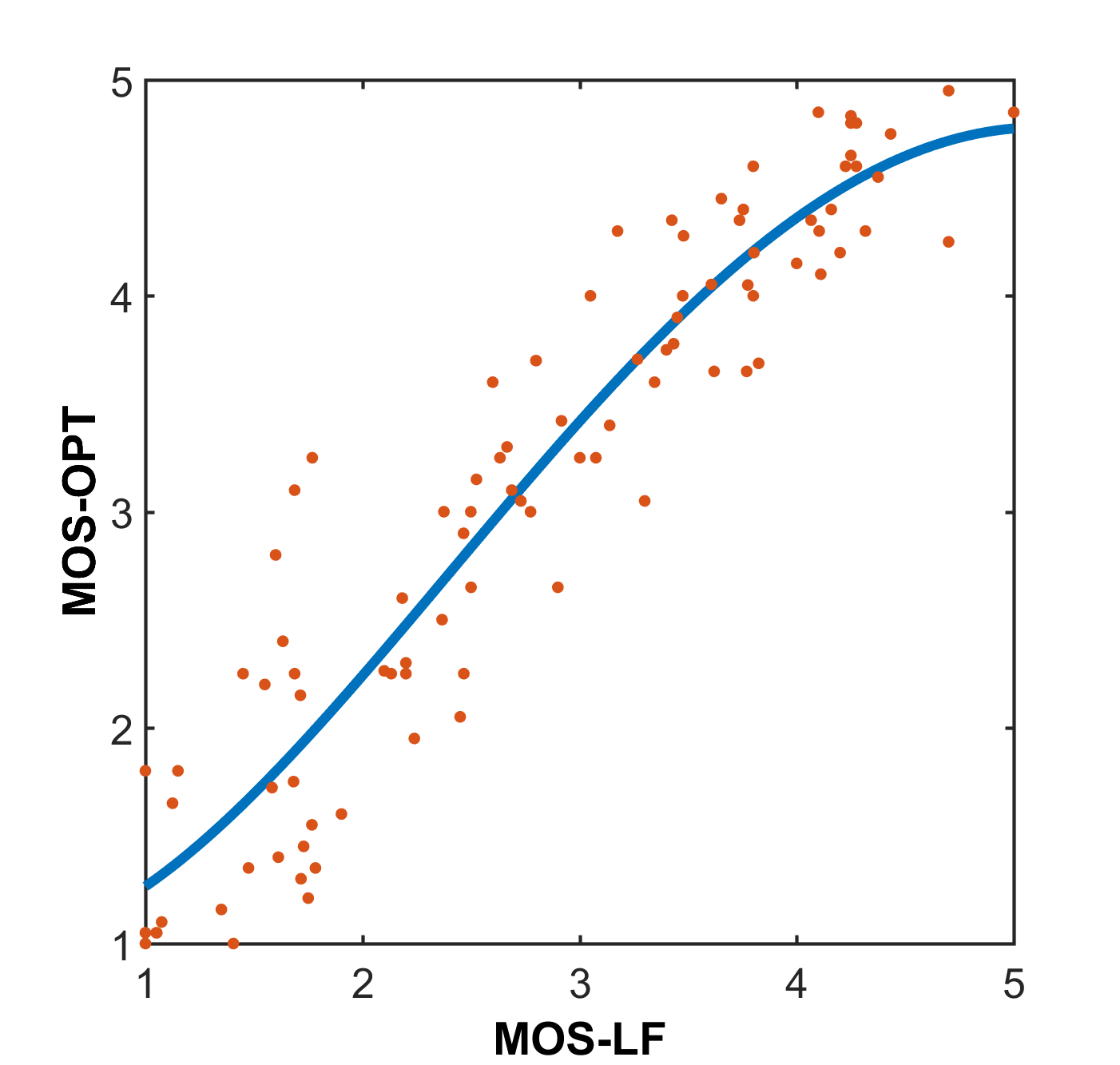}}
	\subfloat[ $MOS_{OPTr}$ vs $MOS_{2Dr}$]
	{\includegraphics[width=0.3\textwidth]{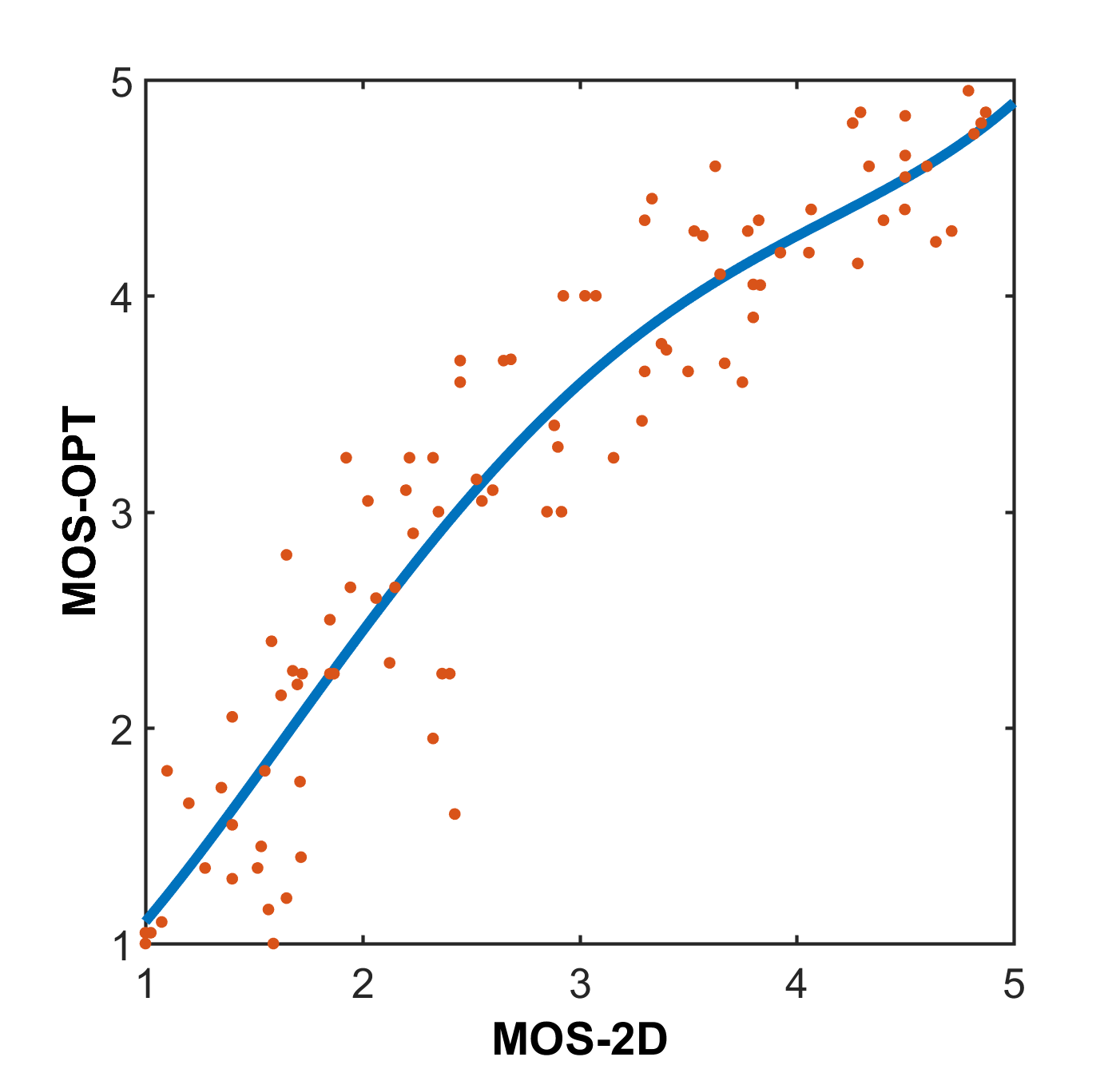}}
	\subfloat[ $MOS_{LFr}$ vs $MOS_{2Dr}$]
	{\includegraphics[width=0.3\textwidth]{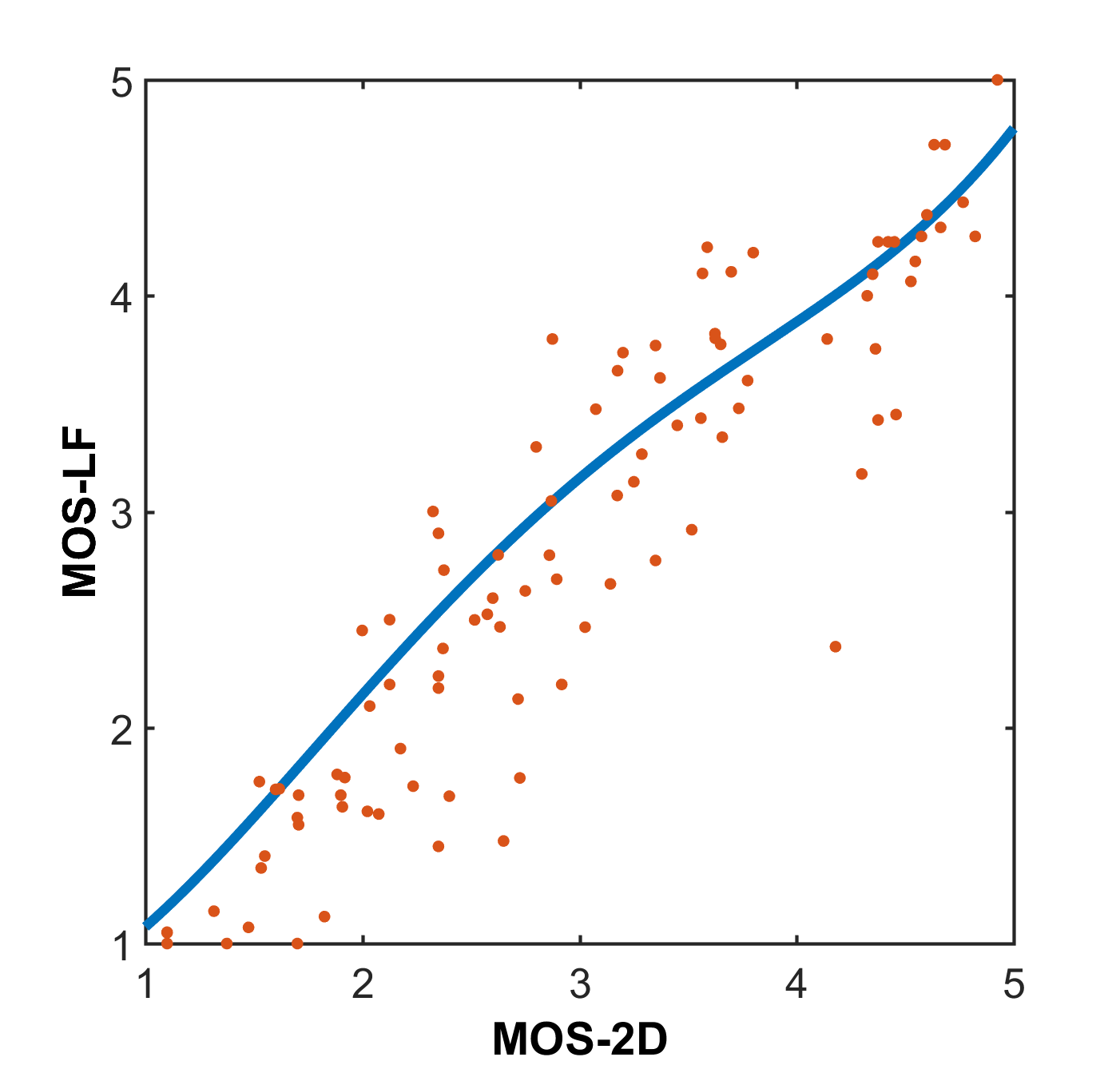}}
	
\caption{Right corner-view MOS comparison for holographic versus Light Field display (a,d), holographic versus 2D display (b,e) and Light Field versus 2D display (c,f).}
\label{fig:allMOSr_OPT-LF-2D}%
\vspace*{-1em}
\end{figure*}

Fig.~\ref{fig:allMOSc_OPT-LF-2D} shows the inter-setup comparison for the center-view. Similar to Fig.~\ref{fig:allMOS_CR-comp}, the MOS results for the optical setup (a,b) and for the light field display (c) were sorted and their order was used to plot the other corresponding MOS for the other setup. The lines (solid and dashed) depict again the 4th-degree polynomial fits for the corresponding data and their confidence intervals. The data shown in Fig.~\ref{fig:allMOSc_OPT-LF-2D}(a, d), represent the comparison between the optical and light field setup. Interestingly, a specific gap exists between the two setups. For each hologram the MOS obtained for the optical setup is typically higher than the MOS for the light field setup - except in the very low quality range. A similar trend can be observed in Fig.~\ref{fig:allMOSc_OPT-LF-2D} (b,e) where the optical setup was compared with a 2D display setup. However, the gap for the mid-range quality holograms (scores $2.5-4$) is slightly larger now. On the other hand, the graphs shown in Fig.\ref{fig:allMOSc_OPT-LF-2D}(c, f), demonstrate a rather close agreement between the MOS of the light field and 2D display setups. A comparison for the right corner view perspective is provided in Fig.~\ref{fig:allMOSr_OPT-LF-2D} and the right corner view scores follow closely the trend found for the center view. However, the level of disagreement between some individual light field and 2D display scores is slightly increased for the corner-views. Unfortunately, we do not have an immediate explanation for this phenomena.

Additionally, for the cases where quality scores of a real holographic setup are not available, while having access to the 2D or light field scores; one can use the fit-functions evaluated and shown as blue lines in the scatter plots of Fig.~\ref{fig:allMOSc_OPT-LF-2D} (d,e) and Fig.~\ref{fig:allMOSr_OPT-LF-2D} (d,e) to predict the scores for the same data reconstructed in optical setup. 

Table \ref{tab:FitFuns_Corrs} shows the coefficients for the 4th-degree polynomials (Fig.~\ref{fig:allMOSc_OPT-LF-2D} and Fig.~\ref{fig:allMOSr_OPT-LF-2D}), which are best fit in a least-squares sense, for the three comparisons between each pair of the display setups. During our experiments, the 4-th degree polynomial showed the lowest regression error while not over-fitting the data when comparing the fitting behaviour of polynomials of degree 1 to 7. In the same table, the Pearson and Spearman correlation coefficients are shown before and after applying each fit function to the data. A logical concern, which arises here, is the robustness of the provided fit functions. For example, in the case a test subject changes its opinion about a hologram. The last column of the Table \ref{tab:FitFuns_Corrs} represents the maximum possible change of the fitted values, if a single subject changes the score for a test condition by $\pm1$ unit. The reported errors are in unit scores as well.  

These results indicate that a high correlation exists between the MOS obtained for the three display systems and, in particular, after polynomial fitting. Both Pearson and Spearman correlation coefficients are very large in the latter case, thus underlining the predictive power of 2D and light field displays with respect to a holographic setup. Nonetheless, it is important to understand these fits cannot be transferred to other display setups and a calibration process will always be needed.
When looking at the non-fitted MOS, it is interesting to observe that the 2D and LF displays are more sensitive to artifacts than the holographic display. This is partially related to the higher quality of the issued 2D and light field displays, but also due to the fact they are displaying numerically reconstructed holograms which contain more coherent speckle noise than the content rendered on the holographic display, for which a partially coherent LED illumination reduced this effect. Also non-optimal optics further reduce naturally the amount of coherent speckle noise. The test subjects were not familiar with the phenomenon of speckle noise and were instructed to ignore it for the 2D and LF displays. Though, it might have influenced their scoring.

\begin{figure*}
	\centering
	\subfloat[$(MOS_{OPTc} - MOS_{LFc})$]
	{\includegraphics[width=0.3\textwidth]{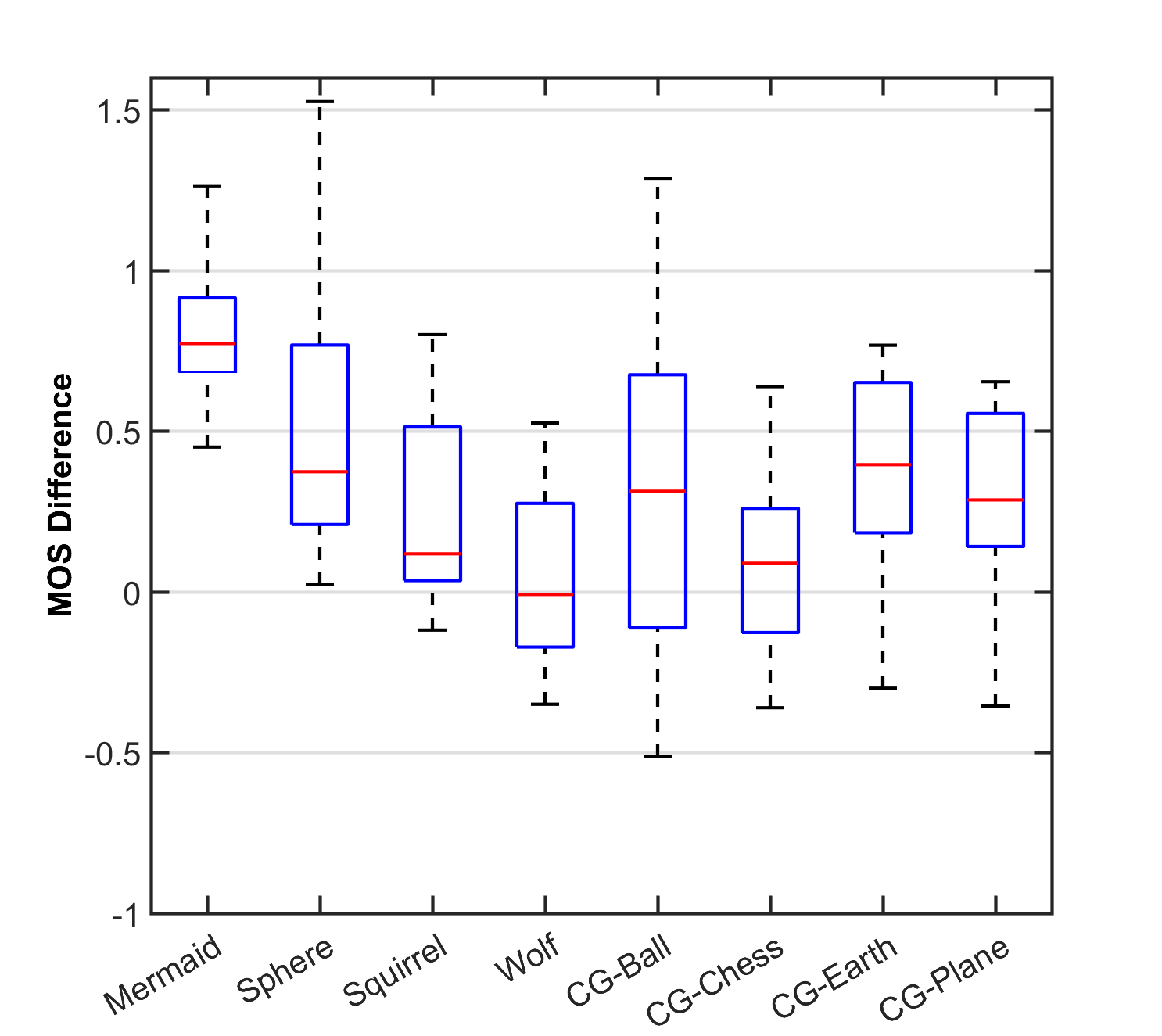}}
	\subfloat[$(MOS_{OPTc} - MOS_{2Dc})$]
	{\includegraphics[width=0.3\textwidth]{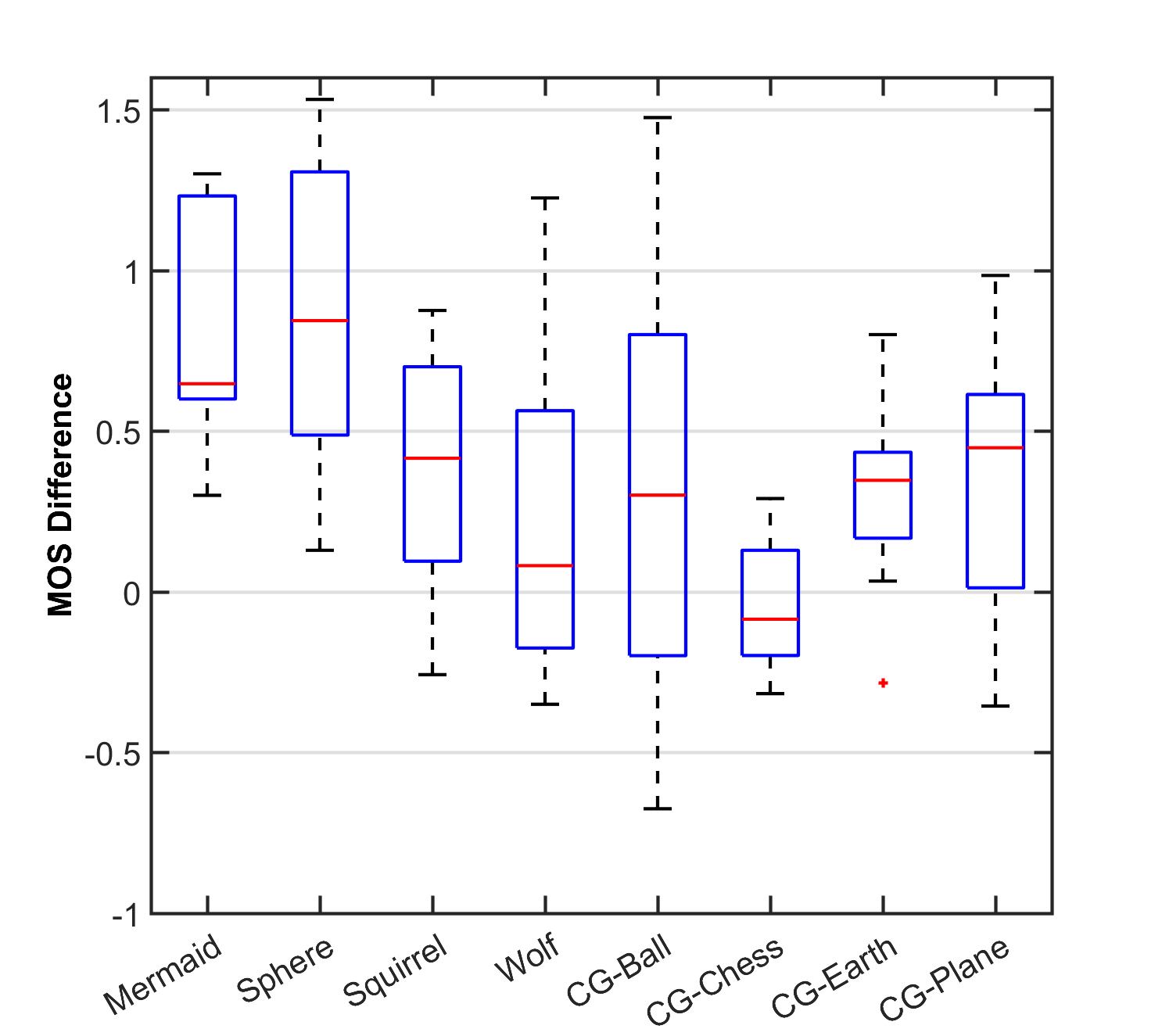}}
	\subfloat[$(MOS_{LFc} - MOS_{2Dc})$]
	{\includegraphics[width=0.3\textwidth]{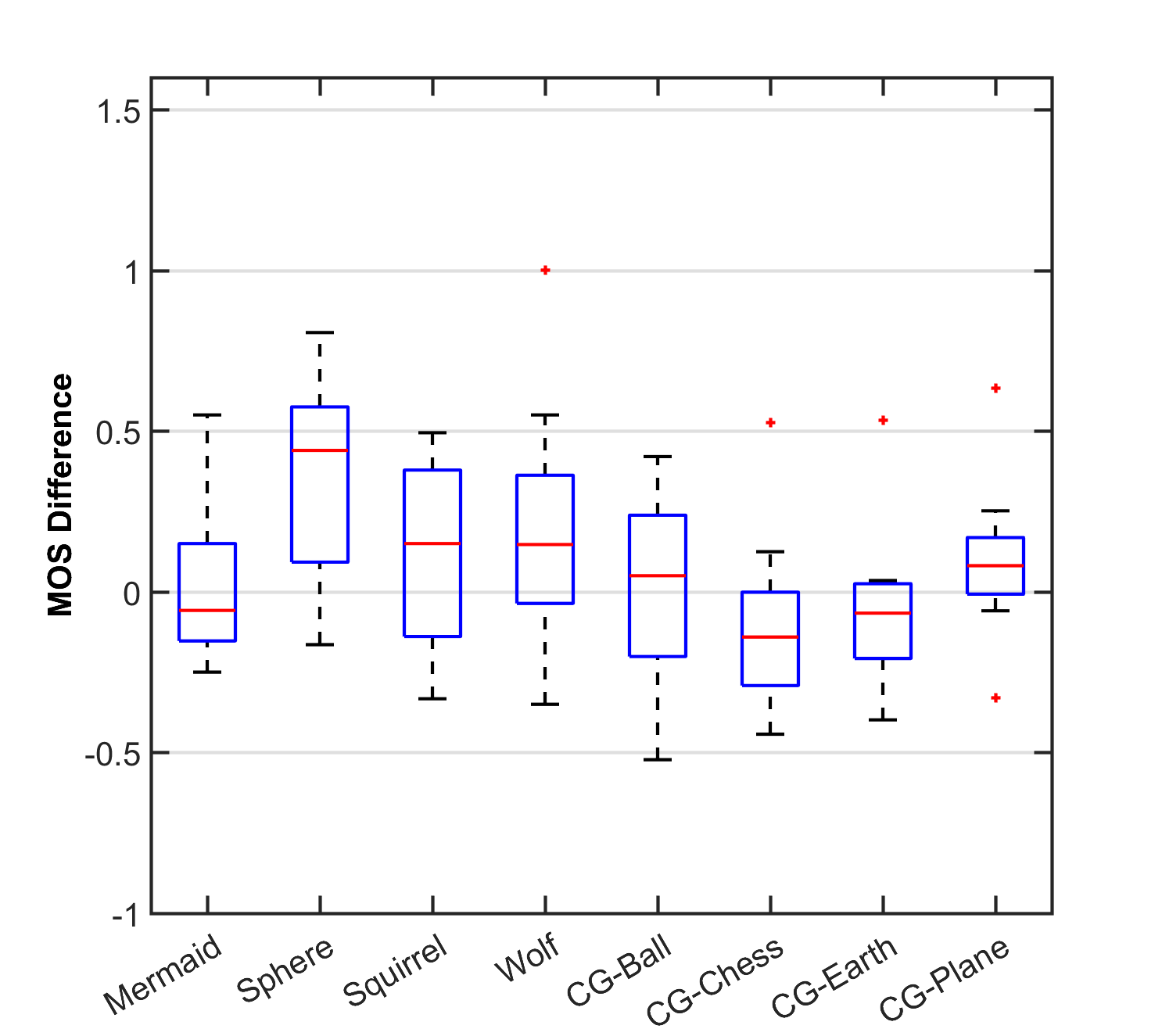}}
	\vspace*{0.15em}
	\subfloat[$(MOS_{OPTr} - MOS_{LFr})$]
	{\includegraphics[width=0.3\textwidth]{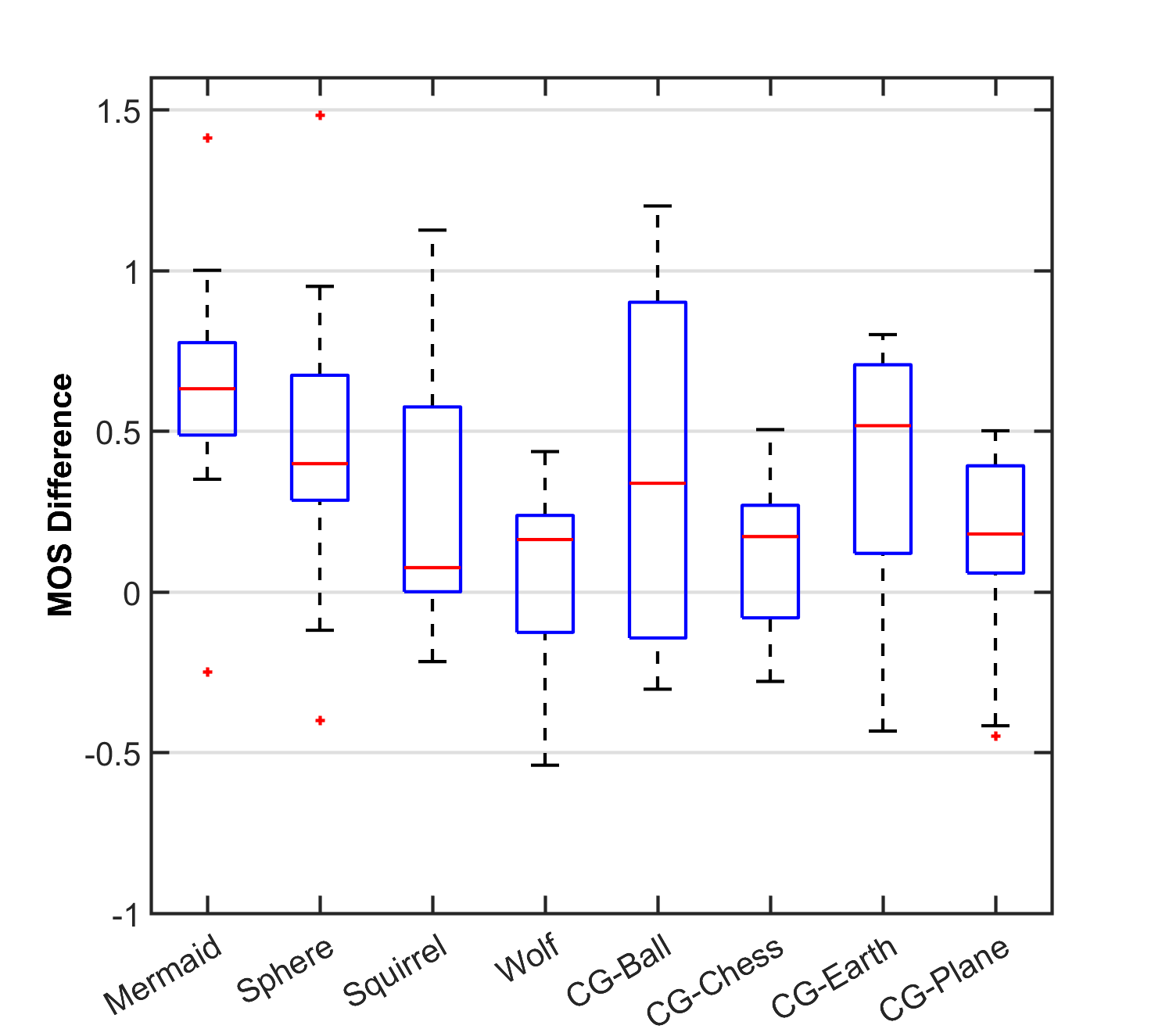}}
	\subfloat[$(MOS_{OPTr} - MOS_{2Dr})$]
	{\includegraphics[width=0.3\textwidth]{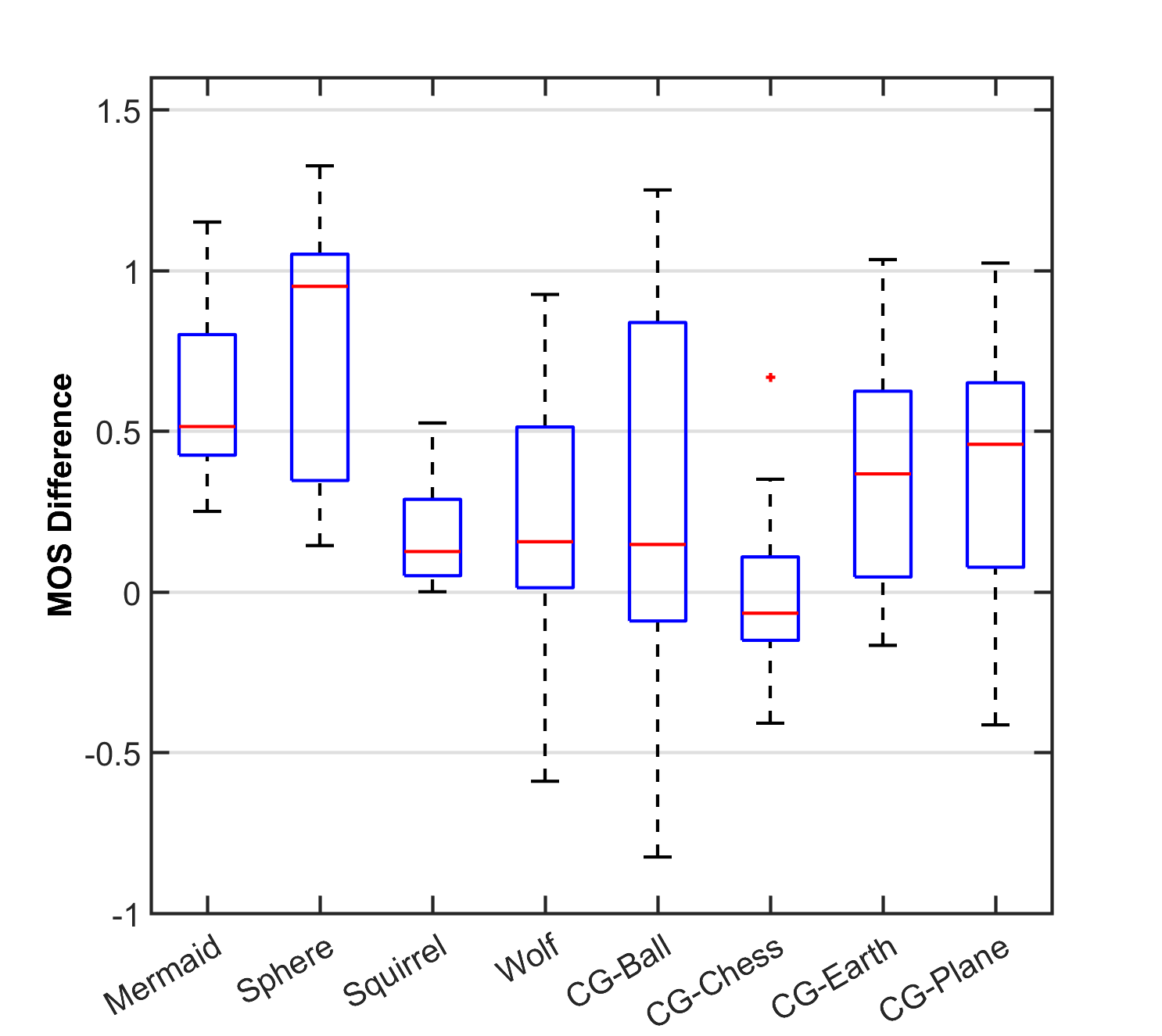}}
	\subfloat[$(MOS_{LFr} - MOS_{2Dr})$]
	{\includegraphics[width=0.3\textwidth]{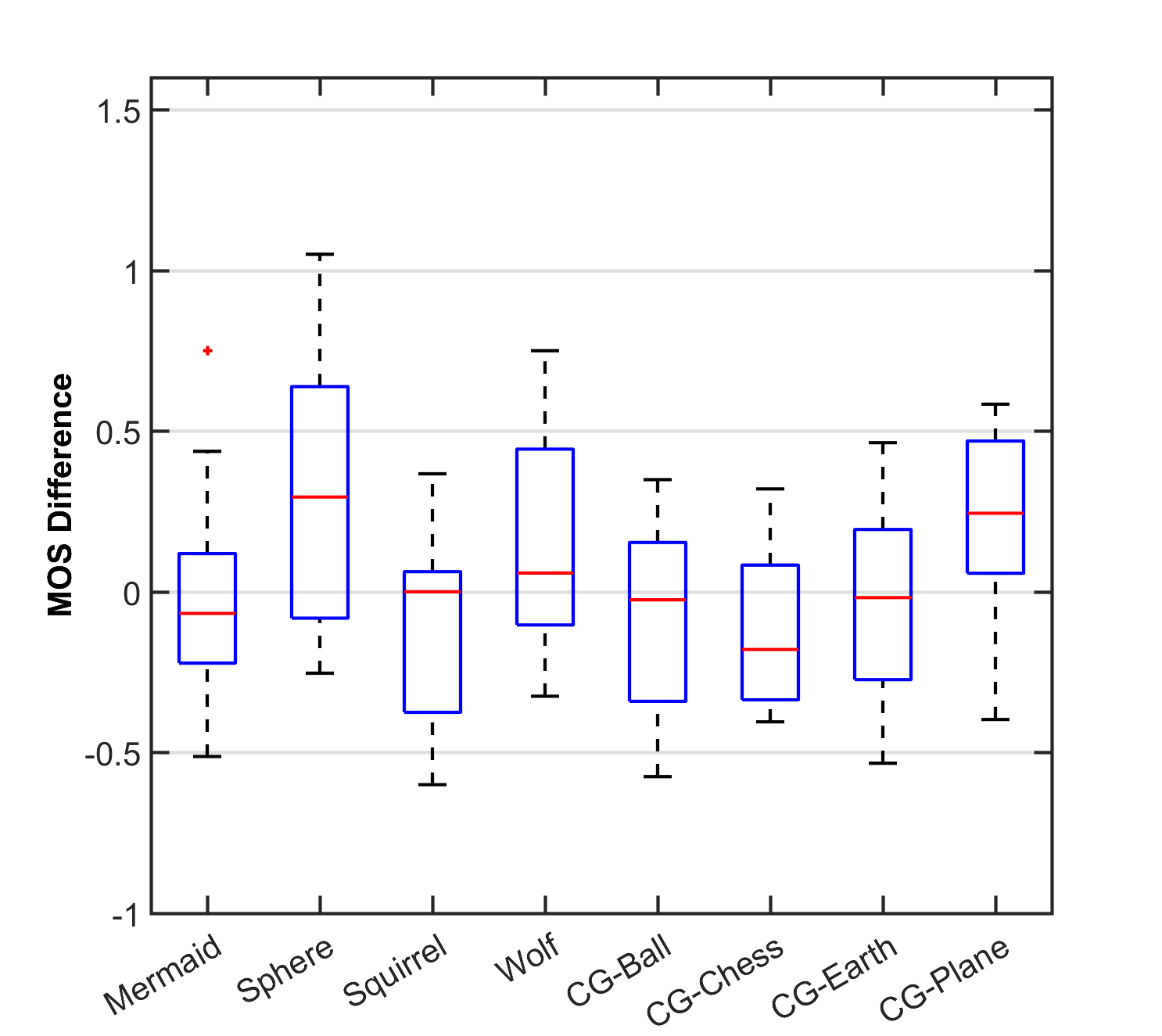}}
		
	\caption{Box plots of the MOS difference per hologram. First-row boxplots(a,b,c) corresponds to the MOS differences for center view and second row (d,e,f) show the MOS differences collected from right-corner view. Column-wise, the boxplots represent MOS diffrences between holographic (optical setup) - light field display, holographic - 2D regular display, and light field-2D display, respectively. }
	\label{fig:BoxplotPerOBJ}%
	\vspace*{-1em}
\end{figure*}

Apart from the overall inter-setup comparisons, it is interesting to analyse the influence of the tested hologram on the scoring behaviour of the test subjects: do responses to the test display systems differ for different holograms? For reasons of brevity, only the difference range (dashed line), the 25\%-75\% interquartile interval and the median difference is provided. Fig.~\ref{fig:BoxplotPerOBJ} represents the inter-setup MOS differences per test hologram (for all test conditions), seperately for the center (a,b,c) and right-corner (d,e,f) views. Here, we consider the median difference between the MOS of each pair of setups as an indicator of the difference between the scores (shown as red line inside each box). The smaller the interquartile range for each boxplot, the higher the certainty on the difference-level (red-line) of the opinions for that object. For example, in Fig.~\ref{fig:BoxplotPerOBJ}.a, the MOS of the optical setup for all distorted versions of "Mermaid" is $\approx 0.75$ larger than the MOS for the light field setup. Considering the small size of the interquartile-range $\approx 0.25$, one may conclude that the shown difference almost equally persists across all the distortion types and distortion levels. When comparing Fig.\ref{fig:BoxplotPerOBJ}.(a,b,c), the general trend related to the MOS differences for the different setups (shown in Fig.~\ref{fig:allMOSc_OPT-LF-2D}), persists for each tested hologram individually. Thereby the MOS values between non-holographic displays are spread less compared to the MOS of the holographic display. The MOS results for "Chess" are the only results that show a rather stable behaviour across all setups.

\begin{figure*}
	\centering
	\subfloat[$(MOS_{OPTc} - MOS_{LFc})$]
	{\includegraphics[width=0.3\textwidth]{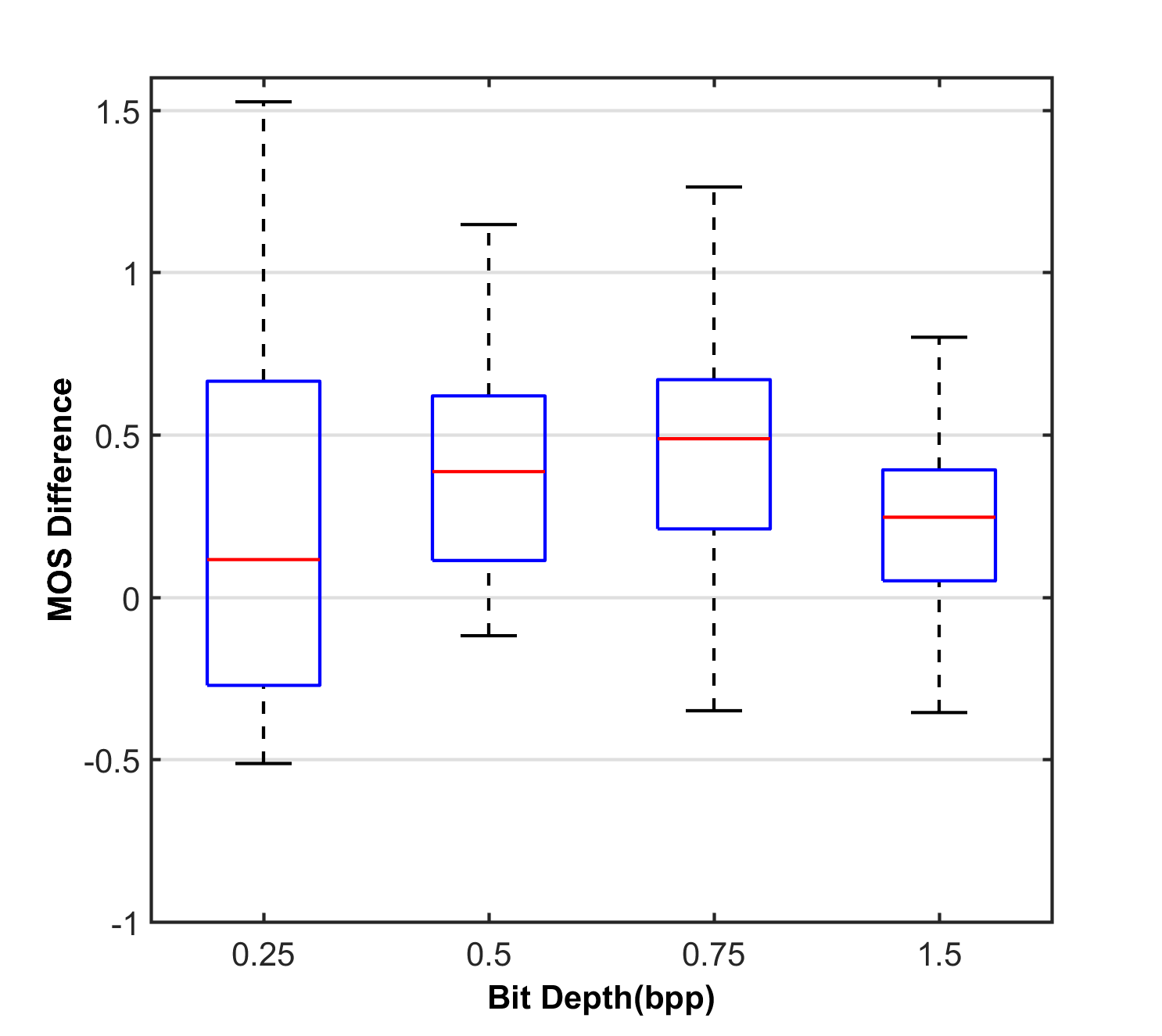}}
	\subfloat[$(MOS_{OPTc} - MOS_{2Dc})$]
	{\includegraphics[width=0.3\textwidth]{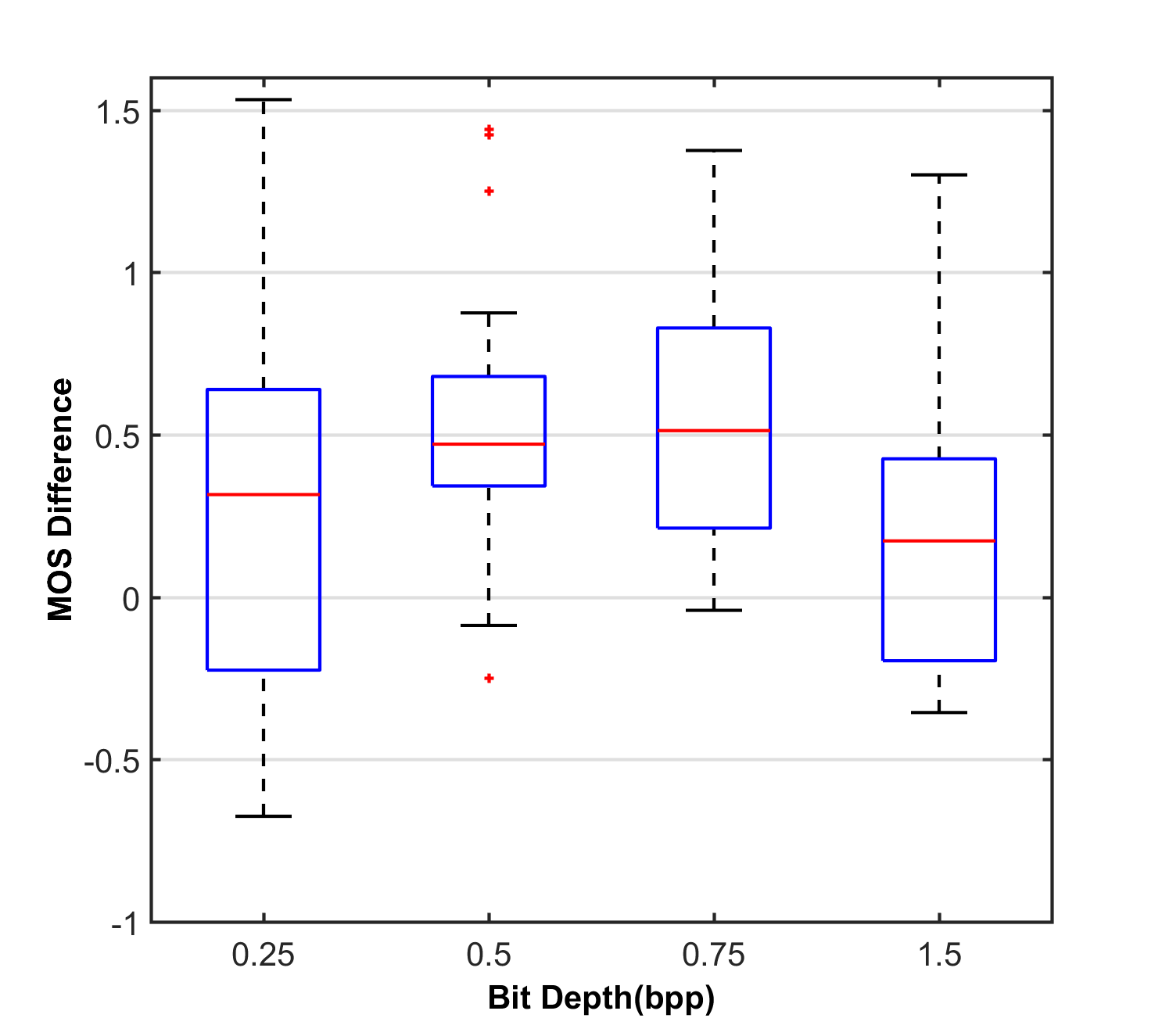}}
	\subfloat[$(MOS_{LFc} - MOS_{2Dc})$]
	{\includegraphics[width=0.3\textwidth]{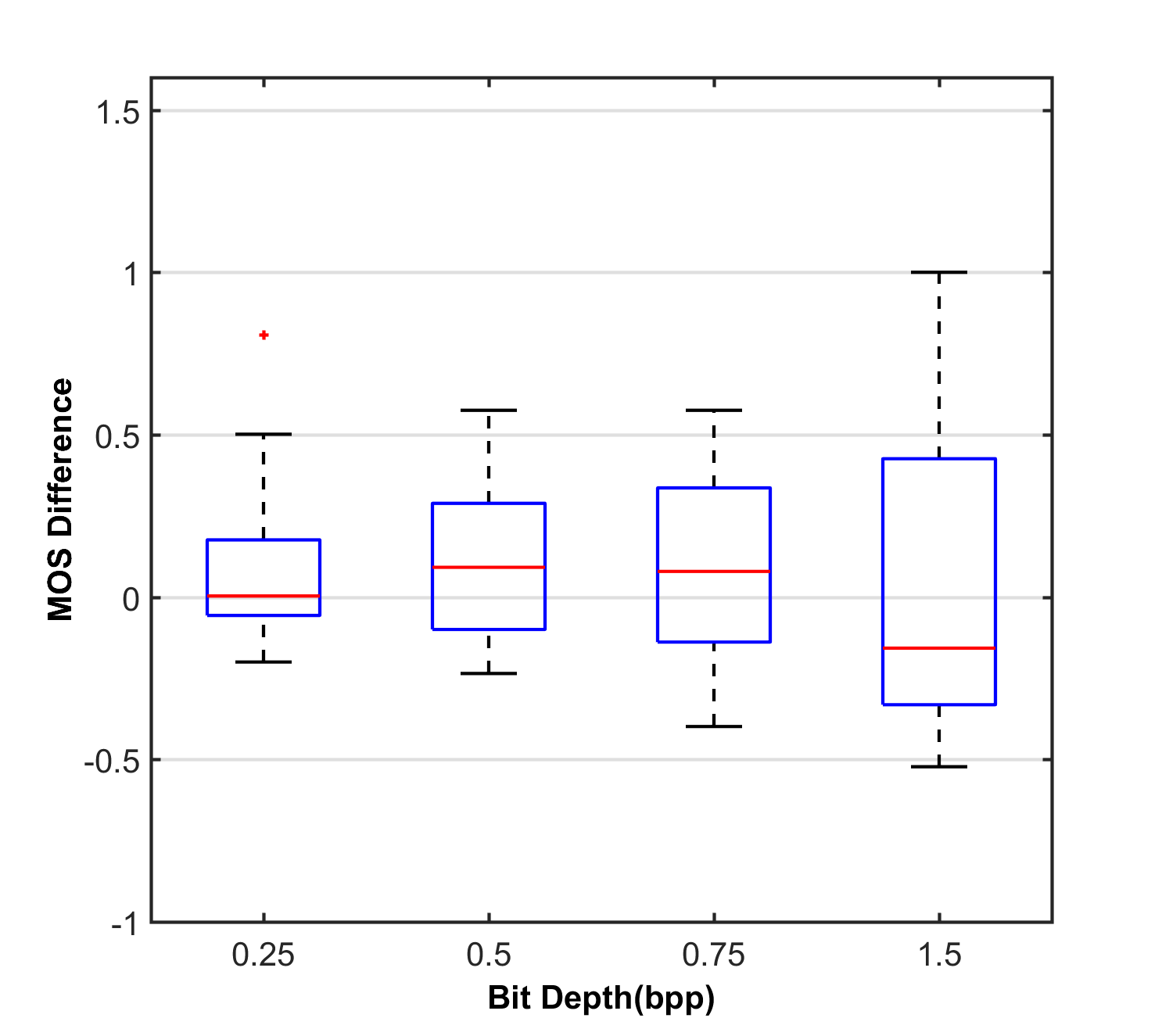}}
	\vspace*{0.1em}
	\subfloat[$(MOS_{OPTr} - MOS_{LFr})$]
	{\includegraphics[width=0.3\textwidth]{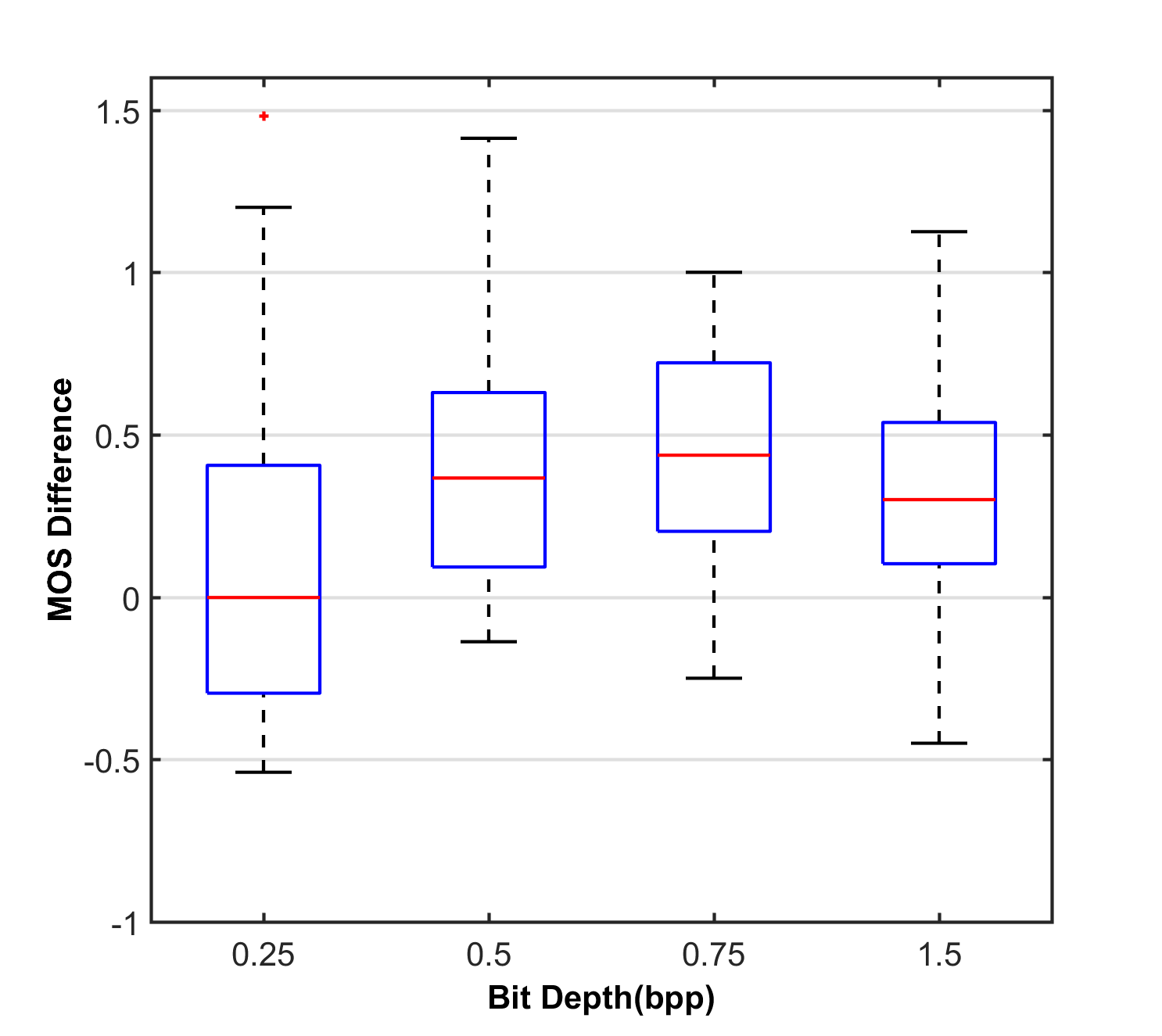}}
	\subfloat[$(MOS_{OPTr} - MOS_{2Dr})$]
	{\includegraphics[width=0.3\textwidth]{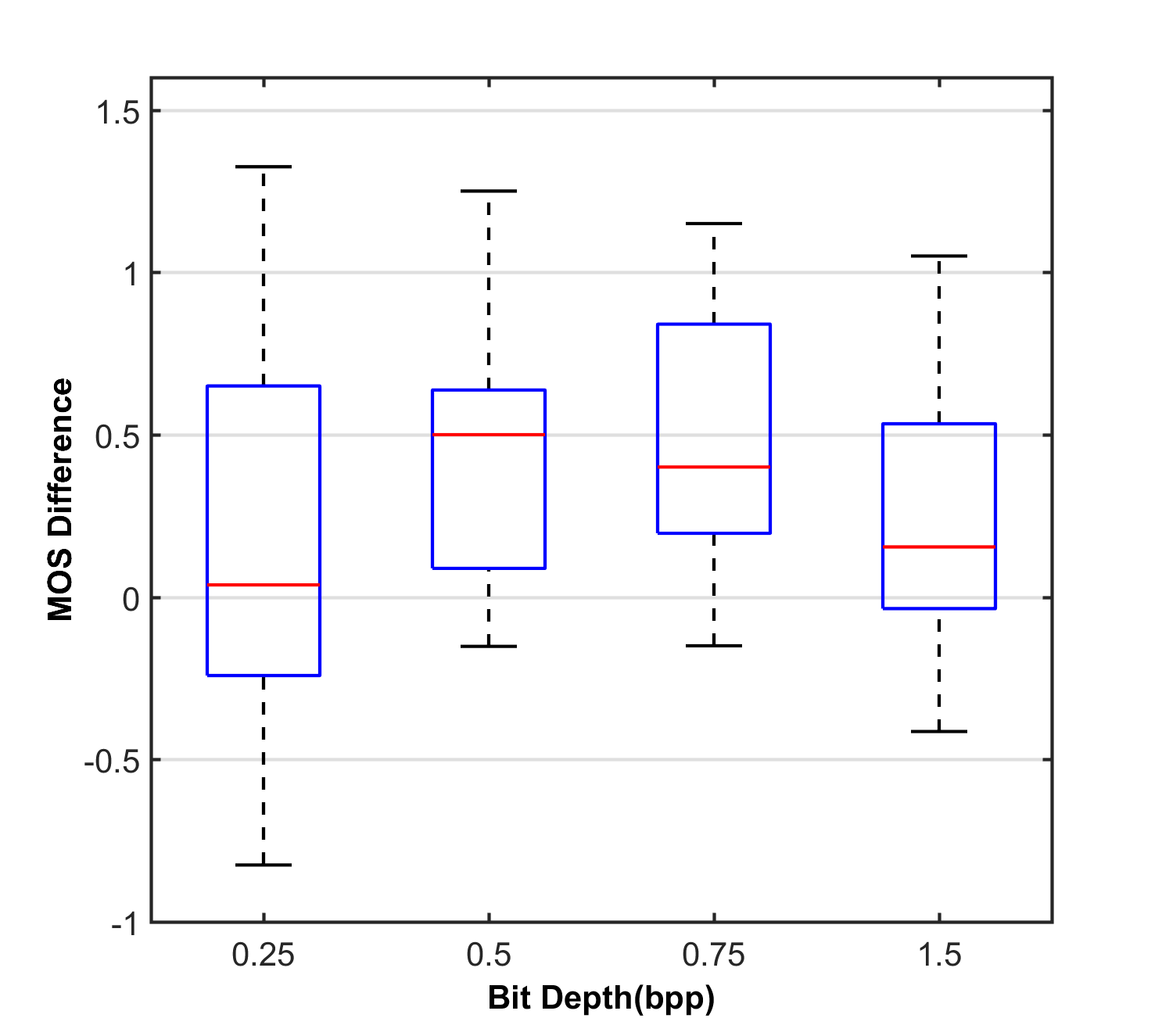}}
	\subfloat[$(MOS_{LFr} - MOS_{2Dr})$]
	{\includegraphics[width=0.3\textwidth]{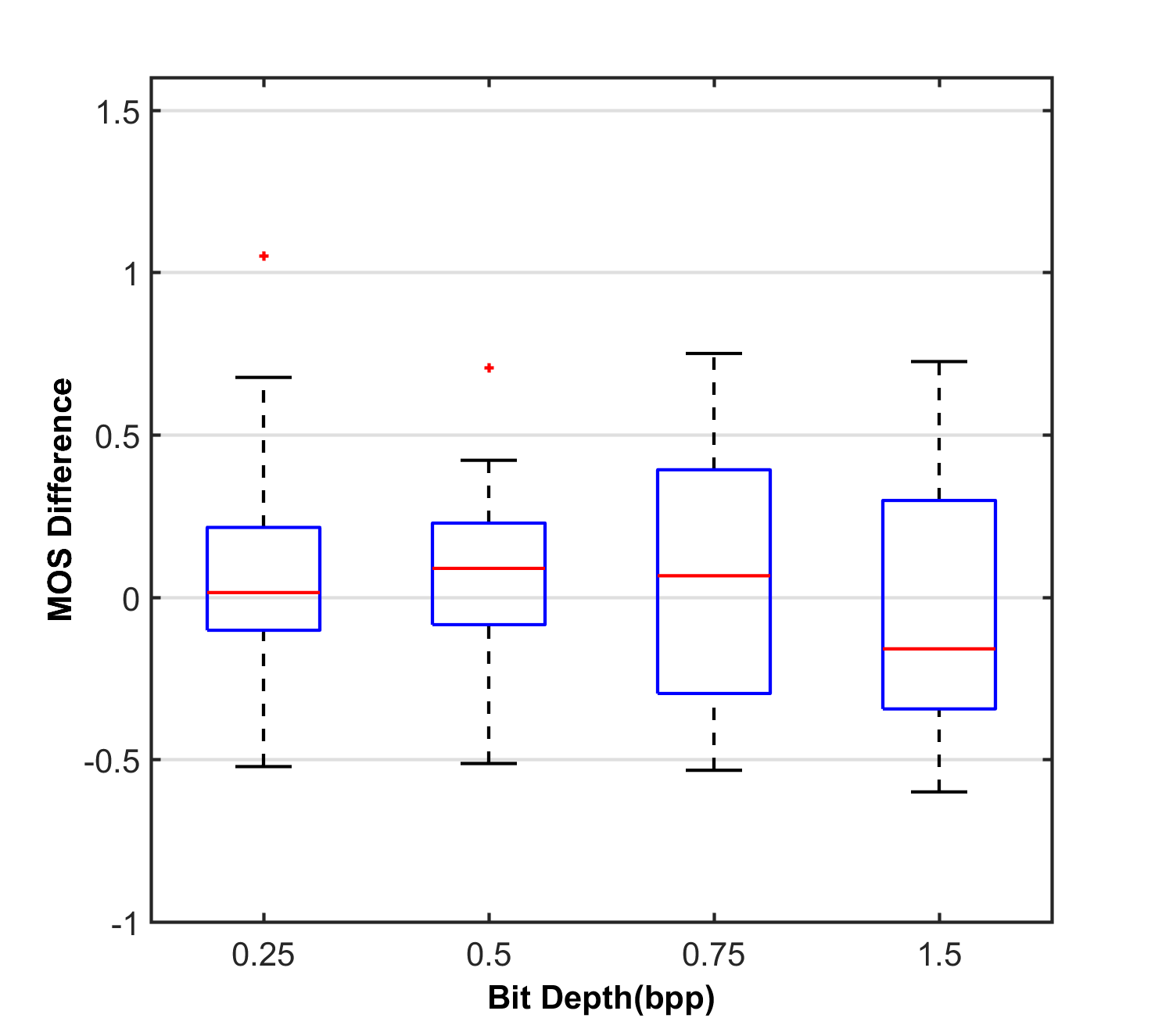}}
	
	\caption{Box plots of the MOS difference per compression level. The first and second rows shows the MOS differences for the center and right-corner view, respectively. }
	\label{fig:BoxplotPerBPP}%
	\vspace*{-1em}
\end{figure*}

 In Fig.~\ref{fig:BoxplotPerBPP}, another categorization of the MOS differences between the three setups and two perspectives is shown. Here, the MOS obtained from the holograms with the same compression level (quality range) are compared across setups. When considering the median differences (red lines), a rather similar trend is recognizable in Fig.~\ref{fig:BoxplotPerBPP} (a,b,d, and e), where the score gap between the holographic display and 2D or LF displays for the bit-depths 0.5 and 0.75 bpp are larger; while the level of disagreement is smaller for the lowest and the highest bit-depth. The certainty level of these results (referring to the size of the interquartile range) increases for higher bit-depths as well. In the case of a direct comparison of LF and 2D display the differences are statistically not relevant.   

\section{Conclusions}
\label{sec:conclusion}
In this paper, we reported the results of a series of comprehensive subjective experiments where, for the first time, a set of digital holograms was designed and created to evaluate various aspects of macroscopic holography. For each hologram 12 distorted versions were generated by compression at different bit-rates using state-of-the-art holographic encoders. Three separate subjective experiments were designed and implemented utilizing a holographic display, a light-field, and a 2D standard monitor. For the subjective tests, a double stimulus, multi-perspective, multi-depth subjective testing methodology was designed and adapted to the characteristics of the utilized displays. 
A total of 120 human subjects participated in the experiments. The acquired quality scores of the reconstructed holograms were compared based on perspective and focal distance. Our results showed no explicit distinction between the scores of holograms when different parts of the encoded objects were in focus. However, with a change of perspective there was a consistent gap between the rated visual qualities. The corner-view generally scored lower than the center-view, especially for mid-range and high distortion levels. 
Further, we compared the scores obtained from a holographic display with the scores from the Light-field and 2D displays and identified another rather consistent and distinctive gap. Our results show that the same distorted holograms rendered on holographic displays appear less distorted to the human eye than it is the case for light-field or 2D display. However, it was demonstrated that the scores on different displays are highly correlated and follow a consistent trend through the quality range. This indicates that numerically reconstructed holograms displayed on light field or 2D displays allow for appropriate predictions on the  perceptual visual quality of holographic displays. For completeness, we also provided fit-functions which map scores from different setups into one another. 
Finally, we are hoping that the provided results plus the annotated database of our holograms, which are publicly available, facilitate a reliable test-bed for designing or improving available holographic processing methods and plenoptic quality metrics, as well as systematic benchmarking operations for digital holograms.


%
%
%
%
%

\section*{Acknowledgment}
Research for this paper received funding from the European Research Council under the European Union’s Seventh Framework Programme (FP7/2007-2013)/ERC Grant Agreement n.617779 (INTERFERE), the Cross-Ministry Giga KOREA Project (GK19D0100, GigaKOREA) and Warsaw University of Technology.

The models “Perforated Ball“ and “Biplane” are courtesy of Manuel Pi\~neiro from GrabCad.com and ScanLAB Projects, respectively.

\ifCLASSOPTIONcaptionsoff
  \newpage
\fi

\bibliographystyle{IEEEtran}
\bibliography{refs}



\begin{IEEEbiography}[{\includegraphics[width=1in,height=1.25in,clip,keepaspectratio]{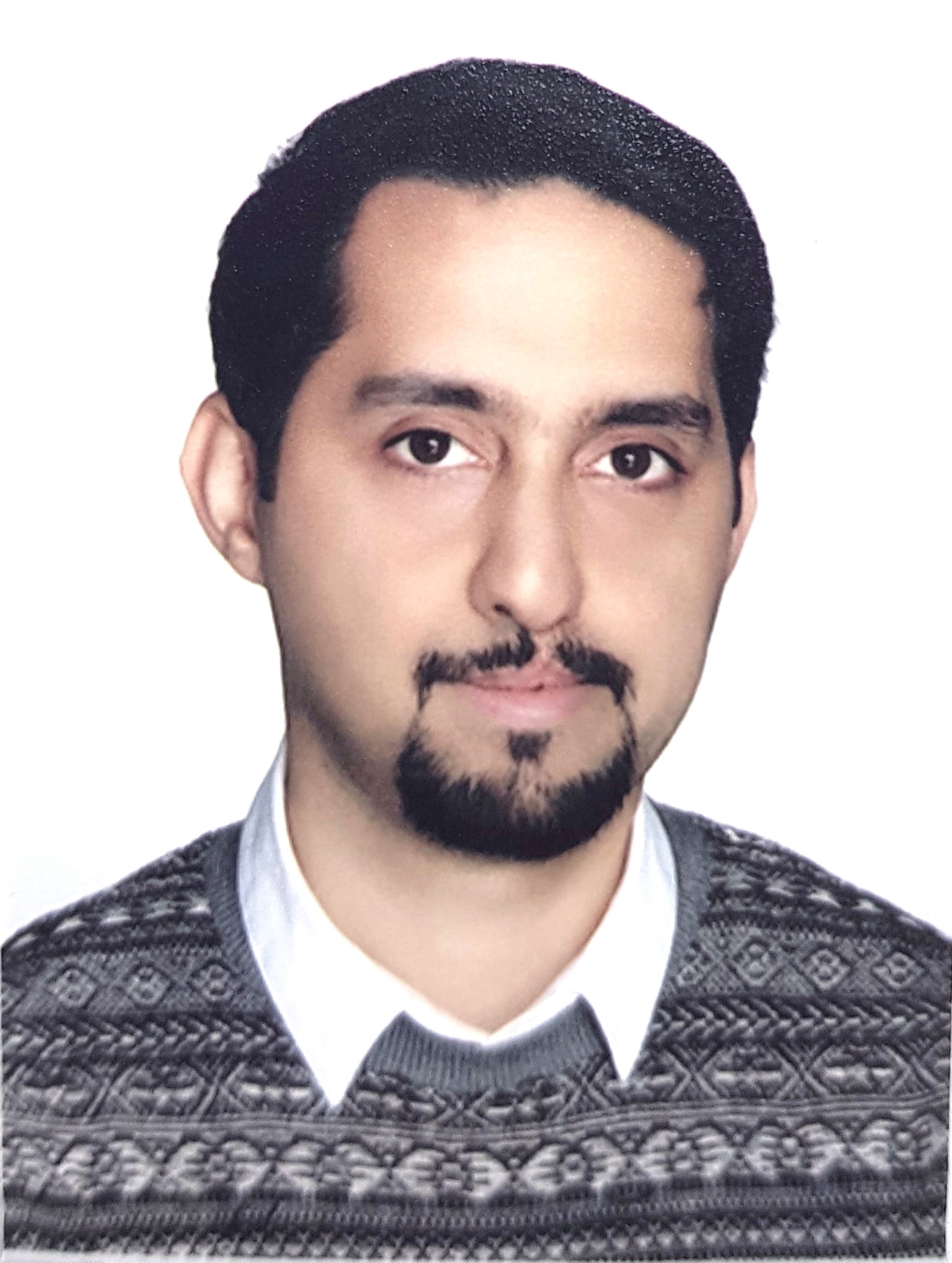}}]{Ayyoub Ahar}
received his B.Sc. degree in applied physics from PNU and the M.Sc. degree in IT-multimedia computing from Multimedia University, Malaysia. He is currently pursuing the Ph.D. degree in electrical engineering with Vrije Universiteit Brussel (VUB), Belgium. He is a member of the INTERFERE, a multidisciplinary research
group at the Department of Electronics and Informatics, VUB. He holds a full scholarship as a part of EU-ERC Consolidator Grant focusing on
digital holography. He is also a Researcher at IMEC, Leuven, Belgium. His current research interests include digital image and video processing, complex data analysis, bio-inspired computing, perceptual
quality assessment and standardization with an emphasis on emerging 3D data modalities, in particular digital holography and light field imaging.
\end{IEEEbiography}

\begin{IEEEbiography}[{\includegraphics[width=1in,height=1.25in,clip,keepaspectratio]{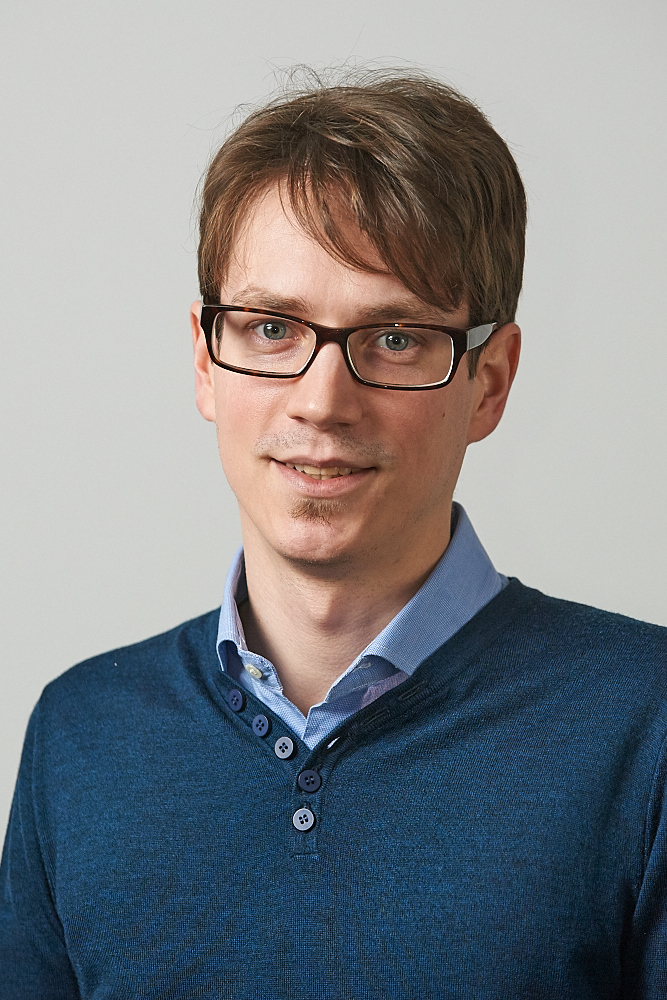}}]{Maksymilian Chlipala}
	 received his PhD degree in area of construction and exploitation of machines from the Faculty of Mechatronics, Warsaw University of Technology, Warsaw, Poland, in 2019. His main research interests are digital holography, holographic displays, Spatial Light Modulators and speckle reduction with LED sources.
\end{IEEEbiography}

\begin{IEEEbiography}[{\includegraphics[width=1in,height=1.25in,clip,keepaspectratio]{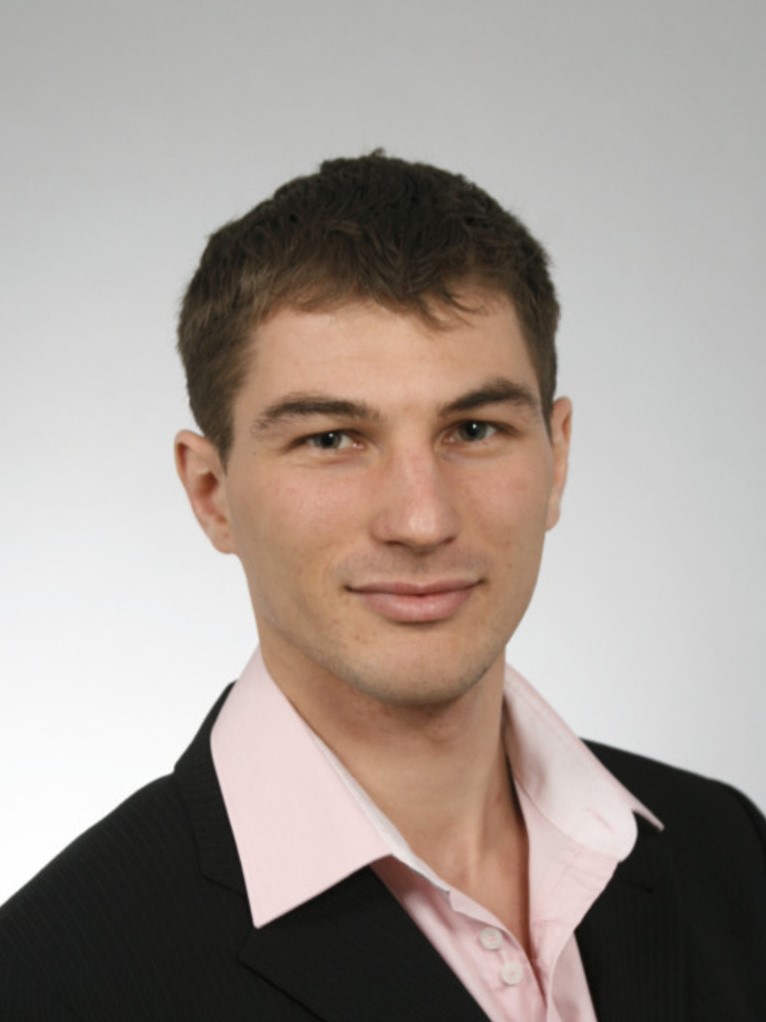}}]{Tobias Birnbaum}
	received his B.Sc. degree in Applied Natural Science (2012) and his Dipl. in Applied Mathemathics (2016) from TU Bergakademie Freiberg, Germany. He is currently working towards a PhD in the field of signal processing of digital holograms at Vrije Universiteit Brussels, Belgium as part of the INTERFERE EU465 project. His main research interests are compression and  post-processing, such as denoising, of digital holograms, as well as compressed sensing, and dictionary learning.
\end{IEEEbiography}

\begin{IEEEbiography}[{\includegraphics[width=1in,height=1.25in,clip,keepaspectratio]{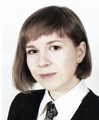}}]{Weronika Zaperty}
	received the M.Sc. degree from Warsaw University of Technology, Warsaw, Poland in 2011, where she is currently working toward the Ph.D. degree in the field of color digital holography. Her scientific research are related to digital
holography, holographic displays and holographic
interferometry.
\end{IEEEbiography}

\begin{IEEEbiography}[{\includegraphics[width=1in,height=1.25in,clip,keepaspectratio]{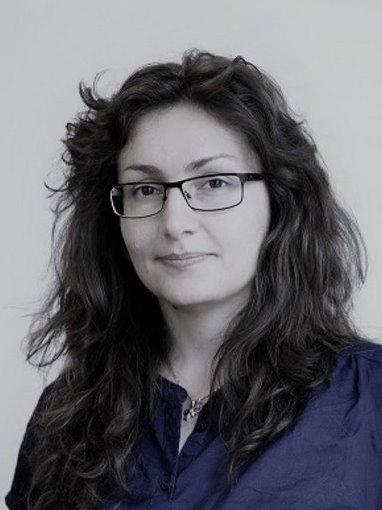}}]{Athanasia Symeonidou}
 received a B.Sc. degree in Physics in 2009 and a M.Sc. degree in Electronic Physics in 2012 from Aristotle University of Thessaloniki, Greece. She currently works towards her Ph.D. in Engineering Sciences at Vrije Universiteit Brussel, Belgium, as a researcher on the ERC project INTERFERE. Her research interests are 3D imaging, digital holography and signal processing, focusing on efficient algorithms for generation and display of high quality computer-generated holograms.
 \end{IEEEbiography}

\begin{IEEEbiography}[{\includegraphics[width=1in,height=1.25in,clip,keepaspectratio]{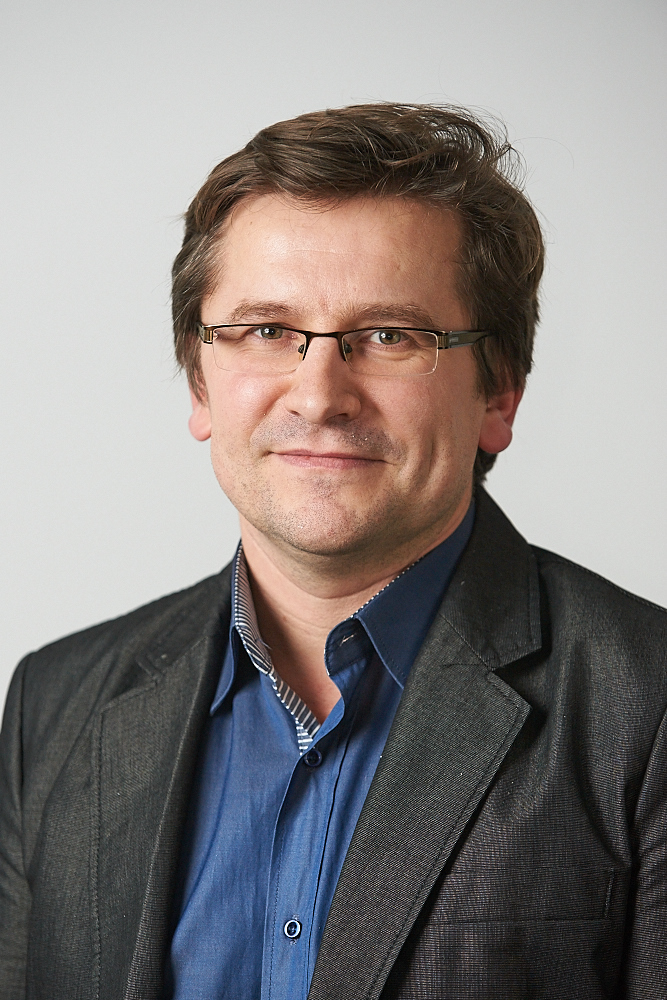}}]{Tomasz Kozacki}
received his Ph.D. in the field of Photonics at Warsaw University of Technology in 2005 and habilitation in 2013, where he also works as a professor. His scientific research are related to digital holography, holographic displays, holographic microscopy, computational diffraction and optical diffraction tomography. He has authored and co-authored more than 30 scientific journal publications and more than 50 conference proceedings.
\end{IEEEbiography}

\begin{IEEEbiography}[{\includegraphics[width=1in,height=1.25in,clip,keepaspectratio]{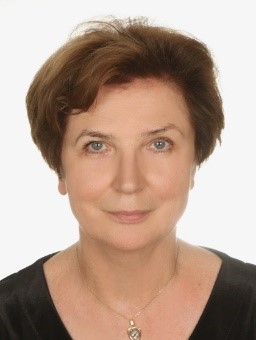}}]{Malgorzata Kujawinska}
PhD DSc., SPIE Fellow, Full Professor of applied optics at Warsaw University of Technology. Expert in full-field optical metrology,  development of novel photonics measurement systems and holographic displays, 3D quantitative imaging in multimedia and biomedical engineering. Author of one monograph, several book chapters and more than 250 papers in international scientific journals. She had been the  SPIE President and vice-President of European Technology Platform Photonics21. The recipient of SPIE 2013 Chandra S. Vikram Award in Optical Metrology.
\end{IEEEbiography}

\begin{IEEEbiography}[{\includegraphics[width=1in,height=1.25in,clip,keepaspectratio]{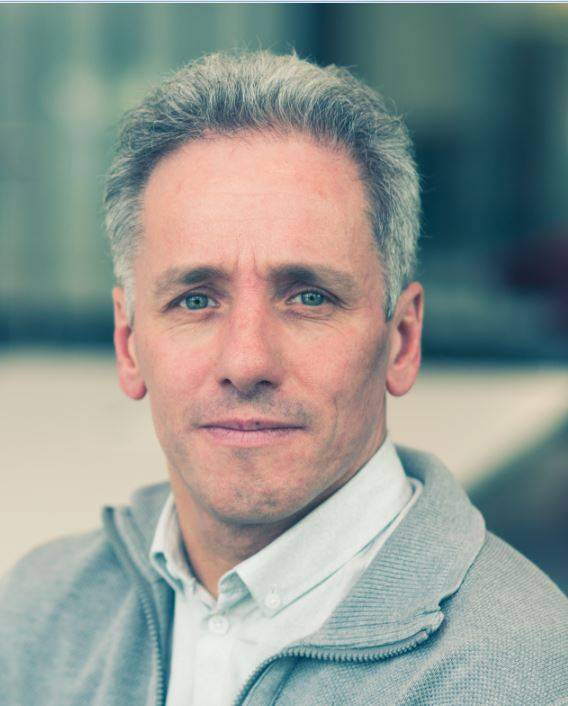}}]{Peter Schelkens}
	holds a professorship with the Department of Electronics and Informatics, Vrije Universiteit Brussel, Belgium and is a Research Group Leader at imec, Belgium. In 2013, he received an EU ERC Consolidator Grant focusing on digital holography. He is the co-editor of the books The JPEG 2000 Suite (Wiley, 2009) and Optical and Digital Image Processing (Wiley, 2011). He is chair of the Coding, Test and Quality subgroup of the ISO/IEC JTC1/ SC29/WG1 (JPEG) standardization committee and Associate Editor for the IEEE Transactions on Circuits and Systems for Video Technology, Signal Processing: Image Communications and JPhys Photonics.
\end{IEEEbiography}





\end{document}